\documentclass[12pt]{amsart}
\usepackage{graphicx}
\usepackage{fourier}
\usepackage{cancel}
\usepackage{verbatim} 
\usepackage{epsfig}
\usepackage{color}

\newtheorem{thm}{Theorem}[section]

\newtheorem{remark}[thm]{Remark}

\usepackage{amsmath}
\usepackage{amsxtra}
\usepackage{amscd}
\usepackage{amsthm}
\usepackage{amsfonts}
\usepackage{amssymb}
\usepackage{eucal}
\newcommand{\bmb}{\left( \begin{array}{rr}}
\newcommand{\enm}{\end{array}\right)}

\newcommand{\C}{{\mathbb C}}

\newcommand{\al}{{\alpha}}

\numberwithin{equation}{section}

\usepackage{tikz}
\usetikzlibrary{patterns}
\usetikzlibrary{calc,arrows,positioning}
\usetikzlibrary{arrows,shadows}
\usetikzlibrary{decorations.pathmorphing}	
\usetikzlibrary{decorations.markings}

\usepackage{pgfplots}

\begin{document}

\title[Arctic Curves of the Twenty-Vertex Model with Domain Wall Boundaries]{Arctic Curves of the Twenty-Vertex Model with Domain Wall Boundaries}
\author{Bryan Debin}
\address{Institut de Recherche en Math\'ematique et Physique,
Universit\'e catholique de Louvain, Louvain-la-Neuve, B-1348, Belgium\hfill
\break e-mail:  bryan.debin@uclouvain.be
}
\author{Philippe Di Francesco} 
\address{
Department of Mathematics, University of Illinois, Urbana, IL 61821, U.S.A. 
and 
Institut de physique th\'eorique, Universit\'e Paris Saclay, 
CEA, CNRS, F-91191 Gif-sur-Yvette, FRANCE\hfill
\break  e-mail: philippe@illinois.edu
}
\author{Emmanuel Guitter}
\address{
Institut de physique th\'eorique, Universit\'e Paris Saclay, 
CEA, CNRS, F-91191 Gif-sur-Yvette, FRANCE
\break  e-mail: emmanuel.guitter@ipht.fr
}

\begin{abstract}
We use the tangent method to compute the arctic curve of the Twenty-Vertex (20V) model with particular domain wall boundary conditions
for a wide set of integrable weights. To this end, we extend to the finite geometry of domain wall boundary conditions the standard 
connection between the bulk 20V and 6V models via the Kagome lattice ice model. This allows to express refined
partition functions of the 20V model in terms of their 6V counterparts, leading to explicit parametric expressions for the various portions
of its arctic curve. The latter displays a large variety of shapes depending on the weights and separates a central liquid phase 
from up to six different frozen phases. A number of numerical simulations are also presented,
which highlight the arctic curve phenomenon and corroborate perfectly the analytic predictions of the tangent method.
We finally compute the arctic curve of the Quarter-turn symmetric Holey Aztec Domino Tiling (QTHADT) model, a problem 
closely related to the 20V model and whose asymptotics may be analyzed via a similar tangent method approach.
Again results for the QTHADT model are found to be in perfect agreement with our numerical simulations.
\end{abstract}

\maketitle
\date{\today}
\tableofcontents

\section{Introduction}
\label{sec:introduction}

\subsection{The 20V model with Domain Wall Boundary Conditions}
\label{sec:defmodel}
\begin{figure}
\begin{center}
\includegraphics[width=15cm]{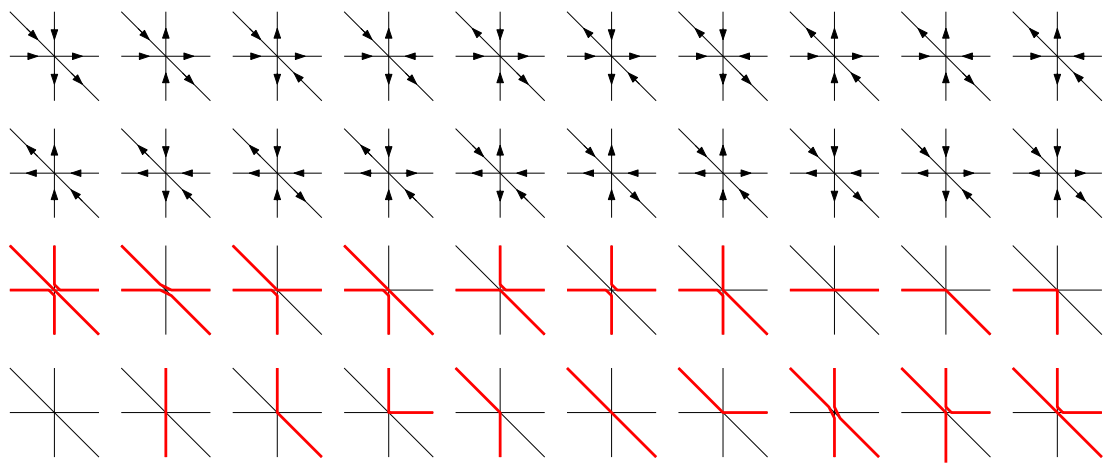}
\end{center}
\caption{\small Top rows: the $20$ possible environments satisfying the ice rule at a node of the triangular lattice.  Bottom rows: the $20$ equivalent 
osculating path configurations.}
\label{fig:twentyV}
\end{figure}

The present paper deals  with the so called \emph{Twenty-Vertex (20V) model} \cite{Kel,Baxter}, an alternative denomination for
the \emph{ice model on the regular triangular lattice}. Recall that the ice model is defined by assigning to each edge of the lattice an orientation satisfying 
the so-called ``ice rule'' that \emph{each node is incident to as many ingoing as outgoing edges}.
In the case of the triangular lattice, the ice rule gives rise to 20 possible environments around a given node, displayed in Figure~\ref{fig:twentyV}, hence the 
alternative denomination of the model. For convenience, we represent the triangular lattice as a square lattice supplemented with a second diagonal within each face. Instead of using edge orientations, we may
alternatively represent the ice model configurations by ``osculating paths'' taking steps along the lattice edges. Paths are obtained by drawing a path step whenever the underlying edge orientation runs 
from North, Northwest or West to East, Southeast or South. These steps are then uniquely concatenated at each node into properly oriented non-crossing but possibly kissing or osculating 
paths, as shown in Figure~\ref{fig:twentyV},
where the underlying orientation may be erased without loss of information. In all generality, configurations of the 20V model are enumerated with Boltzmann weights attached to each node of the lattice,
according to its local environment: the model therefore involves {\it a priori} the data of 20 possible local weights for the 20 possible vertices.

\begin{figure}
\begin{center}
\includegraphics[width=11cm]{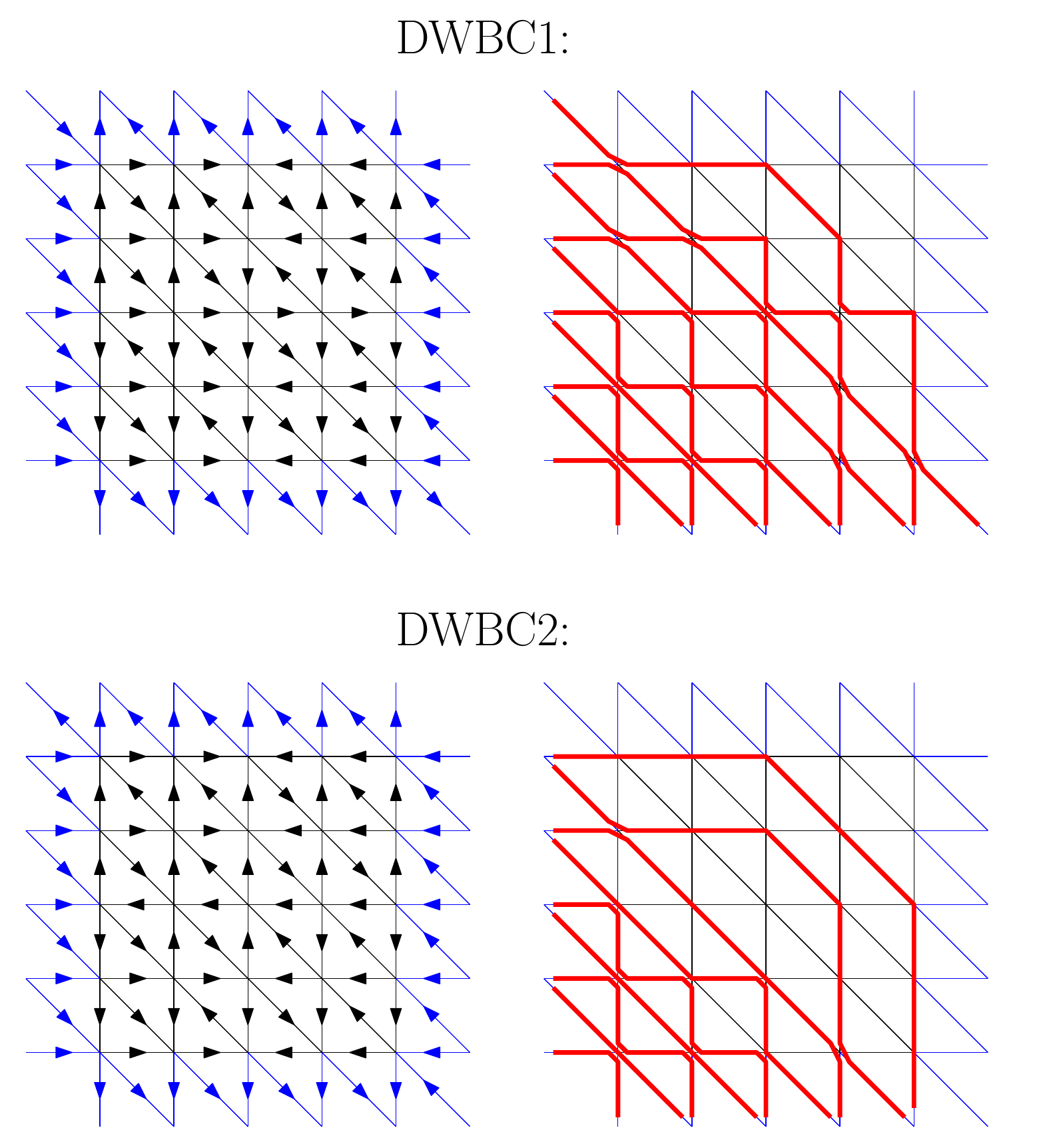}
\end{center}
\caption{\small Two allowed 20V model configurations with DWBC1 (top) or DWBC2 (bottom). In each case the left picture uses the representation in terms of oriented edges and the right one the osculating path representation. The arrows drawn in blue are fixed.}
\label{fig:DWBC}
\end{figure}

So far, most of the results on the 20V model concern its bulk properties, corresponding to local properties of the model defined on the \emph{infinite}
triangular lattice \cite{Kel,Baxter}. For instance, the bulk phase diagram of the model (with restricted values of the vertex weights) was 
established in Ref.~\cite{Kel} while the bulk entropy of the model was obtained in \cite{Baxter}. Here, following \cite{DFG20V},
we consider instead 
the 20V model defined on a \emph{finite} domain of the triangular lattice, with appropriate boundary conditions for which
exact enumeration results may be obtained. At large size and upon rescaling, a sensible limit can be reached, which describes the continuous
behavior of the model in finite geometry. 
Our study concerns more specifically 20V model configurations with Domain Wall Boundary Conditions (DWBC), as defined in \cite{DFG20V}.
The model is defined on an $n\times n$ square portion of 
the square lattice, with nodes at integer coordinates $(i,j)$ for $i,j=1,2,\dots,n$, and edges along all the elementary horizontal segments $(i,j)\to (i+1,j)$ (for $0<i<n$), all
the elementary vertical segments $(i,j)\to (i,j+1)$ (for $0<j<n$), and all the elementary second diagonals $(i,j+1)\to (i+1,j)$ (for $0<i,j<n$). 
The internal edge set is completed by boundary edges with \emph{fixed orientations} according to either of the following two DWBC prescriptions (see Figure~\ref{fig:DWBC}):
\begin{itemize}
\item{for DWBC1 (see Figure~\ref{fig:DWBC}-top), all the (horizontal or diagonal) edges of the left and right boundaries except the (diagonal) lower right one are oriented 
towards the central square, while all the (vertical or diagonal) edges of the top and bottom boundaries except the (diagonal) upper left one are oriented 
away from the central square;}
\item{for DWBC2 (see Figure~\ref{fig:DWBC}-bottom), all the (horizontal or diagonal) edges of the left and right boundaries except the (diagonal) upper left one are oriented 
towards the central square, while all the (vertical or diagonal) edges of the top and bottom boundaries except the (diagonal) lower right one are oriented 
away from the central square.}
\end{itemize}  
Note that DWBC1 and 2 differ only by the orientations of the upper right and lower left diagonal edges, which are opposite in the two settings. As a consequence, 
the configurations of the 20V model with DWBC1 are in one-to-one correspondence with those of the 20V model with DWBC2 
by a simple $180^\circ$ rotation. For instance, the configurations depicted in Figure~\ref{fig:DWBC}-left are image of each other under this rotation.
As a consequence, in the particular case of vertex weights invariant under $180^\circ$ rotation, the two models have the same partition function. In the following,
we will always be in such a situation.

In the alternative osculating path language, our boundary conditions correspond to having all the edges of the left and bottom boundaries occupied
by a path step (with
the upper left and lower right included for DWBC1, excluded for DWBC2), and all the other boundary edges unoccupied.
Note that performing a $180^\circ$ rotation on the edge orientations amounts in the path language to performing both a $180^\circ$ rotation
and then a complementation in which occupied and unoccupied edges are interchanged. 

\subsection{The arctic curve and the tangent method: generalities}
\label{sec:tgmethod}
The 20V model with DWBC1 or 2 is the analog on the triangular lattice of the celebrated 6V model with DWBC, defined 
on an $n\times n$ square portion of the square lattice. For appropriate vertex weights, this latter model is known to exhibit the so-called
 {\it arctic curve phenomenon} \cite{CEP,JPS}:
in the limit of large $n$ (and after rescaling of the coordinates by $1/n$), a typical configuration presents a sharp phase separation between a number of ``frozen'' phases 
adjacent to the square boundaries and in which the node environments are fixed, and a ``liquid'' disordered phase in the center, with
fluctuating node environments. We expect our  20V model with DWBC1 or 2 to exhibit the same phenomenon: frozen phases
where \emph{all nodes have the same environment} should exist in the vicinity of the boundaries, separated by a well-defined 
\emph{arctic curve} from a central liquid region. In the path language, the frozen regions may be empty of all paths or, on the contrary,
maximally filled with all the edges occupied, or also regions with only vertical (resp. horizontal) occupied edges, etc...

The purpose of this paper is to get an explicit expression for the location of the arctic curve of the  20V model with DWBC1 or 2,
with possibly some non-trivial weights attached to the twenty different vertices, and to identify the nature of the
surrounding frozen phases. A number of methods were developed to locate the arctic curve for non-intersecting or osculating 
path problems, usually in the equivalent dimer or tiling language: these methods include the asymptotic study of bulk expectation values 
via the technique of the Kasteleyn operator \cite{KO1,KO2,KOS}, or the machinery of cluster integrable systems of dimers \cite{DFSG,KP}.
Here we will instead recourse to so-called \emph{tangent method} invented by Colomo and Sportiello \cite{COSPO} whose 
implementation is as follows: one of the portion of the arctic curve consists in the separation line between the liquid phase and the ``empty'' region, 
i.e. a region not visited by any path. Clearly, at the microscopic level of the paths, this limit corresponds to the trajectory of the uppermost path.
To get the most likely location of this trajectory, the idea is to force this uppermost path to exit the original $n\times n$ square domain
at some ``escape'' point $A$ along the right boundary by sliding its original endpoint along the horizontal axis to some new distant point $B$ lying to the right of the original square domain (see Figure~\ref{fig:tgmethod1} for an illustration in the case of the 20V model). From $A$ to $B$, the outermost path follows (in the continuous limit of large $n$ and after rescaling) a straight line since the visited region is empty of any other path.
For a fixed endpoint $B$, the most likely position of $A$ is such that the line $(AB)$ is tangent to the original trajectory, hence
to the arctic curve. Indeed the new trajectory of the uppermost path is expected to first follow its original trajectory (due to its steric interaction with 
the other paths), hence to follow the arctic curve until $B$ is in its line of sight, and from then on quit the arctic curve tangentially\footnote{
The tangency property was proved in \cite{DGR} in the case of non-intersecting path models by a convexity argument.}
 in order to attain $B$
via a straight line (since, from this point, the trajectory takes place in a region empty of any other path) passing through the escape point 
$A$. By changing 
the position of $B$ and computing the corresponding most likely escape point $A$, we get
a family of tangent lines whose envelope is a portion of the arctic curve. Other portions of the curve are then obtained by the same technique upon 
using other equivalent path representations of the model. Even though not fully proved at this stage except in a few cases \cite{Agg}, the tangent method
was tested successfully in various models \cite{COSPO,DFLAP,DFGUI,DR,DFG2,DFG3} and has led to a number of new predictions as it is 
quite easy to implement. 

In this paper, we will consider exclusively the 20V model with DWBC1 or 2 with attached vertex weights which are \emph{invariant 
under $180^\circ$ rotation} (in the oriented edge language) around any node. Then, since at the global level,
this transformation interchanges the DWBC1 and DWBC2 prescriptions, the arctic curve of the 20V model with DWBC1
is the image under $180^\circ$ rotation of the arctic curve of the 20V model with DWBC2. Moreover, the two models differ only
by the presence of one more path for the DWBC1 prescription, starting at the upper left and ending at the lower right diagonal edge. 
Even though this path is then precisely the outermost path which probes the location of the arctic curve, we expect its trajectory to
be undistinguishable from that of the path just below in the continuous limit, itself undistinguishable from the trajectory of the outermost
path for the DWBC2 prescription. Otherwise stated, the boundary difference between the two prescriptions is irrelevant in the
continuous limit, so that both lead to the \emph{same arctic curve}\footnote{This argument holds strictly speaking only for the portion of arctic curve delimiting the empty frozen region but can be repeated for the other portions
by use of the appropriate alternative path description.}.
From the discussion above and for vertex weights invariant under $180^\circ$ rotation, we deduce that \emph{the arctic curve itself is symmetric under $180^\circ$ rotation}.

\subsection{Plan of the paper}
The paper is organized as follows: 
Section~\ref{sec:20V6V} explains the connection between the 20V model and the 6V model: first, 
following \cite{Baxter}, we recall some correspondence between the 20V model and a triple of 6V models on sublattices 
of the Kagome lattice (Section~\ref{sec:Kagome}) for some appropriate choice of the vertex weights. 
In practice, this holds for a specific set of \emph{integrable vertex weights} parametrized by one quantum and three spectral parameters. 
For the particular case of DWBC2, we then use the unraveling procedure of \cite{DFG20V} to 
obtain a direct correspondence  between the 20V model and a single, properly weighted, 6V model with DWBC (Section~\ref{sec:square},
Theorem \ref{thm:first}).

This correspondence is refined in Section \ref{sec:onepoint} which deals with so-called ``one-point functions'', which are generating functions
keeping track of the position of the point where the uppermost path hits the right boundary (this point will become the escape point $A$ in the tangent method geometry). A relation between the one-point function of the 20V model with DWBC2 and that of the 6V model with DWBC,
generalizing that of \cite{DFG20V}, is given in 
Section \ref{sec:refined} (Theorem \ref{thm:second}) and proved in Appendix~\ref{sec:6V20Vgeneralrel}. This relation is then used in Section~\ref{sec:onepointasymp} to obtain
the large $n$ asymptotics of the 20V model one-point function, a crucial ingredient in the computation of the arctic curve.

We then discuss in detail in Section~\ref{sec:warmup} the implementation of the tangent method in a simple case where the vertex weights depend
on a single quantum parameter. A first portion of arctic curve, called the ``normal'' portion, is obtained in Section~\ref{sec:normalbranch} (Theorem~\ref{thm:acnormal}) by
a direct application of the recipe described in Section~\ref{sec:tgmethod} and illustrated in Figure~\ref{fig:tgmethod1}, i.e., after moving the endpoint of the uppermost path to some distant positions $B$, (i) finding the 
most likely position of $A$ of the escape point, (ii) getting the equation of the tangent lines $(AB)$ and (iii) deducing the location of their envelope.
This involves computing the partition function of a single path (the escaping part of the uppermost path from $A$ to $B$) with general 20V weights,
using a general transfer matrix formalism detailed in Appendix~\ref{sec:trmat}. 
To get a second portion of arctic curve (Theorem~\ref{thm:acshear}), we recourse to an alternative set of paths describing the 20V model configurations.
Remarkably, as explained in Section~\ref{sec:shearbranch}, a shear transformations maps these new path configurations into 
those of some ``inverted'' 20V model in a modified geometry where the tangent method can still be applied and gives rise to 
 a new ``shear'' portion. The remaining portions of the arctic curve are deduced by 
symmetry arguments. 

This approach is then extended in Section~\ref{sec:general} to the general case where the weights depend on all
parameters, leading the main result of this paper in the form of Theorem~\ref{thm:acgeneral}, which gives a complete description of the arctic curve, made of three portions and their symmetric
counterparts under $180^\circ$ rotation. 
Sections~\ref{sec:bone} and \ref{sec:btwo} are devoted to the computation of the analogue of the ``normal'' and ``shear'' portions in this more general weighting. Section 5.3 presents the computation of a new ``final'' portion obtained along the same lines as the ``shear'' portion after exchanging the role of vertical and horizontal directions. Section~\ref{sec:phases} discusses the nature of the various frozen phases and illustrates our results on the arctic curve by a number
of explicit plots. 

Section~\ref{sec:simul} presents numerical simulations for large typical 20V model configurations. Those are obtained
by a Markov-chain process described in Section~\ref{sec:nummethod}. The resulting patterns are represented in Section~\ref{sec:numresults}
for various values of the parameters and display a perfect agreement with the tangent method results. 

In \cite{DFG20V}, it was shown
that a correspondence exists between the 20V model with DWBC1 or 2 and uniform weights, and a particular domino tiling problem:
the Quarter Turn symmetric Holey Aztec Domino Tiling (QTHADT). We analyze the arctic curve of this model in Section~\ref{sec:QTHADT}.
Its partition function is obtained in Section~\ref{sec:PF} and refined in Section~\ref{sec:RPF}. The later is identified with
a suitably refined 6V partition function in Section~\ref{sec:QTHADT6V}, leading to an explicit arctic curve via the tangent method 
(Section~\ref{sec:QTHADTtg}, Theorem~\ref{thm:acqt}). Again, these results are in perfect agreement with numerical simulations presented in Section~\ref{sec:QTHADTsimul}.

We gather a few concluding remarks in Section~\ref{sec:conclusion}.

\section{The 20V/6V model correspondence: partition functions}
\label{sec:20V6V}

\subsection{From the 20V model to three 6V models on the Kagome lattice} 
\label{sec:Kagome}
\begin{figure}
\begin{center}
\includegraphics[width=13cm]{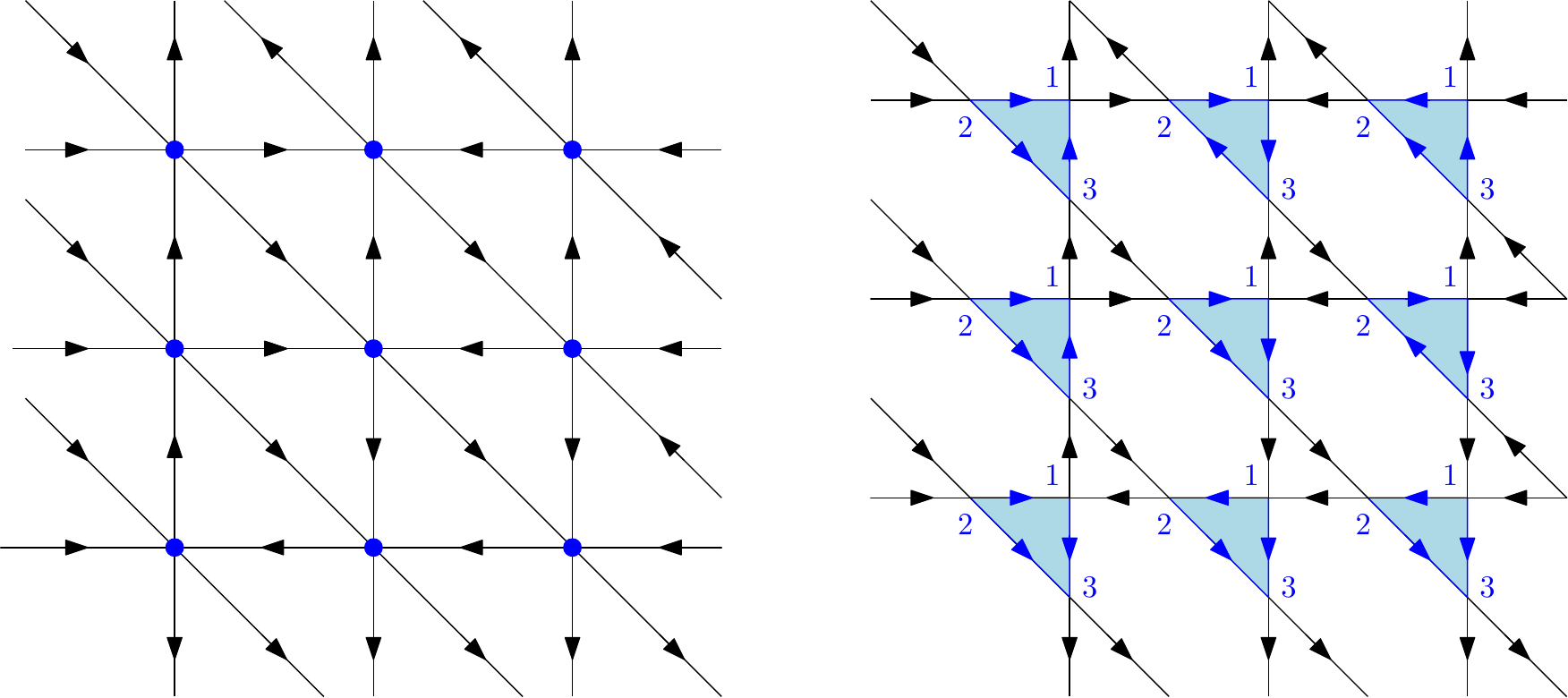}
\end{center}
\caption{\small  The transformation of a piece of triangular lattice into a piece of Kagome lattice. The Kagome lattice is naturally divided into
three sub-lattices $1$, $2$ and $3$ as indicated.}
\label{fig:Kagome}
\end{figure}
\begin{figure}
\begin{center}
\includegraphics[width=10cm]{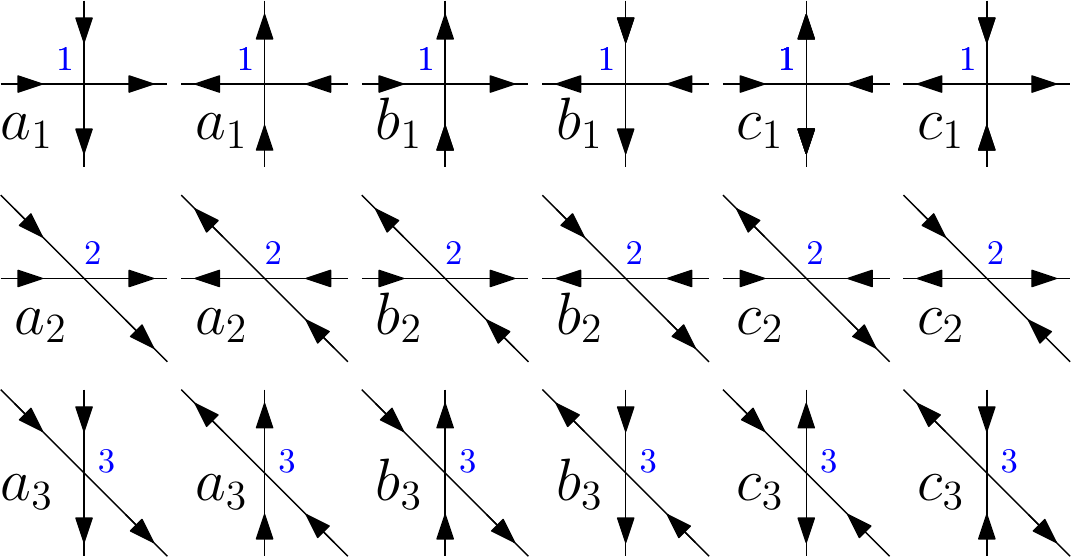}
\end{center}
\caption{\small The weights of the three 6V models on the three sub-lattices $1$, $2$ and $3$ of the Kagome lattice. }
\label{fig:KagomeWeights}
\end{figure}
As noticed in \cite{Baxter}, for appropriate vertex weights, the 20V model may be reformulated as an ice model on the
Kagome lattice obtained by slightly shifting up each horizontal line. 
Some of the triangles (Southwest pointing triangles with black surrounding edges in Figure 3-right) of the original lattice are preserved during this procedure. Their incident edges keep their orientation. New triangles (Northeast pointing triangles with blue surrounding edges in Figure 3-right) 
appear, in correspondence with the nodes of the original triangular lattice. One may then choose an orientation of their incident edges such 
that the ice rule is satisfied at each node of the Kagome lattice. Moreover the maximum number of such orientations is two per newly formed triangle. 
When two orientations are possible, the choice is independent for each of the newly formed triangle.
Each 20V model configuration is thus in correspondence with a number of ice model configurations
on the Kagome lattice, with weights which may be related locally as follows: the Kagome lattice is naturally split into three sub-lattices numbered $1$, $2$ and $3$
(see  Figure~\ref{fig:Kagome}-right).
The ice rule at each tetravalent node of the Kagome lattice leads to six possible vertex environments, hence we are lead to three 6V models to which we assign
the weights of Figure~\ref{fig:KagomeWeights}, where we imposed for convenience that the vertex weights be \emph{invariant under a global reversing of the orientations}.
As a consequence, the weights of the equivalent 20V model, obtained by summing over the possible orientations around the newly formed triangles, are also invariant under 
a global reversing of the orientations, leading to a list of ten weights: $a_1a_2a_3$, $b_1a_2b_3$, $b_1a_2c_3$, $c_1a_2a_3$, $b_1c_2a_3$, $b_1b_2a_3$, $a_1b_2c_3+c_1c_2b_3$, $a_1b_2b_3+c_1c_2c_3$, $c_1b_2b_3+a_1c_2c_3$ and $c_1b_2c_3+a_1c_2b_3$.
\begin{figure}
\begin{center}
\includegraphics[width=6cm]{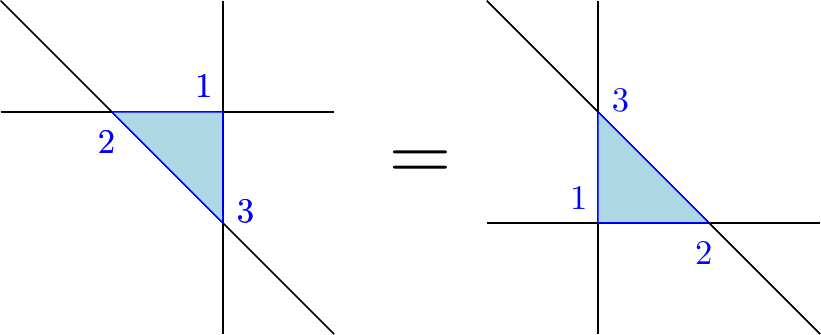}
\end{center}
\caption{\small A schematic picture of the Yang-Baxter condition. The equality must hold for any fixed choice of orientation for the six external edges 
and upon
summation over the three internal edge orientations, subject to the ice rule at each node.}
\label{fig:YB}
\end{figure}
\begin{figure}
\begin{center}
\includegraphics[width=16cm]{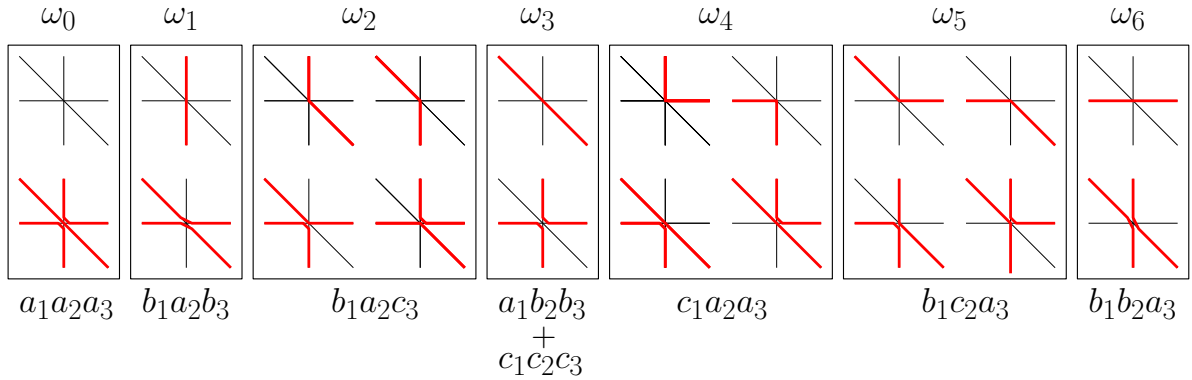}
\end{center}
\caption{\small Weight dictionary for the 20V configurations in the osculating path formulation. The weights are invariant by complementation (exchange of 
empty and occupied edges) and by $180^\circ$ rotation.}
\label{fig:generalweights}
\end{figure}

As a final and crucial restriction on the weights, we demand that the so-called Yang Baxter condition be satisfied, namely that \emph{the same weights}
are obtained for the 20V model if, in our equivalence, we shift the horizontal lines of the triangular lattice down instead of up.
This leads to the identity depicted in Figure~\ref{fig:YB}, and imposes the following extra conditions:
\begin{equation*}
a_1b_2c_3+c_1c_2b_3=b_1a_2c_3\ ,\quad c_1b_2b_3+a_1c_2c_3=c_1a_2a_3\ , \quad c_1b_2c_3+a_1c_2b_3= b_1c_2a_3\ .
\end{equation*}
These conditions are crucial as they will allow us to unravel the 20V model configurations with DWBC1 or DWBC2 into configurations of a single 6V model with standard DWBC 
(corresponding to the Kagome sub-lattice labelled by $1$, see below). 
The set of the 20V model weights is then reduced to a list of seven values, with the dictionary depicted in Figure~\ref{fig:generalweights}:
\begin{equation}
\begin{split}
&\omega_0=a_1a_2a_3,\quad\omega_1=b_1a_2b_3,\quad\omega_2=b_1a_2c_3,\\
&\omega_3=a_1b_2b_3+c_1c_2c_3,\quad\omega_4=c_1a_2a_3,
\quad\omega_5=b_1c_2a_3, \quad\omega_6=b_1b_2a_3\ .\\
\end{split}
\label{eq:weightlist}
\end{equation}
In this paper, we shall discuss exclusively the 20V model with weights of the form \eqref{eq:weightlist} above (or specializations of these expressions). 
In particular, in the osculating path language, these weights are manifestly invariant
both under $180^\circ$ rotation and under complementation (exchange of occupied and un-occupied edge). As a consequence, 
the 20V models with DWBC1 and 2 share the same partition function. 

\subsection{From the 20V model with DWBC1 or 2 to one copy of 6V with DWBC}
\label{sec:square}
A final restriction on the 20V model weights corresponds to choosing the weights of the equivalent three Kagome 6V models among a set of \emph{integrable weights} as
done in \cite{DFG20V}. Introducing the notations:
\begin{equation*}
A(z,w)=z-w\ , \qquad B(z,w)=q^{-2}\, z-q^2\, w\ , \qquad C(u,v)=(q^2-q^{-2})\, \sqrt{z\, w}\ ,
\end{equation*}
we set 
\begin{equation}
\begin{matrix}
a_1=\nu\,A(z,w)\ , \hfill &b_1=\nu\,B(z,w)\ , \hfill &c_1=\nu\,C(z,w)\ , \hfill \\
a_2=\nu\,A(q\, z,q^{-1}\, t)\ , \hfill &b_2=\nu\,B(q\, z,q^{-1}\, t)\ , \hfill &c_2=\nu\,C(q\, z,q^{-1}\, t)\ , \hfill \\
a_3=\nu\,A(q\, t,q^{-1}\, w)\ , \hfill &b_3=\nu\,B(q\, t,q^{-1}\, w)\ , \hfill &c_3=\nu\,C(q\, t,q^{-1}\, w)\ , \hfill \\
\end{matrix}
\label{eq:intweights}
\end{equation}
with  some arbitrary $q\in \C^*$, and arbitrary ``spectral parameters'' $z$, $w$ and $t$, respectively attached to the
horizontal, vertical and diagonal direction. 
The global normalization $\nu$ does not affect the statistics of the model and may be chosen arbitrarily: 
we take $\nu=1/(2{\rm i}t^{1/3})$ for future convenience.

The above homogeneous weights are members of a more general family of inhomogeneous (i.e. node-dependent) integrable weights:
\begin{equation*}
\begin{matrix}
a_1(i,j)=\nu\,A(z_i,w_j)\ , \hfill &b_1(i,j)=\nu\,B(z_i,w_j)\ , \hfill &c_1(i,j)=\nu\,C(z_i,w_j)\ , \hfill \\
a_2(i,k)=\nu\,A(q\, z_i,q^{-1}\, t_k)\ , \hfill &b_2(i,k)=\nu\,B(q\, z_i,q^{-1}\, t_k)\ , \hfill &c_2(i,k)=\nu\,C(q\, z_i,q^{-1}\, t_k)\ , \hfill \\
a_3(k,j)=\nu\,A(q\, t_k,q^{-1}\, w_j)\ , \hfill &b_3(k,j)=\nu\,B(q\, t_k,q^{-1}\, w_j)\ , \hfill &c_3(k,j)=\nu\,C(q\, t_k,q^{-1}\, w_j)\ , \hfill \\
\end{matrix}
\end{equation*}
where a different spectral parameter $z_i$, $w_j$ and $t_k$ is attached to each horizontal, vertical and diagonal line respectively
(the normalization $\nu$ is again arbitrary). Here the horizontal lines are numbered by $i=1,\dots n$ from bottom to top, the vertical lines by
$j=1,2,\dots,n$ from left to right, and the diagonal lines by $k=1,2,\dots,2n-1$ from the lower left to the upper right corner.  
In this setting, the pair $(i,j)$, (respectively $(i,k)$ and $(j,k)$) therefore refers to the node of sub-lattice $1$ (respectively $2$ and $3$)
at the crossing of the $i$-th horizontal and $j$-th vertical lines (respectively of the $i$-th horizontal/$k$-th diagonal and of the $j$-th vertical/$k$-th diagonal
lines).
A crucial property of the above integrable weights is that they \emph{satisfy the Yang Baxter condition at each node for arbitrary spectral parameters}
$z_i$, $w_j$ and $t_k$. This property is used in Appendix~\ref{sec:6V20Vgeneralrel}.

Returning to the situation with homogeneous spectral parameters $z$, $w$ and $t$, we finally 
choose\footnote{This choice corresponds to imposing $z\, w=1$, which may be done without loss of generality since only the ratios of weights matter
for the statistics of the model.}
\begin{equation*}
q={\rm e}^{{\rm i}\, \eta}\ , \quad z={\rm e}^{{\rm i}\,(\eta+\lambda)}\ , \quad w={\rm e}^{-{\rm i}\, (\eta+\lambda)}\ , \quad t={\rm e}^{{\rm i}\, \mu}\ , \qquad (\eta, \lambda, \mu \in \C).
\end{equation*}
With this parametrization, the 20V model weights of \eqref{eq:weightlist}, with the choice \eqref{eq:intweights}, are eventually given by 
\begin{equation}
\begin{split}
\omega_0&=\sin(\lambda+\eta)\sin\left(\frac{\lambda+3\eta+\mu}{2}\right)\sin\left(\frac{\lambda+3\eta-\mu}{2}\right)\\
\omega_1&= \sin(\lambda-\eta)\sin\left(\frac{\lambda-\eta+\mu}{2}\right)\sin\left(\frac{\lambda+3\eta-\mu}{2}\right)\\
\omega_2&=\sin(2\eta)\sin(\lambda-\eta)\sin\left(\frac{\lambda+3\eta-\mu}{2}\right)\\
\omega_3&=\sin(2\eta)^3+\sin(\lambda+\eta)\sin\left(\frac{\lambda-\eta+\mu}{2}\right)\sin\left(\frac{\lambda-\eta-\mu}{2}\right)\\
\omega_4&=\sin(2\eta)\sin\left(\frac{\lambda+3\eta+\mu}{2}\right)\sin\left(\frac{\lambda+3\eta-\mu}{2}\right)\\
\omega_5&=\sin(2\eta)\sin(\lambda-\eta)\sin\left(\frac{\lambda+3\eta+\mu}{2}\right)\\
\omega_6&=\sin(\lambda-\eta)\sin\left(\frac{\lambda+3\eta+\mu}{2}\right)\sin\left(\frac{\lambda-\eta-\mu}{2}\right)\ ,
\end{split}
\label{eq:explweights}
\end{equation}
a parametrization equivalent to that of Kelland in \cite{Kel}.
We will finally restrict our choice to \emph{real} values
of the angles $\eta$, $\lambda$ and $\mu$, which implies in particular that the three Kagome 6V models are in the so-called
disordered phase \cite{Baxter}. The range of these angles is taken so as to ensure that all $\omega$'s are positive, namely:
\begin{equation}
 0<\eta<\lambda<\pi-\eta,\quad \eta-\lambda<\mu<\lambda-\eta\ .
 \label{eq:admissible}
 \end{equation}
Note as a consequence that $\eta<\frac{\pi}{2}$ and $\lambda+3\eta>\mu$.

As for the 6V model on the Kagome sub-lattice labelled $1$, it has weights $(a_1,b_1,c_1)=(a,b,c)/t^{1/3}$,
where we recognize the standard 6V-model weight parametrization in the disordered phase:
\begin{equation}
a=\sin(\lambda+\eta)\ , \quad b= \sin(\lambda-\eta)\ , \quad c=\sin(2 \eta)\ .
\label{eq:abc}
\end{equation}

\begin{figure}
\begin{center}
\includegraphics[width=16cm]{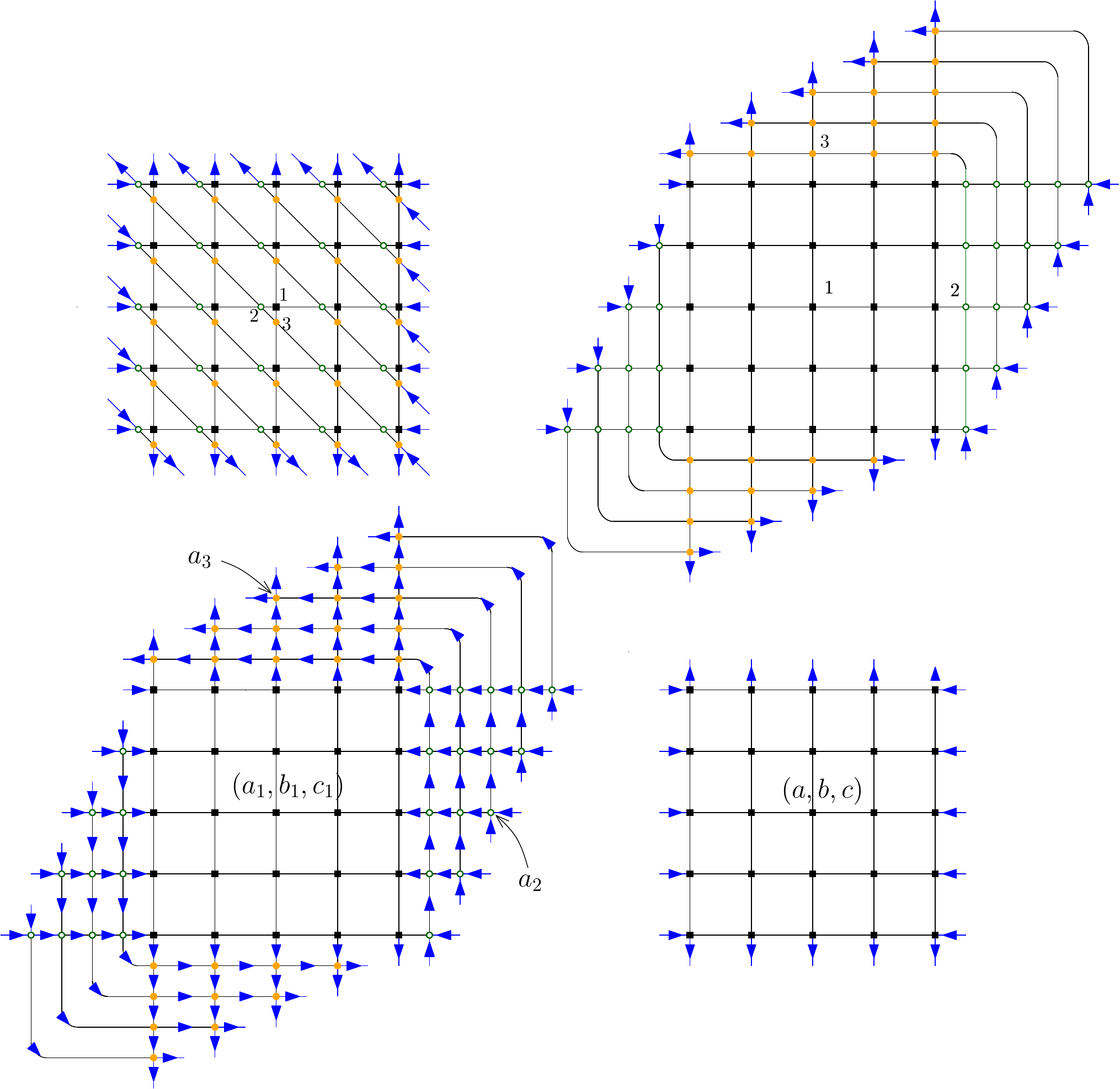}
\end{center}
\caption{\small The unraveling of a 20V model configuration with DWBC2 (upper left) into a 6V model configuration with DWBC on a square grid (lower right). 
Thanks to the Yang-Baxter relation, the diagonal lines may be expelled out of the central square grid (upper right). This has the effect of isolating the vertices according their type. The vertices of sub-lattice 1 stay in the central square region while the vertices of sub-lattice $2$ and $3$ are expelled, respectively to the left and the right, and to the top and the bottom. Due to the DWBC2 prescription and to the ice rule at each node, all the orientations of the expelled edges
are fixed (lower left) so that all the nodes of sub-lattice $2$ (respectively $3$) receive the weight $a_2$ (respectively $a_3$), leading to a global multiplicative factor
$(a_2 a_3)^{n^2}$. The remaining central configuration is a 6V model configuration with DWBC and weights $(a_1,b_1,c_1)$ or equivalently $(a,b,c)$ of
\eqref{eq:abc} up to a global factor $(1/t^{1/3})^{n^2}$. This leads to the identity \eqref{eq:Z20Z6}.}
\label{fig:unraveling}
\end{figure}
As explained in \cite{DFG20V} and depicted in Figure~\ref{fig:unraveling}, the Yang-Baxter condition allows to unravel the configurations of the 20V model with DWBC1 or 2 and with 
the integrable weights \eqref{eq:intweights} into configurations of a single 6V model on the Kagome sub-lattice $1$. As a consequence, 
the partition function for either DWBC1 or  DWBC2 (denoted $Z^{20V}$ as the two are equal) with the weights \eqref{eq:explweights} and that for the 6V model 
with weights $(a,b,c)$ of \eqref{eq:abc} above (denoted $Z^{6V}_{n}[a,b,c]$) are then directly proportional, namely:
\begin{thm}{see Ref.~\cite{DFG20V}}
\label{thm:first}
\begin{equation}
Z^{20V}_n =\left(\frac{a_2 a_3}{t^{1/3}}\right)^{n^2} Z^{6V}_{n}[a,b,c]= \left(\sin\left(\frac{\lambda+3\eta-\mu}{2}\right)\sin\left(\frac{\lambda+3\eta+\mu}{2}\right)\right)^{n^2} Z^{6V}_{n}[a,b,c]\ .
\label{eq:Z20Z6}
\end{equation} 
\end{thm}

\section{The 20V/6V model correspondence: one point functions}
\label{sec:onepoint}

\def\Ell{L}

\subsection{Refined enumeration}
\label{sec:refined}
The use of the tangent method requires a refined enumeration of the 20V model configurations where we keep track of the position where the uppermost path
hits the right boundary, i.e.\ the vertical line $j=n$. As explained in Appendix~\ref{sec:6V20Vgeneralrel}, this enumeration may be obtained by changing 
$w\to w\, \theta$ for the spectral parameter of the last column ($j=n$). Here, we will concentrate on the DWBC2 prescription, which turns out to 
lead to simpler enumeration formulas. Note that the uppermost path corresponds in this case to the path starting at the horizontal edge
$(0,n)\to(1,n)$ and ending at the vertical edge $(n,1)\to(n,0)$.
The net result is best expressed upon introducing the following refined partition function:
\begin{equation*}
Z^{20V_{BC2}}_n(\tau)=\sum_{\Ell=1}^{n} Z^{20V_{BC2}}_{n;\Ell}\, \tau^{\Ell-1}
\end{equation*}
where $Z^{20V_{BC2}}_{n;\Ell}$ denotes the partition function of the 20V model configurations with DWBC2 for which
the uppermost path hits the vertical line $j=n$ at position $(n,\Ell)$. Note that the step just before the hitting point is
either a horizontal or a diagonal step 
and we call 
$Z^{20V_{BC2}\, \mbox{--}}_{n;\Ell}$ and  $Z^{20V_{BC2}\, {\scriptscriptstyle{\diagdown}}}_{n;\Ell}$ the corresponding restricted refined partition functions,
as well as:
\begin{equation*}
 Z^{20V_{BC2}\,  \mbox{--}}_n(\tau)=\sum_{\Ell=1}^{n} Z^{20V_{BC2}\, \mbox{--}}_{n;\Ell}\, \tau^{\Ell-1}\ , \qquad
Z^{20V_{BC2}\, {\scriptscriptstyle{\diagdown}}}_n(\tau)=\sum_{\Ell=1}^{n} Z^{20V_{BC2}\, {\scriptscriptstyle{\diagdown}}}_{n;\Ell}\, \tau^{\Ell-1}
\end{equation*}
with the obvious sum rule:
\begin{equation*}
Z^{20V_{BC2}}_n(\tau)=Z^{20V_{BC2}\,  \mbox{--}}_n(\tau)+Z^{20V_{BC2}\, {\scriptscriptstyle{\diagdown}}}_n(\tau)\ .
\end{equation*}
As for the 6V model with DWBC configurations, we consider a similar refinement as follows: as is well-known, configurations
of the 6V model with DWBC are in bijection with configurations of osculating paths. Those are obtained as for the 20V model 
by drawing path steps along the edges oriented East or South and by connecting them uniquely into non-crossing but possibly kissing 
well-oriented paths (going from the West boundary to the South boundary of the square grid). We then
denote by $Z^{6V}_{n;\Ell}$ the partition function for those configurations 
where the uppermost path (starting at the horizontal edge
$(0,n)\to(1,n)$ and ending at the vertical edge $(n,1)\to(n,0)$)
hits the line $j=n$ at position $(n,\Ell)$. Note that this hitting point is necessarily preceded by a horizontal step. We finally set
\begin{equation*}
Z^{6V}_n(\sigma)=\sum_{\Ell=1}^{n} Z^{6V}_{n;\Ell}\, \sigma^{\Ell-1}\ .
\end{equation*}
From now on, it will be implicitly assumed that all the 6V model partition functions are evaluated with the weights 
$(a,b,c)$ of Eq.~\eqref{eq:abc}.
With these notations, we may prove the following identity, which generalizes \eqref{eq:Z20Z6} (see Appendix~\ref{sec:6V20Vgeneralrel} for a detailed proof):
\begin{thm}
\label{thm:second}
The refined partition functions $Z^{20V_{BC2}\,\mbox{--}}_n(\tau)$ and $Z^{20V_{BC2}{\scriptscriptstyle{\diagdown}}}_n(\tau)$ of the 20V model with DWBC2
are related to the refined partition function $Z^{6V}_n(\sigma)$ of the 6V model with DWBC via
\begin{equation}
Z^{20V_{BC2}\,\mbox{--}}_n(\tau) +g(\sigma)Z^{20V_{BC2}{\scriptscriptstyle{\diagdown}}}_n(\tau)=
\left(\frac{a_2a_3}{t^{1/3}}\right)^{n^2}\ Z^{6V}_n(\sigma)\ ,
\label{eq:genrel}
\end{equation}
where
\begin{equation}
\begin{split} 
\tau&=\sigma\ 
\ \frac{\sigma\, \sin(\lambda-\eta)\sin\left(\frac{\lambda+3\eta-\mu}{2}\right)-\sin(\lambda+\eta)\sin\left(\frac{\lambda-\eta-\mu}{2}\right)
}{\sigma\,  \sin(\lambda-\eta) \sin\left(\frac{\lambda-\eta-\mu}{2}\right)-\sin(\lambda+\eta)\sin\left(\frac{\lambda-5\eta-\mu}{2}\right)} \times  \frac{\sin\left(\frac{\lambda+3\eta+\mu}{2}\right)}{\sin\left(\frac{\lambda-\eta+\mu}{2}\right)}\ ,\\
g(\sigma)&=\frac{ \sigma \sin(2\eta)\sin\left(\frac{\lambda+3\eta+\mu}{2}\right)}{\sigma\,  \sin(\lambda-\eta) \sin\left(\frac{\lambda-\eta-\mu}{2}\right)-\sin(\lambda+\eta)\sin\left(\frac{\lambda-5\eta-\mu}{2}\right)}\ .\\
\label{eq:taugofsigma}
\end{split}
\end{equation}
\end{thm}
\noindent Note that, for $\sigma=1$, we have $\tau=1$ and $g(\sigma)=1$ as expected so as to recover \eqref{eq:Z20Z6}.

\begin{remark}
If we insist on having a strict proportionality relation between $Z^{20V_{BC2}}_n(\tau)$ and $Z^{6V}_n(\sigma)$, we have to demand that $g(\sigma)=1$ for all $\sigma$, with the easily checked property:
\begin{equation*}
g(\sigma)=1 \ \hbox{for all}\ \sigma \Leftrightarrow \mu=\lambda-5\eta.
\end{equation*}
From the general expressions \eqref{eq:explweights}, this latter relation implies $\omega_4=\omega_2$, $\omega_5=\omega_6=\omega_3$, hence reduces in practice the number of weights to 
four values $\omega_0$, $\omega_1$,$\omega_2$, $\omega_3$.
We then have a strict proportionality relation:
\begin{equation*}
Z^{20V_{BC2}}_n(\tau) \underset{\mu=\lambda-5\eta}{=} \left(\sin(4\eta)\sin(\lambda-\eta)\right)^{n^2} Z^{6V}_n(\sigma) 
\end{equation*}
with $\tau$ and $\sigma$ related via
\begin{equation}
\tau =\frac{2\cos(2\eta)\sin(\lambda-\eta)\, \sigma-\sin(\lambda+\eta)}{\sin(\lambda-3\eta)}\ .
\label{eq:tausigma}
\end{equation}
\end{remark}

Another useful refined enumeration of the 20V model configurations corresponds to keeping track of the position where the uppermost path
leaves the upper boundary, i.e.\ the horizontal line $i=n$. This enumeration may be obtained by now changing 
$z\to z\, \tilde{\theta}$ for the spectral parameter of the top line ($i=n$). We now introduce the refined partition function:
\begin{equation*}
\tilde{Z}^{20V_{BC2}}_n(\tilde{\tau})=\sum_{\Ell=1}^{n} \tilde{Z}^{20V_{BC2}}_{n;\Ell}\, \tilde{\tau}^{\Ell-1}
\end{equation*}
where $\tilde{Z}^{20V_{BC2}}_{n;\Ell}$ denotes the partition function of the 20V model configurations with DWBC2 for which
the uppermost path leaves the horizontal line $i=n$ at position $(\Ell,n)$. The step just after the leaving point is
either vertical or a diagonal step 
and we call 
$\tilde{Z}^{20V_{BC2}\, \vert}_{n;\Ell}$ and  $\tilde{Z}^{20V_{BC2}\, {\scriptscriptstyle{\diagdown}}}_{n;\Ell}$ the corresponding restricted partition functions.
With obvious notations, we now have the sum rule:
\begin{equation*}
\tilde{Z}^{20V_{BC2}}_n(\tilde{\tau})=\tilde{Z}^{20V_{BC2}\,  \vert}_n(\tilde{\tau})+\tilde{Z}^{20V_{BC2}\, {\scriptscriptstyle{\diagdown}}}_n(\tilde{\tau})\ .
\end{equation*}
Without any further calculation, we note that, as clearly seen in the osculating path formulation, the present refined enumeration is identical to the previous one up to a symmetry $x\leftrightarrow y$. As
apparent in Figure~\ref{fig:generalweights}, this
symmetry amounts to exchanging the weights $\omega_2\leftrightarrow \omega_5$ and $\omega_1\leftrightarrow \omega_6$, leaving the other weights unchanged. 
From their explicit expressions \eqref{eq:explweights}, this amounts precisely to performing the transformation $\mu\leftrightarrow -\mu$, leaving $\eta$ and $\lambda$ unchanged.
We immediately deduce:
\begin{thm}  
The refined partition functions $\tilde{Z}^{20V_{BC2}\,\vert}_n(\tilde{\tau}) $ and $\tilde{Z}^{20V_{BC2}{\scriptscriptstyle{\diagdown}}}_n(\tilde{\tau})$ of the 20V model with DWBC2
are related to the refined partition function $Z^{6V}_n(\sigma)$ of the 6V model with DWBC via
\begin{equation}
\tilde{Z}^{20V_{BC2}\,\vert}_n(\tilde{\tau}) +\tilde{g}(\sigma)\tilde{Z}^{20V_{BC2}{\scriptscriptstyle{\diagdown}}}_n(\tilde{\tau})=
\left(\frac{a_2a_3}{t^{1/3}}\right)^{n^2}\ Z^{6V}_n(\sigma)
\label{eq:genrelbis}
\end{equation}
with
\begin{equation}
\begin{split} 
\tilde{\tau}&=\sigma\ 
\ \frac{\sigma\, \sin(\lambda-\eta)\sin\left(\frac{\lambda+3\eta+\mu}{2}\right)-\sin(\lambda+\eta)\sin\left(\frac{\lambda-\eta+\mu}{2}\right)
}{\sigma\,  \sin(\lambda-\eta) \sin\left(\frac{\lambda-\eta+\mu}{2}\right)-\sin(\lambda+\eta)\sin\left(\frac{\lambda-5\eta+\mu}{2}\right)} \times  \frac{\sin\left(\frac{\lambda+3\eta-\mu}{2}\right)}{\sin\left(\frac{\lambda-\eta-\mu}{2}\right)}\ ,\\
\tilde{g}(\sigma)&=\frac{ \sigma \sin(2\eta)\sin\left(\frac{\lambda+3\eta-\mu}{2}\right)}{\sigma\,  \sin(\lambda-\eta) \sin\left(\frac{\lambda-\eta+\mu}{2}\right)-\sin(\lambda+\eta)\sin\left(\frac{\lambda-5\eta+\mu}{2}\right)}\ .\\
\end{split}
\label{eq:taugofsigmabis}
\end{equation}
\end{thm}
\begin{remark}
Again a strict proportionality relation between $\tilde{Z}^{20V_{BC2}}_n(\tilde{\tau})$ and $Z^{6V}_n(\sigma)$ is obtained whenever $\tilde{g}(\sigma)=1$ for all $\sigma$, 
with 
\begin{equation*}
\tilde{g}(\sigma)=1  \ \hbox{for all}\ \sigma  \Rightarrow \mu=-\lambda+5\eta
\end{equation*}
in which case
\begin{equation*}
\tilde{Z}^{20V_{BC2}}_n(\tilde{\tau}) \underset{\mu=-\lambda+5\eta}{=} \left(\sin(4\eta)\sin(\lambda-\eta)\right)^{n^2} Z^{6V}_n(\sigma) 
\end{equation*}
with $\tilde{\tau}$ and $\sigma$ related via
\begin{equation*}
\tilde{\tau} =\frac{2\cos(2\eta)\sin(\lambda-\eta)\, \sigma-\sin(\lambda+\eta)}{\sin(\lambda-3\eta)}
\end{equation*}
as in \eqref{eq:tausigma}.
\end{remark}

In the particular case $\mu=0$, we have $\omega_2=\omega_5$ and $\omega_1= \omega_6$ and the 20V model with DWBC1 or 2 is symmetric under $x\leftrightarrow y$.  If moreover $\lambda=5\eta$, we have the 
proportionality relations
\begin{equation*}
Z^{20V_{BC2}}_n(\tau) \underset{\mu=0,\ \lambda=5\eta}{=} \tilde{Z}^{20V_{BC2}}_n(\tau) \underset{\mu=0,\ \lambda=5\eta}{=}(\sin(4\eta))^{2 n^2} Z^{6V}_n(\sigma)\ , \quad
\tau= 1+4 \cos^2(2\eta)\, (\sigma-1).
\end{equation*} 
This case corresponds to a situation in which $\omega_1=\omega_2=\omega_3=\omega_4=\omega_5=\omega_6=\sin (2\eta) \sin^2 (4\eta)$
while $\omega_0=\sin (6\eta) \sin^2 (4\eta)$
and will be studied in detail in Section~\ref{sec:warmup}.

\subsection{Asymptotics of one-point functions}
\label{sec:onepointasymp}
The refined \emph{one-point functions} $H^{20V_{BC2}}_n(\tau)$ and $H^{6V}_n(\sigma)$ are simply defined as normalized refined partition functions via:
\begin{equation*}
H^{20V_{BC2}}_n(\tau)=\frac{Z^{20V_{BC2}}_n(\tau)}{Z^{20V}_n}\ , \qquad H^{6V}_n(\sigma)=\frac{Z^{6V}_n(\sigma)}{Z^{6V}_n}
\end{equation*}
(recall that all the 6V model partition functions are implicitly evaluated with the weights $(a,b,c)$ of Eq.~\eqref{eq:abc}).
We also introduce the restricted refined one point functions 
\begin{equation*}
H^{20V_{BC2}\,\mbox{--}}_n(\tau)=\frac{Z^{20V_{BC2}\,\mbox{--}}_n(\tau)}{Z^{20V}_n}\ , \qquad  
H^{20V_{BC2}{\scriptscriptstyle{\diagdown}}}_n(\tau)=\frac{Z^{20V_{BC2}{\scriptscriptstyle{\diagdown}}}_n(\tau)}{Z^{20V}_n}
\end{equation*}
which satisfy
\begin{equation}
\begin{split}
& H^{20V_{BC2}\,\mbox{--}}_n(\tau) + H^{20V_{BC2}{\scriptscriptstyle{\diagdown}}}_n(\tau)= H^{20V_{BC2}}_n(\tau)\\
& H^{20V_{BC2}\,\mbox{--}}_n(\tau) +g(\sigma)H^{20V_{BC2}{\scriptscriptstyle{\diagdown}}}_n(\tau)= H^{6V}_n(\sigma)\\
\end{split}
\label{eq:onepointrel}
\end{equation}
with $\tau$ and $g(\sigma)$ as in \eqref{eq:taugofsigma}, 
as well as their tilde counterparts obtained via the change $\mu\to -\mu$.
 
In the limit of large $n$, the asymptotics for the one-point function $H^{6V}_n(\sigma)$ of the 6V model with DWBC and the weights $(a,b,c)$ of \eqref{eq:abc} above is characterized by 
the function
\begin{equation*}
f(\sigma)=\lim_{n\to \infty}\frac{1}{n}{\rm Log}\left( H^{6V}_n(\sigma)\right)
\end{equation*}
whose expression is known \cite{CP2009} and may be given in parametric form as:
\begin{equation}
\begin{split}
f(\sigma(\xi))&={\rm Log}\left(\frac{\sin(\alpha(\lambda-\eta))\sin(\xi+\lambda-\eta)\sin(\alpha \xi)}{\alpha \sin(\lambda-\eta)\sin(\alpha(\xi+\lambda-\eta))\sin(\xi)}\right)\\
\sigma(\xi)&=\frac{\sin(\lambda+\eta)\sin(\xi+\lambda-\eta)}{\sin(\lambda-\eta)\sin(\xi+\lambda+\eta)}\\
\end{split}
\label{eq:fsigma}
\end{equation}
where
\begin{equation}
\alpha=\frac{\pi}{\pi-2\eta}\ .
\label{eq:valalpha}
\end{equation}

Using the correspondence \eqref{eq:taugofsigma}, the parametrization $\sigma(\xi)$ of $\sigma$ translates into the following
parametrization $\tau(\xi)$ of $\tau$:
\begin{equation}
\tau(\xi)=\frac{\sin(\lambda+\eta)\sin\left(\frac{\lambda+3\eta+\mu}{2}\right)\sin(\xi+\lambda-\eta)\sin\left(
\xi+\frac{\lambda-\eta+\mu}{2}\right)}{\sin(\lambda-\eta)\sin\left(\frac{\lambda-\eta+\mu}{2}\right)\sin(\xi+\lambda+\eta)\sin\left(
\xi+\frac{\lambda+3\eta+\mu}{2}\right)}\ .
\label{eq:tauparamgen}
\end{equation}
Since $g(\sigma)$ is bounded independently of $n$, the relations \eqref{eq:onepointrel} imply that
\begin{equation*}
\lim_{n\to \infty}\frac{1}{n}{\rm Log}\left( H^{20V_{BC2}}_n\left(\tau(\xi)\right)\right)=f(\sigma(\xi))\ .
\end{equation*}

Let us now introduce for future use the function
\begin{equation}
r(\tau)=\tau \frac{d}{d\tau}\left(\lim_{n\to \infty}\frac{1}{n}{\rm Log}\left( H^{20V_{BC2}}_n(\tau)\right)\right)\ .
\label{eq:defr}
\end{equation} 
The function $r(\tau)$ is given in parametric form by the above parametrization \eqref{eq:tauparamgen} of $\tau$ and
the following parametrization for $r(\tau)$: 
\begin{equation}
\begin{split}
r(\tau(\xi))&=\frac{\tau(\xi)}{\partial_\xi \tau(\xi)}\partial_\xi f(\sigma(\xi))\\
&=\left(\cot(\xi+\lambda-\eta)-\cot(\xi)+\alpha\cot(\alpha\, \xi)
-\alpha\cot(\alpha(\xi+\lambda-\eta))\right)\\
&\ \ \times
\frac{\sin(\xi+\lambda+\eta)\sin(\xi+\lambda-\eta)\sin\left(
\xi+\frac{\lambda-\eta+\mu}{2}\right)\sin\left(
\xi+\frac{\lambda+3\eta+\mu}{2}\right)}{\sin(2\eta)\left(\sin(\xi+\lambda+\eta)\sin(\xi+\lambda-\eta)+\sin\left(
\xi+\frac{\lambda-\eta+\mu}{2}\right)\sin\left(
\xi+\frac{\lambda+3\eta+\mu}{2}\right)\right)}
\\
\end{split}
\label{eq:valrgen}
\end{equation}
with $\alpha$ as in \eqref{eq:valalpha}.
\medskip

\begin{remark}
In the particular case $\mu=\lambda-5 \eta$, using the correspondence \eqref{eq:tausigma}, the parametrization $\tau(\xi)$ of $\ \tau$ simplifies into:
\begin{equation}
\tau(\xi) \underset{\mu=\lambda-5\eta}{=}\frac{\sin(\lambda+\eta)\sin(\xi+\lambda-3\eta)}{\sin(\lambda-3\eta)\sin(\xi+\lambda+\eta)}
\label{eq:tauparam}
\end{equation}
and that of $r(\tau)$ into:
\begin{equation}
r(\tau(\xi)) \underset{\mu=\lambda-5\eta}{=}\left(\cot(\xi+\lambda-\eta)-\cot(\xi)+\alpha\cot(\alpha\, \xi)
-\alpha\cot(\alpha(\xi+\lambda-\eta))\right)
\frac{\sin(\xi+\lambda+\eta)\sin(\xi+\lambda-3\eta)}{\sin(4\eta)}\ .
\label{eq:valr}
\end{equation}
\end{remark}

\section{The case $\mu=\lambda-5\eta=0$}
\label{sec:warmup}
We now turn to the explicit computation of the arctic curve of the 20V model with DWBC1 or 2.
As a warmup, we start with the simple case 
\begin{equation*}
\mu=0 ,\quad \lambda=5\eta
\end{equation*}
where the weights thus depend on a single ``angle'' $\eta$. From their general expression \eqref{eq:explweights},
it is easily checked that, \emph{up to a global normalization factor} $\sin (2\eta) \sin^2 (4\eta)$, the weights for the various vertex
environments are all equal to $1$ except for the empty/full vertex with weight\footnote{
We use the notation $\varpi_0$ to recall that  a global normalization factor $\sin (2\eta) \sin^2 (4\eta)$ has been factored out in all
the weights, i.e. $\omega_0=\sin (2\eta) \sin^2 (4\eta)\times \varpi_0$ while $\omega_i=\sin (2\eta) \sin^2 (4\eta)\times 1$ for
$i=1$ to $6$.}
\begin{equation*}
\varpi_0=\frac{\sin(6\eta)}{\sin(2\eta)}=1+2\cos(4\eta)\ .
\end{equation*}
In particular, the weights are invariant under the transformation $x\leftrightarrow y$, which implies that the arctic curve is symmetric under
this transformation. In the particular case $\eta=\pi/8$, all the weights are equal to $1$.
\medskip

\begin{figure}
\begin{center}
\includegraphics[width=12cm]{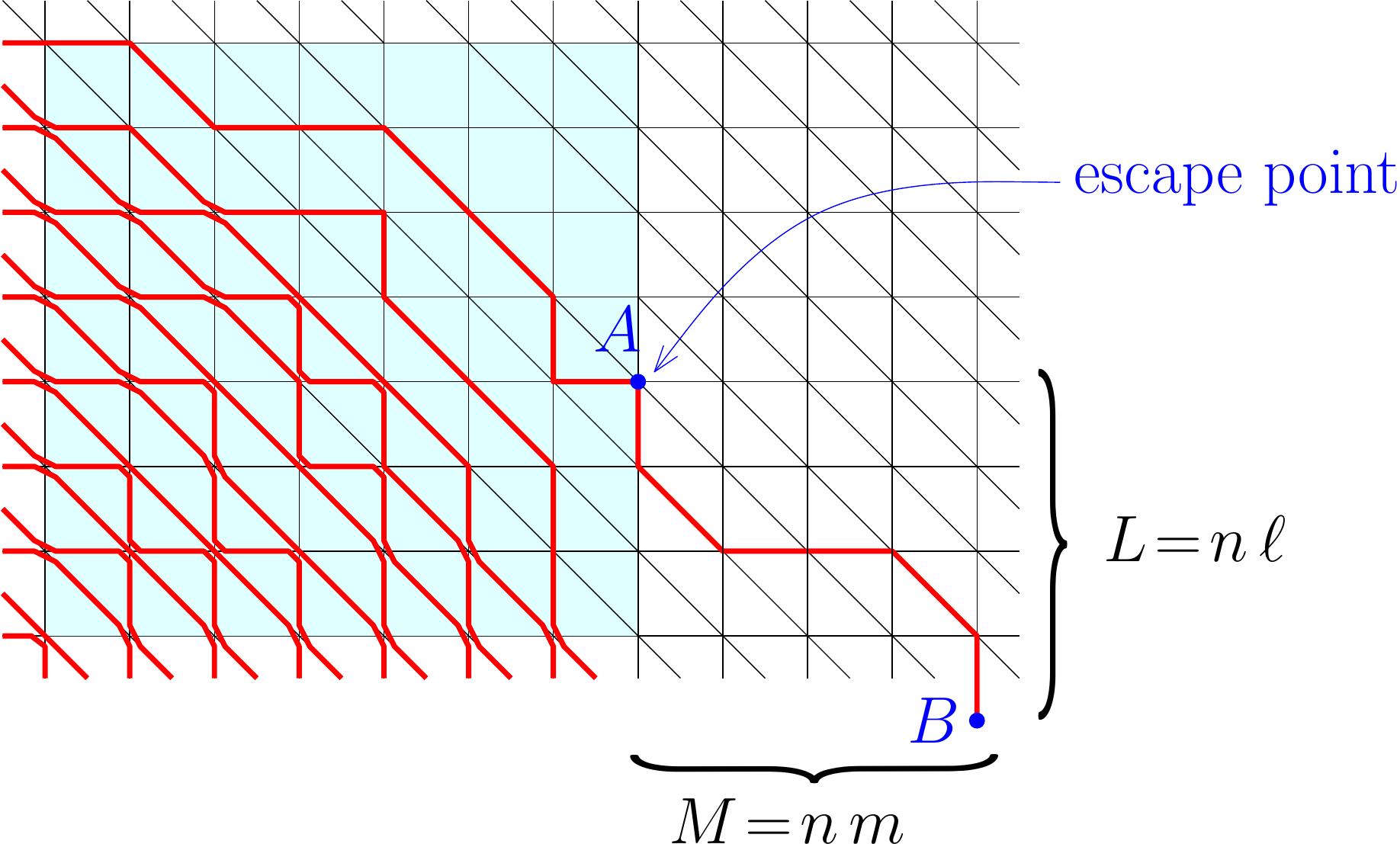}
\end{center}
\caption{\small A sample configuration of the 20V model with DWBC2 in which the uppermost path, starting at the horizontal edge
$(0,n)\to(1,n)$, exits the domain $1< X,Y < n$ (in light blue) at the escape point $A$ with position $(n,L)=(n,n\ell)$ and reaches a shifted endpoint $B$ at position 
$(n+M,0)=(n(1+m),0)$ via a final vertical edge.}
\label{fig:tgmethod1}
\end{figure}
\subsection{The tangent method in its simplest flavor}
\label{sec:normalbranch}
We again consider the slightly simpler DWBC2 prescription. The main result of this section is the identification of a first portion of the arctic curve:
\begin{thm} 
\label{thm:acnormal}
The portion of arctic curve for the 20V model with DWBC2 at $\mu=0$ and $\lambda=5\eta$, as predicted by the direct application of the tangent method, has the following parametric equation:
\begin{equation*}
x(\xi)=1+\frac{\partial_\xi R_0(\xi)}{\partial_\xi S_0(\xi)}\ ,\qquad y(\xi)=R_0(\xi)- S_0(\xi) \frac{\partial_\xi R_0(\xi)}{\partial_\xi S_0(\xi)}\ , \qquad \xi \in [0,\pi-6\eta]\ .
\end{equation*}
where
\begin{equation*}
\begin{split}
R_0(\xi)&=\left(\cot(\xi+4\eta)-\cot(\xi)+\alpha\cot(\alpha\, \xi)
-\alpha\cot(\alpha(\xi+4\eta))\right)
\frac{\sin(\xi+6\eta)\sin(\xi+2\eta)}{\sin(4\eta)}\ ,\\
S_0(\xi)&=\frac{\sin(\xi+6\eta)\sin(\xi+2\eta)}{\sin(\xi)\sin(\xi+4\eta)}\ ,\\
\end{split}
\end{equation*}
and $\alpha=\pi/(\pi-2\eta)$.
\end{thm}
To get the above result, we use the direct tangent method setting, with a geometry where the uppermost path, starting at the horizontal edge
$(0,n)\to(1,n)$ and originally ending at the vertical edge $(n,1)\to(n,0)$, now exits the square domain $1< X,Y < n$
 (in the original coordinate system $(X,Y)$ of the square lattice) and reaches a shifted endpoint at position
$(n+M,0)=(n(1+m),0)$ (i.e. ends with a vertical edge $(n+M,1)\to (n+M,0)$) for some positive $M$ (see Figure \ref{fig:tgmethod1} for an illustration).
We define by convention the escape point as the point, with position $(n,L)=(n,n\ell)$, where the uppermost path \emph{hits the right vertical boundary for
the first time}, even if the path makes a number of vertical step before it eventually leaves the originally accessible domain $1\leq X,Y \leq n$. Note that our choice of escape point rather
than the slightly more natural choice of the point where the path eventually leaves this originally accessible domain, i.e.\ has $X> n$ for the first time, makes in practice
no difference in the scaling limit of large $n$ and turns out to be simpler for explicit computations.
In rescaled coordinates $(x,y)=(X/n,Y/n)$, the escape point and endpoint have respective positions $(1,\ell)$ and $(1+m,0)$.

The most likely escape point position $\ell$ for a given $m$ is obtained by maximizing with respect to $\ell$ the partition function of those configurations 
having a fixed value of $\ell$, namely the quantity\footnote{Here and throughout the paper, the notation $F(x)\vert_{x^p}$ refers to the 
coefficient of $x^p$ in the series expansion of $F(x)$ in the variable $x$.}
\begin{equation*}
H^{20V_{BC2}}_n(\tau)\vert_{\tau^{n\ell}} Y^{20V}_{(n,n\ell)\to (n(1+m),0)}\, \varpi_0^{n\ell-1}\ ,
\end{equation*}
where $Y^{20V}_{(n,n\ell)\to (n(1+m),0)}$ denotes the partition function for a single path from $(n,n\ell)$ to $(n(1+m),0)$ 
(starting possibly with a number of vertical steps as just discussed) with, at each node along the path,  \emph{the same weight as that of the corresponding 20V configuration}.
More precisely, as discussed in Appendix~\ref{sec:trmat},  each node in the empty space around the escaping path also receives a weight $\varpi_0$ per empty vertex. 
Factoring those weights, the remaining effective weight for the nodes visited by the path is $1/\varpi_0$. Paths in $Y^{20V}_{(n,n\ell)\to (n(1+m),0)}$ are
therefore enumerated with a weight $1/\varpi_0$ per visited node. To be fully consistent,  the contribution $1$ to $H^{20V_{BC2}}_n(\tau)\vert_{\tau^{n\ell}}$ of the part of the uppermost path
going from $(n,n\ell)$ to $(n,0)$ must first be replaced by that of a background segment of empty vertices, since this portion of path is no longer present and replaced by a portion of path enumerated by 
$Y^{20V}_{(n,n\ell)\to (n(1+m),0)}$. This background segment is made of $n\ell-1$ empty vertices, hence the final factor $\varpi_0^{n\ell-1}$ in the above quantity to be maximized.

Introducing the large $n$ asymptotics
\begin{equation*}
 Y^{20V}_{(n,n\ell)\to (n(1+m),0)}\underset{n\to \infty}{\sim} {\rm e}^{n\, S(\ell,m)}
\end{equation*}
and writing $H^{20V_{BC2}}_n(\tau)\vert_{\tau^{n\ell}}=\frac{1}{2{\rm i}\pi} \oint d\tau H^{20V_{BC2}}_n(\tau)/\tau^{n\ell+1}$ so as to evaluate this latter quantity by a saddle point
method, 
the extremization conditions over $\tau$ (saddle point condition) and $\ell$ (most likely escape point condition) read respectively:
\begin{equation*}
\begin{split}
& \frac{d}{d\tau} \left( -\ell\, {\rm Log}\,\tau + f(\sigma(\tau))\right)=0\\
& \frac{d}{d\ell}  \left(-\ell\, {\rm Log}\,\tau  + S(\ell,m) +\ell\, {\rm Log}\,\varpi_0\right)=0\\
\end{split}
\end{equation*}
or equivalently, using the function $r(\tau)$ introduced in \eqref{eq:defr}: 
\begin{equation*}
\ell=r(\tau), \quad \tau=\varpi_0\, {\rm e}^{\frac{d}{d\ell} S(\ell,m)}\ .
\end{equation*}
The corresponding tangent line passing trough $(1,\ell)$ and $(1+m,0)$ has equation
$y+\frac{\ell}{m}(x-1)-\ell=0$, hence for the most likely escape point:
\begin{equation}
y+s(\tau) (x-1)-r(\tau)=0
\label{eq:tgline}
\end{equation}
where the ``slope'' $s(\tau)$ is given by 
\begin{equation}
s(\tau)=\frac{r(\tau)}{m}\ \hbox{with $m:=m(\tau)$ solution of}\ \tau=\varpi_0\, {\rm e}^{\frac{d}{d\ell} S(\ell,m)}\ \hbox{at}\ \ell=r(\tau)\ .
\label{eq:eqfors}
\end{equation}
Note that, from \eqref{eq:tgline}, the actual slope of the tangent line in the $(x,y)$ plane is $-s(\tau)$, which, from the underlying geometry,
must run from $0$ (horizontal line $y=1$) to $-\infty$ (vertical line $x=1$). The quantity $s(\tau)$ must therefore span the interval $[0,+\infty]$.
In the simple case at hand with $\mu=\lambda-5\eta=0$, we have from \eqref{eq:tauparamgen} (or from the simpler expression \eqref{eq:tauparam}):
\begin{equation*}
\begin{split}
\tau(\xi)&=\frac{\sin(6\eta)\sin(\xi+2\eta)}{\sin(2\eta)\sin(\xi+6\eta)}\\
r(\tau(\xi))&=\left(\cot(\xi+4\eta)-\cot(\xi)+\alpha\cot(\alpha\, \xi)
-\alpha\cot(\alpha(\xi+4\eta))\right)
\frac{\sin(\xi+6\eta)\sin(\xi+2\eta)}{\sin(4\eta)}\ .\\
\end{split}
\end{equation*}
As for $S(\ell,m)$, it may be obtained via a transfer matrix approach\footnote{Here the use of transfer matrix approach is not
fully necessary but we use it as it will cover all the more involved cases encountered in this paper.} upon following
the step by step evolution of the escape path, as described in Appendix \ref{sec:trmat}.
For our simple case where all vertex weights are equal to $1$ except for $\varpi_0$, the values of $\beta$ in this Appendix are to be chosen as $\beta_1=\beta_2=\beta_3=1$, while the values of $\alpha$ are 
$\alpha_1=\alpha_2=\alpha_3=\frac{1}{\varpi_0}$,  $\alpha_4=\alpha_5=\alpha_6=0$. This leads to\footnote{Here the quantity $n p_3$ may be simply interpreted 
as the number of diagonal steps in the path.}
\begin{equation*}
\begin{split}
S(\ell,m)=S(\ell,m,p_3)&:=(\ell+m-p_3){\rm Log}(\ell+m-p_3)-(\ell-p_3){\rm Log}(\ell-p_3)\\  & \   -(m-p_3){\rm Log}(m-p_3)
-p_3\, {\rm Log}\, p_3
+(\ell+m-p_3) {\rm Log}\left(\frac{1}{\varpi_0}\right)\\
\end{split}
\end{equation*}
taken at the value of $p_3$ which maximizes $S(\ell,m,p_3)$ at fixed $\ell$ and $m$.
Writing the new extremization condition $\partial_{p_3} S(\ell,m,p_3)=0$ and solving \eqref{eq:eqfors} yields
\begin{equation*}
s(\tau)= \frac{\tau(1+\varpi_0)}{(\tau-1)(\tau+\varpi_0)}
=\frac{4 \tau \cos^2(2\eta)}{(\tau-1)(\tau+1+2\cos(4\eta))}\ .
\end{equation*}
Using the parametrization \eqref{eq:tauparam} for $\tau$, we get immediately 
\begin{equation*}
s(\tau(\xi))=\frac{\sin(\xi+6\eta)\sin(\xi+2\eta)}{\sin(\xi)\sin(\xi+4\eta)}
\end{equation*}
so that we end up with a family of tangent lines parametrized by $\xi$ with equations:
\begin{equation*}
\begin{split}
0=F(x,y;\xi)&:=y+s(\tau(\xi)) (x-1)-r(\tau(\xi))\\&=
y+\frac{\sin(\xi+6\eta)\sin(\xi+2\eta)}{\sin(\xi)\sin(\xi+4\eta)}(x-1)\\
& \quad -\left(\cot(\xi+4\eta)-\cot(\xi)+\alpha\cot(\alpha\, \xi)
-\alpha\cot(\alpha(\xi+4\eta))\right)
\frac{\sin(\xi+6\eta)\sin(\xi+2\eta)}{\sin(4\eta)}\ .\\
\end{split}
\end{equation*}
Here $\xi$ runs over the range $[0,\pi-6\eta]$ to guarantee that $s(\tau(\xi))$ spans the interval $[0,+\infty]$, as dictated by the tangent method in the present geometry.

\begin{figure}
\begin{center}
\includegraphics[width=10cm]{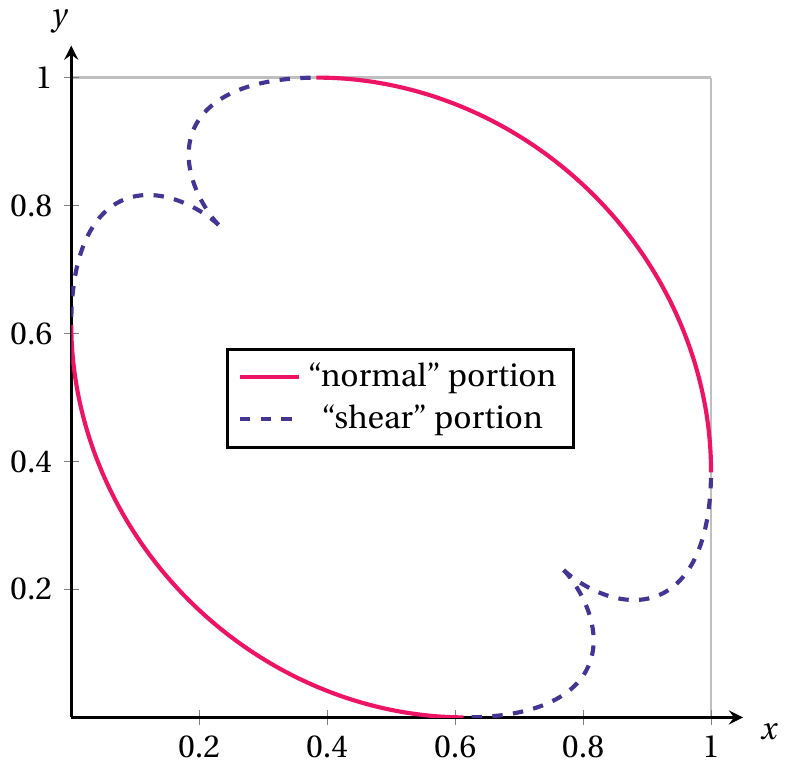}
\end{center}
\caption{\small The arctic curve for $\mu=\lambda-5\eta=0$, here at $\eta=\pi/8$. It is formed of a ``normal'' portion (solid curve tangent to
the $x=1$ line) and its
symmetric portion under $180^\circ$ rotation, and by a ``shear'' portion (dashed curve tangent to the $x=1$ line) and its symmetric portions
under $180^\circ$ rotation and $x\leftrightarrow y$ symmetry.}
\label{fig:ACpisurhuit}
\end{figure}

The corresponding portion of arctic curve (hereafter called the ``normal'' portion) is obtained by solving the system of
equations $F(x,y;\xi)=\partial_\xi F(x,y;\xi)=0$,
which is linear in the variables $x$ and $y$, hence straightforwardly yields a parametric expression $(x(\xi),y(\xi))$:
\begin{equation*}
x(\xi)=1+\frac{\partial_\xi r(\tau(\xi))}{\partial_\xi s(\tau(\xi))}\ ,\qquad y(\xi)=r(\tau(\xi))- s(\tau(\xi)) \frac{\partial_\xi r(\tau(\xi))}{\partial_\xi s(\tau(\xi))}\ , \qquad \xi \in [0,\pi-6\eta]\ .
\end{equation*}
This completes the proof of Theorem~\ref{thm:acnormal} with the identification $R_0(\xi)=r(\tau(\xi))$ and $S_0(\xi)=s(\tau(\xi))$. We do not give more explicit expressions here as the formulas are quite involved and not particularly illuminating.
A plot of this portion of arctic curve is depicted in Figure~\ref{fig:ACpisurhuit}.

Strictly speaking, the above expressions hold for the 20V model with DWBC2 only. As explained in Section~\ref{sec:tgmethod}, 
we expect however that \emph{the very same portion is found for DWBC1} since the two boundary conditions differ only by the presence of one extra path, a difference which is irrelevant in the continuous limit.
From the DWBC1/2 symmetry under rotation by $180^\circ$, this implies that the portion $(x(\xi),y(\xi))_{\xi\in [0,\pi-6\eta]}$
has a symmetric portion $(1-x(\xi),1-y(\xi))_{\xi\in [0,\pi-6\eta]}$.

Note that each portion of curve is invariant under $x\leftrightarrow y$, in agreement with the fact that, for $\mu=\lambda-5\eta=0$, 
the model has this symmetry. In the above parametric formulation, this symmetry is a consequence of the easily checked property:
\begin{equation*}
F(x,y;\xi)=  \frac{\sin(\xi+6\eta)\sin(\xi+2\eta)}{\sin(\xi)\sin(\xi+4\eta)}\, F(y,x;\pi-6\eta-\xi)\ .
\end{equation*}

\subsection{The tangent method with the shear trick}
\label{sec:shearbranch}

\begin{figure}
\begin{center}
\includegraphics[width=12cm]{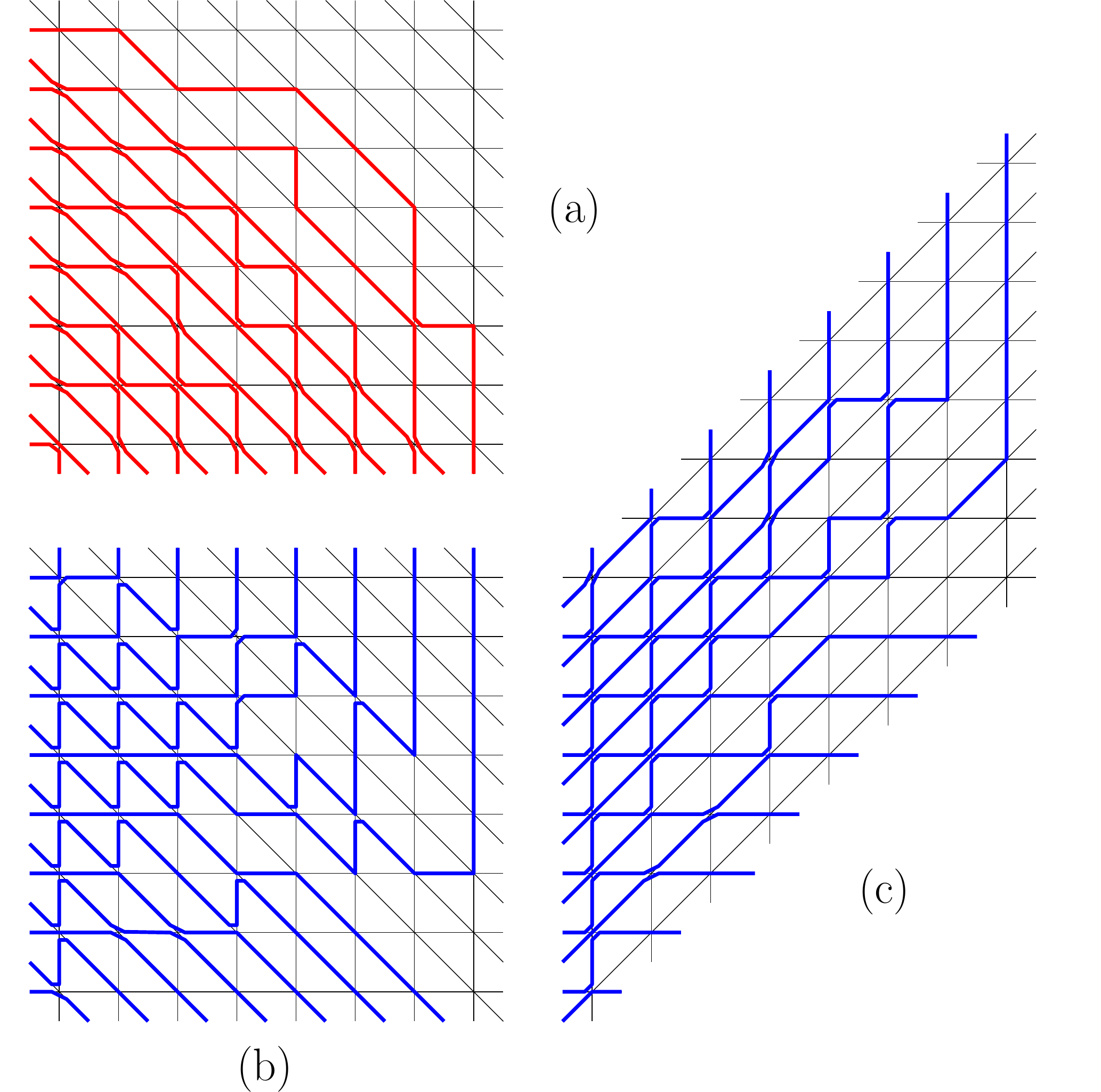}
\end{center}
\caption{\small  Starting from a configuration of the 20V model with DWBC2 (a), we first  
interchange the occupied and non-occupied \emph{vertical} edges (b) and then perform a global shear of the lattice (c) so as
to transform the original square domain into a rhombus. The re-connection of the occupied steps into osculating paths performed in (b) is such that in (c), 
all the paths are oriented from lower left to upper right at each node.}
\label{fig:shear}
\end{figure}

\begin{figure}
\begin{center}
\includegraphics[width=12cm]{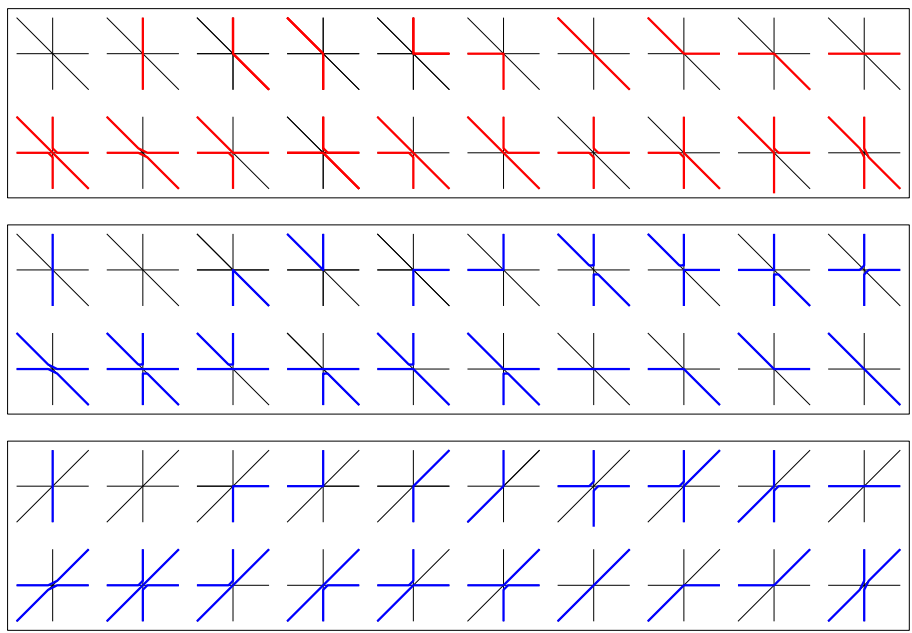}
\end{center}
\caption{\small At the level of a node, the transformation of Figure~\ref{fig:shear} maps the vertices of
the 20V model (top set) to vertices of some inverted 20V model (bottom set) where the paths are
now oriented from South, Southwest or West to East, Northeast or North. We have also indicated the intermediate set of vertices (middle set) 
before shear. }
\label{fig:sheartransf}
\end{figure}

The main result of this section is the identification of a second portion of the arctic curve via what we shall call the ``shear trick'':
\begin{thm} 
\label{thm:acshear}
The portion of arctic curve for the 20V model with DWBC2 at $\mu=0$ and $\lambda=5\eta$, as predicted by applying the tangent method to
a different set of paths and using the ``shear trick'' has the following parametric equation:
\begin{equation*}
x(\xi)=1+\frac{\partial_\xi \bar{R}_0(\xi)}{\partial_\xi \bar{S}_0(\xi)}\ ,\qquad y(\xi)=\bar{R}_0(\xi)- \bar{S}_0(\xi) \frac{\partial_\xi \bar{R}_0(\xi)}{\partial_\xi \bar{S}_0(\xi)}\ , \qquad \xi \in [-2\eta,0]\ .
\end{equation*}
where
\begin{equation*}
\begin{split}
\bar{R}_0(\xi)&=\left(\cot(\xi+4\eta)-\cot(\xi)+\alpha\cot(\alpha\, \xi)
-\alpha\cot(\alpha(\xi+4\eta))\right)
\frac{\sin(\xi+6\eta)\sin(\xi+2\eta)}{\sin(4\eta)}\ ,\\
\bar{S}_0(\xi)&=\left(\frac{\sin(\xi+6\eta)\sin(\xi+2\eta)}{\sin(\xi)\sin(\xi+4\eta)}-\frac{\sin(\xi+6\eta)\sin(\xi+2\eta)}{\sin(\xi-2\eta)\sin(\xi+2\eta)}  \right)\ ,\\
\end{split}
\end{equation*}
and $\alpha=\pi/(\pi-2\eta)$.
\end{thm}

To compute the missing portions of the arctic curve, we recourse to a different family of osculating paths, obtained as follows:
starting from a given configuration of osculating paths as in Figure~\ref{fig:DWBC}
for DWBC2, the new osculating path
configuration is obtained by (i) \emph{exchanging the occupied and non-occupied vertical edges} and (ii) \emph{performing a global shear} of the lattice so as
to recover the same vertices as those of the original 20V model but with diagonals now in the other direction
(see Figure~\ref{fig:shear}). For a given set of occupied edges after transformations (i) and (ii), the elementary steps are re-connected
at each node in the unique way which creates osculating paths all
oriented from South, Southwest or West to East, Northeast or North. As depicted in Figure~\ref{fig:sheartransf}, the transformations 
(i) and (ii), together with the above connection prescription, creates node environments which form some
``inverted'' 20V model, with a set of vertices identical to those of the original 20V model, up to a left-right symmetry.
Note that this symmetry of vertex configurations holds globally but that a given vertex is \emph{not} transformed into its left-right symmetric vertex: 
for instance, the empty vertex is mapped onto the vertex with exactly two occupied vertical edges and conversely.

Note the following important changes for our new inverted 20V model with respect to the original DWBC2 setting:
\begin{itemize}
\item{The inverted 20V model has modified weights inherited from its pre-image:  the non trivial $\varpi_0=1+2\cos(4\eta)$ weight is now attached 
to the vertex with two occupied vertical edges or its complementary vertex (left column of Figure~\ref{fig:sheartransf}).}
\item{The domain spanned by the paths is no longer a square but a rhombus (see Figure~\ref{fig:shear}-(c)).}
\item{The DWBC2 prescription is replaced by the following boundary conditions: ordering the paths from bottom to top according to their starting (diagonal or horizontal) 
edge on the left boundary, the $(n-1)$ first paths (i.e. the lower ``half'' of the paths) now have their final step at the first $n-1$ successive horizontal edges along the lower (diagonal) boundary
of the rhombus while the $n$ last paths (i.e. the upper half of the paths) have their final step at the successive vertical edges along on the upper (diagonal) boundary of the rhombus 
(see Figure~\ref{fig:shear}-(c)). }
\end{itemize} 

\begin{figure}
\begin{center}
\includegraphics[width=10cm]{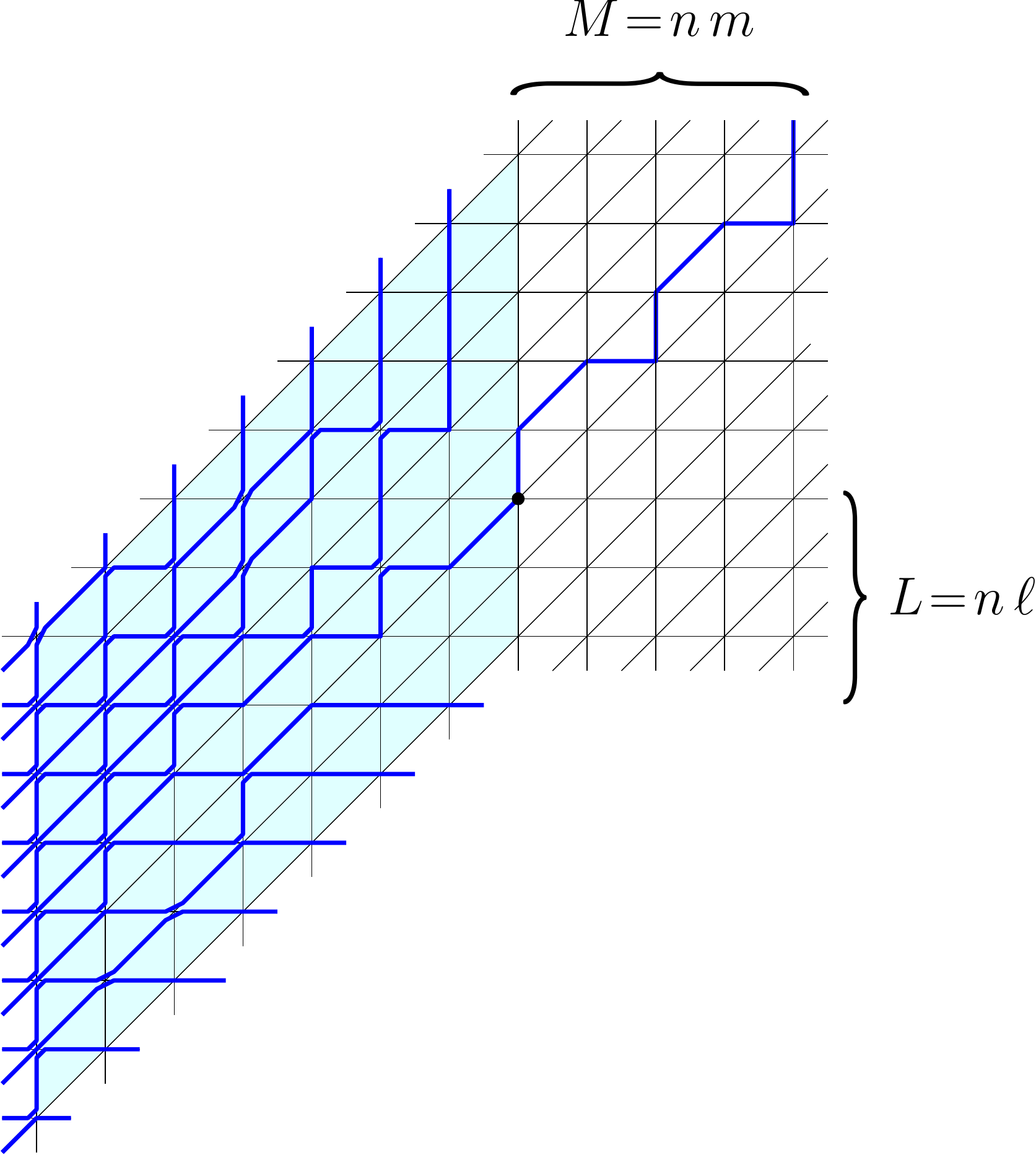}
\end{center}
\caption{\small A sample configuration of the inverted 20V model in which the $n$-th path (counted from the bottom)
exits the domain $1<X,(1+Y-X)<n$ (in light blue) at position $(n,n+L-1)=(n, n(1+\ell)-1)$ and reaches a shifted endpoint at position 
$(n+M,2n)=(n(1+m),2n)$ via a final vertical edge.}
\label{fig:tgmethod2}
\end{figure}

The transformation (i) above was designed to create, in the new geometry, a region empty of paths in the vicinity of the lower-right corner of the rhombus. The desired new portion of arctic
curve is obtained in the scaling limit as the frontier between this empty region and that occupied by the paths \emph{reaching the upper boundary
of the rhombus}.
To obtain this frontier by the tangent method, we again use a geometry where the outermost of these paths, namely the path ending 
at the rightmost vertical edge of the upper boundary of the rhombus (i.e.\ the $n$-th path, whose endpoint is originally at position $(n,2n)$) 
exits the originally accessible domain (with a rhombus shape) on the vertical right boundary at some position\footnote{The escape point is defined with the same convention as before as the position where the escape path hits the right boundary of the original rhombic domain for the first time.} $(n,n+L-1)=(n,n(1+\ell)-1)$  in the original coordinate 
system $(X,Y)$  and reaches its shifted endpoint at position $(n+M,2n)=(n(1+m),2n)$ (see Figure \ref{fig:tgmethod2}). In rescaled coordinates 
$(x,y)=(X/n,Y/n)$, this corresponds 
respectively to position $(1,1+\ell)$) for the escape point and $(1+m,2)$ for the endpoint.

As before, the most likely escape point position $\ell$ for a given $m$ is obtained by maximizing the appropriate quantity, namely:
\begin{equation*}
H^{20V_{BC2}}_n(\tau)\vert_{\tau^{n\ell}} \frac{\bar{Y}^{20V}_{(n,n+n\ell-1)\to (n(1+m),2n)}}{\varpi_0^{n-n\ell}}
\end{equation*}
with respect to $\ell$. The first term $H^{20V_{BC2}}_n(\tau)\vert_{\tau^{n\ell}}$ corresponds to the contribution of the new osculating paths inside the rhombus.
Here we implicitly use the one-to-one correspondence between our two families of osculating paths, which clearly has the following property:
the position of the hitting point of the outermost path of the second family is $(n,n+L-1)=(n,n(1+\ell)-1)$ if and only if the position of the hitting
point of the uppermost path of the first family for the corresponding DWBC2 path configuration is $(n,n\ell)$. The second term 
$\bar{Y}^{20V}_{(n,n+n\ell-1)\to (n(1+m),2n)}$ denotes the partition function for a single path from $(n,n+n\ell-1)$ to $(n(1+m),2n)$ 
(starting possibly with a number of vertical steps) \emph{in the inverted 20V model setting} with, at each node along the path, the same weight as that of the 
corresponding inverted 20V configuration. As before, we view this path as embedded in a background of empty vertices, which now receive the weight $1$
in the inverted 20V model setting. To be be fully consistent, we must finally replace in $H^{20V_{BC2}}_n(\tau)\vert_{\tau^{n\ell}}$ the part of path
going (in the inverted setting) from $(n,n+n\ell-1)$ to $(n,2n)$ by a set of $n-n\ell$ empty vertices since this portion of path is no longer present and replaced by a portion of path enumerated by 
$\bar{Y}^{20V}_{(n,n+n\ell-1)\to (n(1+m),2n)}$. This replacement of $n-\ell$ vertices originally with two adjacent vertical edges (each enumerated with 
an inverted 20V weight $\varpi_0$ in $H^{20V_{BC2}}_n(\tau)\vert_{\tau^{n\ell}}$) 
by $n-n\ell$ empty vertices (new inverted 20V weight $1$) leads to the denominator $\varpi_0^{n-n\ell}$.

To compute the large $n$ asymptotics of $\bar{Y}^{20V}_{(n,n+n\ell-1)\to (n(1+m),2n)}$, namely
\begin{equation*}
\bar{Y}^{20V}_{(n,n+n\ell-1)\to (n(1+m),2n)}\underset{n\to \infty}{\sim} {\rm e}^{n\, \bar{S}(\ell,m)}\ ,
\end{equation*}
we recourse as before to the transfer matrix approach of Appendix \ref{sec:trmat} in a version adapted to the present ``shear'' geometry:
this corresponds to setting $\alpha_1=\varpi_0$, $\alpha_2=1$, $\alpha_3=2-\varpi_0$, $\alpha_4=1-\varpi_0$ and $\alpha_5=\alpha_6=0$
and, as explained in Appendix \ref{sec:trmat}, performing the change $\ell\to 1-\ell$ as the height 
difference for the path beyond the escape point is now measured from the top, hence equal to $1-\ell$ instead of $\ell$. 
We therefore have
\begin{equation*}
\begin{split}
\bar{S}(\ell,m)&=\bar{S}(\ell,m,p_3,p_4)\\
& :=((1-\ell)+m-p_3-2p_4){\rm Log}((1-\ell)+m-p_3-2p_4)-p_3\, {\rm Log}\, p_3-p_4\, {\rm Log}\, p_4\\ &-((1-\ell)-p_3-2p_4){\rm Log}((1-\ell)-p_3-2p_4) 
 -(m-p_3-p_4){\rm Log}(m-p_3-p_4)\\& 
+((1-\ell)-p_3-2p_4) {\rm Log}\, \varpi_0 +p_3 {\rm Log}(2-\varpi_0)+
p_4 {\rm Log}(1-\varpi_0)\\
\end{split}
\end{equation*}
taken at the values of $p_3$ and $p_4$ which maximize $\bar{S}(\ell,m,p_3,p_4)$ at fixed $\ell$ and $m$.

The extremization conditions over $\tau$ and $\ell$ now read
\begin{equation*}
\begin{split}
& \frac{d}{d\tau} \left( -\ell\, {\rm Log}\,\tau + f(\sigma(\tau))\right)=0\\
& \frac{d}{d\ell}  \left(-\ell\, {\rm Log}\,\tau  + \bar{S}(\ell,m) -(1-\ell)\, {\rm Log}\,\varpi_0\right)=0\\
\end{split}
\end{equation*}
or equivalently
\begin{equation*}
\ell=r(\tau), \quad \tau=\varpi_0\, {\rm e}^{\frac{d}{d\ell} \bar{S}(\ell,m)}\ .
\end{equation*}
In the new geometry of Figure~\ref{fig:tgmethod2}, the corresponding tangent line passing trough $(1,1+\ell)$ and $(1+m,2)$ has equation
$y+\frac{\ell-1}{m}(x-1)-\ell-1=0$. Therefore, \emph{after performing the inverse shear transformation $y\to y-x$} to bring
back the path in the original geometry of a square domain, the equation of the tangent line reads: $y+\left(1-\frac{1-\ell}{m}\right)(x-1)- \ell=0$.  
For the most likely escape point, this yields 
the tangent line equation:
\begin{equation}
y+\bar{s}(\tau) (x-1)-r(\tau)=0
\label{eq:tglinebis}
\end{equation}
where the slope $\bar{s}(\tau)$ is now given by 
\begin{equation}
\bar{s}(\tau)=1-\frac{1-r(\tau)}{m}\ \hbox{with $m:=m(\tau)$ solution of}\ \tau=\varpi_0\, {\rm e}^{\frac{d}{d\ell} \bar{S}(\ell,m)}\ \hbox{at}\ \ell=r(\tau)\ .
\label{eq:eqforsbar}
\end{equation}
Writing the extremization conditions $\partial_{p_3} \bar{S}(\ell,m,p_3,p_4)=\partial_{p_4} \bar{S}(\ell,m,p_3,p_4)=0$ and solving \eqref{eq:eqforsbar}
yields now
\begin{equation*}
\bar{s}(\tau)= \frac{\tau(1+\varpi_0)}{(\tau-1)(\tau+\varpi_0)}+\frac{\varpi_0}{\tau(1-\varpi_0)+\varpi_0}\ \Rightarrow
\bar{s}(\tau(\xi))=\frac{\sin(\xi+6\eta)\sin(\xi+2\eta)}{\sin(\xi)\sin(\xi+4\eta)}-\frac{\sin(\xi+6\eta)\sin(\xi+2\eta)}{\sin(\xi-2\eta)\sin(\xi+2\eta)}\ .
\end{equation*}
We end up with a new family of tangent lines parametrized by $\xi$ via
\begin{equation*}
\begin{split}
0=\bar{F}(x,y;\xi)&:=y+\bar{s}(\tau(\xi)) (x-1)-r(\tau(\xi))\\&=
y+\left(\frac{\sin(\xi+6\eta)\sin(\xi+2\eta)}{\sin(\xi)\sin(\xi+4\eta)}-\frac{\sin(\xi+6\eta)\sin(\xi+2\eta)}{\sin(\xi-2\eta)\sin(\xi+2\eta)}  \right)(x-1)\\
& \quad -\left(\cot(\xi+4\eta)-\cot(\xi)+\alpha\cot(\alpha\, \xi)
-\alpha\cot(\alpha(\xi+4\eta))\right)
\frac{\sin(\xi+6\eta)\sin(\xi+2\eta)}{\sin(4\eta)}\ .\\
\end{split}
\end{equation*}
The parameter $\xi$ now spans the range $[-2\eta,0]$. The range $[-2\eta,-\eta]$ corresponds to $\bar{s}(\tau(\xi))$ decreasing from 
$1$ to $0$, while the range $[-\eta,0]$ corresponds to $\bar{s}(\tau(\xi))$ decreasing from $0$ to $-\infty$, as dictated by the tangent
method in the present geometry.

The corresponding new portion of arctic curve, hereafter referred to as the ``shear'' portion, is obtained by solving the linear system $\bar{F}(x,y,\xi)=\partial_\xi \bar{F}(x,y,\xi)=0$
which yields again a parametric expression $({x}(\xi),{y}(\xi))$
\begin{equation*}
{x}(\xi)=1+\frac{\partial_\xi r(\tau(\xi))}{\partial_\xi \bar{s}(\tau(\xi))}\ ,\qquad {y}(\xi)=r(\tau(\xi))- \bar{s}(\tau(\xi)) \frac{\partial_\xi r(\tau(\xi))}{\partial_\xi \bar{s}(\tau(\xi))}\ , \qquad \xi \in [-2\eta,0]\ .
\end{equation*}

This completes the proof of Theorem~\ref{thm:acshear} with the identifications $\bar{R}_0(\xi)=r(\tau(\xi))$ and
$\bar{S}_0(\xi)=\bar{s}(\tau(\xi))$. This branch of the arctic curve is the same for DWBC2 and DWBC1 and the symmetry DWBC1/2 under rotation by $180^\circ$ implies that the portion
$({x}(\xi),{y}(\xi))_{\xi\in [-2\eta,0]}$
has a symmetric portion $(1-{x}(\xi),1-{y}(\xi))_{\xi\in [-2\eta,0]}$. Finally, the $x\leftrightarrow y$ symmetry 
of the problem (since the 20V weights have this symmetry) implies the existence of two symmetric portions
$({y}(\xi),{x}(\xi))_{\xi\in [-2\eta,0]}$ and $(1-{y}(\xi),1-{x}(\xi))_{\xi\in [-2\eta,0]}$. This leads to a total of four 
portions which, together with the two previously computed portions, span the entire arctic curve (see Figure~\ref{fig:ACpisurhuit} for an illustration).

\medskip

To conclude this section, let us discuss the ``uniform'' case where all the weights are equal to $1$,
as easily obtained by setting $\eta=\pi/8$ in the above expressions.
In this case the relation between $\tau$ and $\sigma$ is simply
\begin{equation*}
\tau=2\sigma-1\ .
\end{equation*}
The solution for the ``normal'' portion of the arctic curve in given explicitly by $(x(\xi),y(\xi))=(x_{\pi/8}(\xi),y_{\pi/8}(\xi))$
with
\begin{equation*}
\begin{split}
x_{\pi/8}(\xi)&=\frac{1}{18}\left(3\left(5 \cos\left(\frac{2\xi}{3}\right)+\cos\left(\frac{10\xi}{3}\right)\right)
-\sqrt{3}\left(5 \sin\left(\frac{2\xi}{3}\right)-\sin\left(\frac{10\xi}{3}\right)\right)\right)\\
y_{\pi/8}(\xi)&=\frac{1}{18}\left(\sqrt{3}\left(5 \cos\left(\frac{2\xi}{3}\right)-\cos\left(\frac{10\xi}{3}\right)\right)
+3\left(5 \sin\left(\frac{2\xi}{3}\right)+\sin\left(\frac{10\xi}{3}\right)\right)\right)\\
\end{split}
\end{equation*}
with $\xi\in [0,\pi/4]$ (see Figure~\ref{fig:ACpisurhuit}).
This describes a portion of some algebraic curve with equation
\begin{equation}
\begin{split}
3^{11}(x^2+y^2)^5+3^9\, 10(x^2+y^2)^4-& 3^6\, 5(x^2+y^2)^3+6^2\, 20(73(x^2+y^2)^2-5^4x^2y^2)\\
& -2^8\, 15(x^2+y^2)-2^{12}=0\ .\\
\end{split}
\label{eq:ACuniform}
\end{equation}

As for the ``shear'' portion, it is given by 
\begin{equation*}
({x}(\xi),{y}(\xi))=(x_{\pi/8}(\xi),y_{\pi/8}(\xi)-x_{\pi/8}(\xi)+1)
\end{equation*}
with $\xi\in[-\pi/4,0]$. Otherwise stated, in the uniform case, the ``shear'' portion is obtained from the analytic continuation of the
``normal'' portion by a simple shear transformation sending the $y=0$ line onto the line $x+y=1$. We will comment further on this in Section~\ref{sec:ASMDPP}. Note finally that, as explained in \cite{DFG20V}, configuration of the 20V model with DWBC1 or 2 are in bijection with 
so-called Alternating Phase Matrices (APM). The above arctic curve is therefore also the limit shape of large APM's.

\section{The general case}
\label{sec:general}

Let us now extend the previous study and derive the arctic curve for the general case of weights given by
\eqref{eq:explweights}.
The results of this section are summarized in the following:
\begin{thm} 
\label{thm:acgeneral}
The arctic curve for the 20V model with DWBC2 at arbitrary admissible values of the parameters $\eta$, $\lambda$ and $\mu$ is
made generically of three portions, denoted ``normal'', ``shear'' and ``final'' together with their images under $180^\circ$ rotation.
The three branches have respectively parametric equations:
\begin{equation*}
\begin{split}
\hbox{\rm Normal:}&\qquad x_n(\xi)=1+\frac{\partial_\xi {R}_n(\xi)}{\partial_\xi {S}_n(\xi)}\ ,\qquad y_n(\xi)={R}_n(\xi)- {S}_n(\xi) \frac{\partial_\xi {R}_n(\xi)}{\partial_\xi {S}_n(\xi)}\ , \qquad \xi \in [0,\pi-\lambda-\eta]\\
\hbox{\rm Shear:}&\qquad x_s(\xi)=1+\frac{\partial_\xi {R}_s(\xi)}{\partial_\xi {S}_s(\xi)}\ ,\qquad y_s(\xi)={R}_s(\xi)- {S}_s(\xi) \frac{\partial_\xi {R}_s(\xi)}{\partial_\xi {S}_s(\xi)}\ , \qquad \xi \in \left[-\frac{\lambda-\eta+\mu}{2},0\right]\\
\hbox{\rm Final:}&\qquad x_f(\xi)={R}_f(\xi)- {S}_f(\xi) \frac{\partial_\xi {R}_f(\xi)}{\partial_\xi {S}_f(\xi)}\ ,\qquad 
y_f(\xi)=1+\frac{\partial_\xi {R}_f(\xi)}{\partial_\xi {S}_f(\xi)}\ , \qquad \xi \in \left[-\frac{\lambda-\eta-\mu}{2},0\right]\\
\end{split}
\end{equation*}
where
\begin{equation*}
\begin{split}
{R}_n(\xi)&={R}_s(\xi)=
\left(\cot(\xi+\lambda-\eta)-\cot(\xi)+\alpha\cot(\alpha\, \xi)
-\alpha\cot(\alpha(\xi+\lambda-\eta))\right)\\
&\ \ \times
\frac{\sin(\xi+\lambda+\eta)\sin(\xi+\lambda-\eta)\sin\left(
\xi+\frac{\lambda-\eta+\mu}{2}\right)\sin\left(
\xi+\frac{\lambda+3\eta+\mu}{2}\right)}{\sin(2\eta)\left(\sin(\xi+\lambda+\eta)\sin(\xi+\lambda-\eta)+\sin\left(
\xi+\frac{\lambda-\eta+\mu}{2}\right)\sin\left(
\xi+\frac{\lambda+3\eta+\mu}{2}\right)\right)}\ ,
\\
{R}_f(\xi)
&=\left(\cot(\xi+\lambda-\eta)-\cot(\xi)+\alpha\cot(\alpha\, \xi)
-\alpha\cot(\alpha(\xi+\lambda-\eta))\right)\\
&\ \ \times
\frac{\sin(\xi+\lambda+\eta)\sin(\xi+\lambda-\eta)\sin\left(
\xi+\frac{\lambda-\eta-\mu}{2}\right)\sin\left(
\xi+\frac{\lambda+3\eta-\mu}{2}\right)}{\sin(2\eta)\left(\sin(\xi+\lambda+\eta)\sin(\xi+\lambda-\eta)+\sin\left(
\xi+\frac{\lambda-\eta-\mu}{2}\right)\sin\left(
\xi+\frac{\lambda+3\eta-\mu}{2}\right)\right)}\ ,
\\
\end{split}
\end{equation*}
\begin{equation*}
\begin{split}
{S}_n(\xi)&=\frac{\sin(\xi+\lambda+\eta)\sin(\xi+\lambda-\eta)\left(\sin(\xi)\sin(\xi+2\eta)+\sin\left(\xi+\frac{\lambda-\eta+\mu}{2}\right)
\sin\left(\xi+\frac{\lambda+3\eta+\mu}{2}\right)\right)}{\sin(\xi)\sin(\xi+2\eta)\left(\sin(\xi+\lambda+\eta)\sin(\xi+\lambda-\eta)+\sin\left(\xi+\frac{\lambda-\eta+\mu}{2}\right)
\sin\left(\xi+\frac{\lambda+3\eta+\mu}{2}\right)\right)}\ ,\\
{S}_s(\xi)&=\frac{\sin(\xi+\lambda+\eta)\sin(\xi+\lambda-\eta)\sin\left(
2\xi+\frac{\lambda-\eta+\mu}{2}\right)\sin\left(
\frac{\lambda+3\eta+\mu}{2}\right)}{\sin(2\eta-\xi)\sin(\xi)\left(\sin(\xi+\lambda+\eta)\sin(\xi+\lambda-\eta)+\sin\left(
\xi+\frac{\lambda-\eta+\mu}{2}\right)\sin\left(
\xi+\frac{\lambda+3\eta+\mu}{2}\right)\right)}
\ , \\
{S}_f(\xi)&=\frac{\sin(\xi+\lambda+\eta)\sin(\xi+\lambda-\eta)\sin\left(
2\xi+\frac{\lambda-\eta-\mu}{2}\right)\sin\left(
\frac{\lambda+3\eta-\mu}{2}\right)}{\sin(2\eta-\xi)\sin(\xi)\left(\sin(\xi+\lambda+\eta)\sin(\xi+\lambda-\eta)+\sin\left(
\xi+\frac{\lambda-\eta-\mu}{2}\right)\sin\left(
\xi+\frac{\lambda+3\eta-\mu}{2}\right)\right)}
\ , \\
\end{split}
\end{equation*}
and $\alpha=\pi/(\pi-2\eta)$.
\end{thm}
The three following sub-sections are devoted to the derivation of these three sets of parametric expressions.

\subsection{The ``normal'' portion of the arctic curve and its symmetric portion} 
\label{sec:bone}
The first branch of the arctic curve is obtained exactly as before as the envelope of the tangent 
lines with equation \eqref{eq:tgline} with $r(\tau)$ as is \eqref{eq:defr} and $s(\tau)$ now given by
\begin{equation}
s(\tau)=\frac{r(\tau)}{m}\ \hbox{with $m:=m(\tau)$ solution of}\ \tau=\frac{\omega_0}{\omega_1}\, {\rm e}^{\frac{d}{d\ell} S(\ell,m)}\ \hbox{at}\ \ell=r(\tau)
\label{eq:neweqforsgen}
\end{equation}
since the quantity to maximize with respect to $\ell$ is now
\begin{equation*}
H^{20V_{BC2}}_n(\tau)\vert_{\tau^{n\ell}} Y^{20V}_{(n,n\ell)\to (n(1+m),0)}\, \left(\frac{\omega_0}{\omega_1}\right)^{n\ell-1}
\end{equation*}
with the last factor corresponding, as before, to the replacement of $n\ell-1$ vertices traversed by a vertical line (weight $\omega_1$) by  $n\ell-1$ empty vertices
(weight $\omega_0$).

In the presence of the seven weights $\omega_0,\dots,\omega_6$, the more involved value of $S(\ell,m)$ is again obtained by the transfer matrix approach of Appendix \ref{sec:trmat}.
Its expression may be written as 
\begingroup
\allowdisplaybreaks
\begin{align*}
S(\ell,m)&=S(\ell,m,p_3,p_4,p_5,p_6)\\
& :=(\ell+m-p_3-2p_4-2p_5-3p_6){\rm Log}(\ell+m-p_3-2p_4-2p_5-3p_6)\\&
-p_3\, {\rm Log}\, p_3-p_4\, {\rm Log}\, p_4-p_5\, {\rm Log}\, p_5-p_6\, {\rm Log}\, p_6\\ 
&-(\ell-p_3-2p_4-p_5-2p_6){\rm Log}(\ell-p_3-2p_4-p_5-2p_6) \\&
 -(m-p_3-p_4-2p_5-2p_6){\rm Log}(m-p_3-p_4-2p_5-2p_6)\\& 
+(\ell-p_3-2p_4-p_5-2p_6) {\rm Log}\left(\frac{\omega_1}{\omega_0}\right) +(m-p_3-p_4-2p_5-2p_6) {\rm Log}\left(\frac{\omega_6}{\omega_0}\right)\\&+p_3 {\rm Log}\left(\frac{\omega_0\omega_3+\omega_4^2-\omega_1\omega_6}{\omega_0^2}\right)+
p_4 {\rm Log}\left(\frac{\omega_2^2-\omega_1\omega_3}{\omega_0^2}\right)+p_5 {\rm Log}\left(\frac{\omega_5^2-\omega_6\omega_3}{\omega_0^2}\right)\
\\& +p_6 {\rm Log}\left(\frac{2\omega_2\omega_4\omega_5+\omega_1\omega_6\omega_3-\omega_3\omega_4^2-\omega_1\omega_5^2-\omega_6\omega_2^2}{\omega_0^3}\right)\\
\end{align*}
\endgroup
taken at the values of $p_3, \dots, p_6$ which maximize $S(\ell,m,p_3,p_4,p_5,p_6)$ at fixed $\ell$ and $m$.
Writing $\partial_{p_i} S(\ell,m,p_3,p_4,p_5,p_6)=0$ for $i=3,\dots,6$ and solving \eqref{eq:neweqforsgen}
yields now the parametric expression for $s(\tau)$:
\begin{equation*}
s(\tau(\xi))=\frac{\sin(\xi+\lambda+\eta)\sin(\xi+\lambda-\eta)\left(\sin(\xi)\sin(\xi+2\eta)+\sin\left(\xi+\frac{\lambda-\eta+\mu}{2}\right)
\sin\left(\xi+\frac{\lambda+3\eta+\mu}{2}\right)\right)}{\sin(\xi)\sin(\xi+2\eta)\left(\sin(\xi+\lambda+\eta)\sin(\xi+\lambda-\eta)+\sin\left(\xi+\frac{\lambda-\eta+\mu}{2}\right)
\sin\left(\xi+\frac{\lambda+3\eta+\mu}{2}\right)\right)}
\end{equation*}
with $\tau=\tau(\xi)$ as in \eqref{eq:tauparamgen}.
We end up with the parametric equation for the tangent lines
\begin{equation*}
0=F(x,y;\xi):=y+s(\tau(\xi)) (x-1)-r(\tau(\xi))
\end{equation*}
with $s(\tau(\xi))$ as above and with the general expression $r(\tau(\xi))$ of \eqref{eq:valrgen}, while $\xi$ now runs over $[0,\pi-\lambda-\eta]$.
As before, the corresponding portion of arctic curve has the parametric expression $(x(\xi),y(\xi))$ with
\begin{equation*}
x(\xi)=1+\frac{\partial_\xi r(\tau(\xi))}{\partial_\xi s(\tau(\xi))}\ ,\qquad y(\xi)=r(\tau(\xi))- s(\tau(\xi)) \frac{\partial_\xi r(\tau(\xi))}{\partial_\xi s(\tau(\xi))}\ , \qquad \xi \in [0,\pi-\lambda-\eta]\ .
\end{equation*}
This completes the proof of the first branch of arctic curve in Theorem~\ref{thm:acgeneral} with the identifications  $(x_n(\xi),y_n(\xi))=(x(\xi),y(\xi))$,
$R_n(\xi)=r(\tau(\xi))$ and $S_n(\xi)=s(\tau(\xi))$.
From the symmetry DWBC1/2 under rotation by $180^\circ$, this ``normal'' portion of arctic curve $(x(\xi),y(\xi))_{\xi\in [0,\pi-\lambda-\eta]}$
has a symmetric portion $(1-x(\xi),1-y(\xi))_{\xi\in [0,\pi-\lambda-\eta]}$.

\subsection{The ``shear'' portion of the arctic curve and its symmetric portion}
\label{sec:btwo}
The second branch of the arctic curve is obtained as in Section~\ref{sec:shearbranch} thanks to the shear trick.
The equation of tangent lines is again given by \eqref{eq:tglinebis} with $r(\tau)$ as is \eqref{eq:defr} and $\bar{s}(\tau)$ now given by
\begin{equation*}
\bar{s}(\tau)=1-\frac{1-r(\tau)}{m}\ \hbox{with $m:=m(\tau)$ solution of}\ \tau=\frac{\omega_0}{\omega_1}\, {\rm e}^{\frac{d}{d\ell} \bar{S}(\ell,m)}\ \hbox{at}\ \ell=r(\tau)
\end{equation*}
since the quantity to maximize is now 
\begin{equation*}
H^{20V_{BC2}}_n(\tau)\vert_{\tau^{n\ell}}\bar{Y}^{20V}_{(n,n+n\ell-1)\to (n(1+m),2n)}\left(\frac{\omega_1}{\omega_0}\right)^{n-n\ell}\ .
\end{equation*}
\begin{figure}
\begin{center}
\includegraphics[width=16cm]{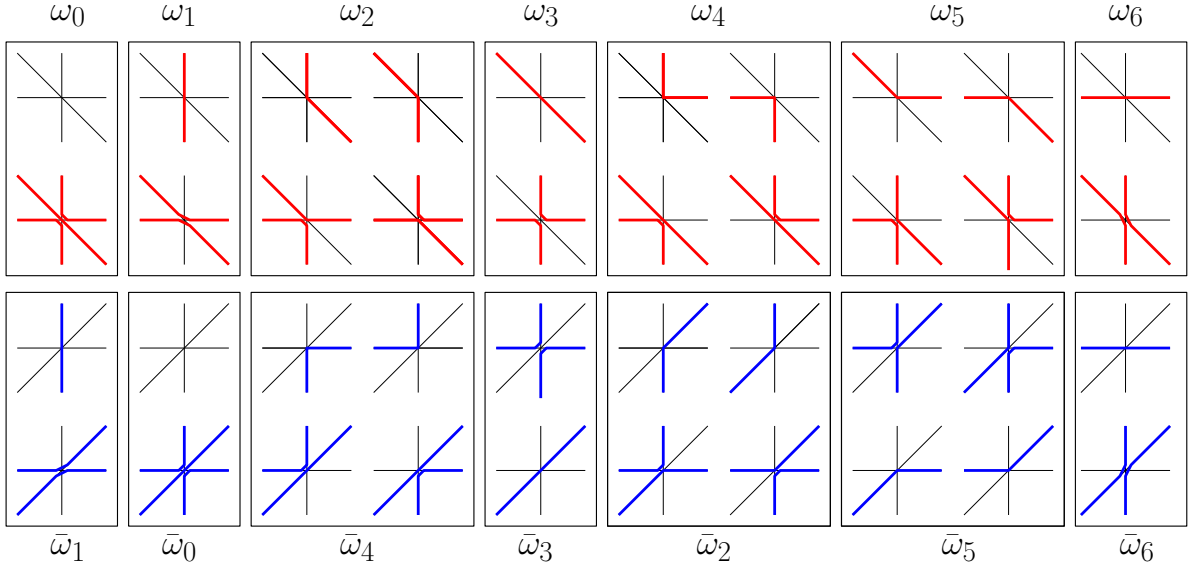}
\end{center}
\caption{\small  The weights of the twenty vertex configurations of the inverted 20V model as inherited from their counterparts in the original 20V model $\omega_i$, $i=1,\dots,6$.
For convenience, we may denote the new weights by $\bar{\omega}_i$, $i=1,\dots,6$ where the indexing is chosen so as to mimic the original model after an up-down reversal. 
}
\label{fig:shearweightsgen}
\end{figure}
The value of $\bar{S}(\ell,m)$ is obtained by the transfer matrix approach of Appendix \ref{sec:trmat} in the ``shear'' geometry. For the escape path, this
geometry is simply obtained from the ``normal'' geometry by a simple up-down symmetry (since the path now goes up), together with a change of the original 20V model weights $\omega_i$ into inverted 20V 
weights $\bar{\omega}_i$. As apparent in Figure~\ref{fig:shearweightsgen}, using for the weights $\bar{\omega}_i$ the appropriate labelling consistent with the up-down symmetry, 
we have $\bar{\omega}_0=\omega_1$, $\bar{\omega}_1=\omega_0$, $\bar{\omega}_2=\omega_4$, $\bar{\omega}_3=\omega_3$, $\bar{\omega}_4=\omega_2$,
$\bar{\omega}_5=\omega_5$ and $\bar{\omega}_6=\omega_6$. In practice, we must therefore simply perform in the expressions for the ``normal'' geometry 
the changes $\omega_0\leftrightarrow \omega_1$ and $\omega_2\leftrightarrow \omega_4$, together with the substitution $\ell\to 1-\ell$.
To summarize, $\bar{S}(\ell,m)$
may be written as 
\begingroup
\allowdisplaybreaks
\begin{align*}
\bar{S}(\ell,m)&=\bar{S}(\ell,m,p_3,p_4,p_5,p_6)\\
& :=((1-\ell)+m-p_3-2p_4-2p_5-3p_6){\rm Log}((1-\ell)+m-p_3-2p_4-2p_5-3p_6)\\&
-p_3\, {\rm Log}\, p_3-p_4\, {\rm Log}\, p_4-p_5\, {\rm Log}\, p_5-p_6\, {\rm Log}\, p_6\\ 
&-((1-\ell)-p_3-2p_4-p_5-2p_6){\rm Log}((1-\ell)-p_3-2p_4-p_5-2p_6) \\&
 -(m-p_3-p_4-2p_5-2p_6){\rm Log}(m-p_3-p_4-2p_5-2p_6)\\& 
+((1-\ell)-p_3-2p_4-p_5-2p_6) {\rm Log}\left(\frac{\omega_0}{\omega_1}\right) +(m-p_3-p_4-2p_5-2p_6) {\rm Log}\left(\frac{\omega_6}{\omega_1}\right)\\&+p_3 {\rm Log}\left(\frac{\omega_1\omega_3+\omega_2^2-\omega_0\omega_6}{\omega_1^2}\right)+
p_4 {\rm Log}\left(\frac{\omega_4^2-\omega_0\omega_3}{\omega_1^2}\right)+p_5 {\rm Log}\left(\frac{\omega_5^2-\omega_6\omega_3}{\omega_1^2}\right)\
\\& +p_6 {\rm Log}\left(\frac{2\omega_4\omega_2\omega_5+\omega_0\omega_6\omega_3-\omega_3\omega_2^2-\omega_0\omega_5^2-\omega_6\omega_4^2}{\omega_1^3}\right)\\
\end{align*}
\endgroup
taken at the values of $p_3, \dots, p_6$ which maximize $\bar{S}(\ell,m,p_3,p_4,p_5,p_6)$ at fixed $\ell$ and $m$.
The extremization conditions now lead to 
\begin{equation*}
\bar{s}(\tau(\xi))=\frac{\sin(\xi+\lambda+\eta)\sin(\xi+\lambda-\eta)\sin\left(
2\xi+\frac{\lambda-\eta+\mu}{2}\right)\sin\left(
\frac{\lambda+3\eta+\mu}{2}\right)}{\sin(2\eta-\xi)\sin(\xi)\left(\sin(\xi+\lambda+\eta)\sin(\xi+\lambda-\eta)+\sin\left(
\xi+\frac{\lambda-\eta+\mu}{2}\right)\sin\left(
\xi+\frac{\lambda+3\eta+\mu}{2}\right)\right)}
\end{equation*}
with $\tau(\xi)$ as in \eqref{eq:tauparamgen}.
We end up with the parametric equation for the tangent lines
\begin{equation*}
0=\bar{F}(x,y;\xi):=y+\bar{s}(\tau(\xi)) (x-1)-r(\tau(\xi))
\end{equation*}
with $\bar{s}(\tau(\xi))$ as above and with the same general expression \eqref{eq:valrgen} for $r(\tau(\xi))$ as for the ``normal'' portion, while $\xi$ now runs over $\left[-\frac{\lambda-\eta+\mu}{2},0\right]$.
The corresponding portion of arctic curve has the parametric expression $(\bar{x}(\xi),\bar{y}(\xi))$ with
\begin{equation*}
{x}(\xi)=1+\frac{\partial_\xi r(\tau(\xi))}{\partial_\xi \bar{s}(\tau(\xi))}\ ,\qquad {y}(\xi)=r(\tau(\xi))- \bar{s}(\tau(\xi)) \frac{\partial_\xi r(\tau(\xi))}{\partial_\xi \bar{s}(\tau(\xi))}\ , 
\qquad \xi \in \left[-\frac{\lambda-\eta+\mu}{2},0\right]\ .
\end{equation*}
This completes the proof of the second branch of arctic curve in Theorem~\ref{thm:acgeneral} with the identifications  $(x_s(\xi),y_s(\xi))=(x(\xi),y(\xi))$,
$R_s(\xi)=r(\tau(\xi))$ and $S_s(\xi)=\bar{s}(\tau(\xi))$.
From the symmetry DWBC1/2 under rotation by $180^\circ$, this ``shear'' portion of arctic curve $({x}(\xi),{y}(\xi))_{\xi\in \left[-\frac{\lambda-\eta+\mu}{2},0\right]}$
has a symmetric portion $(1-{x}(\xi),1-{y}(\xi))_{\xi\in \left[-\frac{\lambda-\eta+\mu}{2},0\right]}$.
As opposed to Section~\ref{sec:warmup}, the arctic curve does not have the symmetry $x\leftrightarrow y$ since the weights explicitly break
this symmetry in general. Still, as discussed just below, the last portions of arctic curve may easily be obtained via some symmetry arguments. 

\subsection{The ``final'' portion of the arctic curve and its symmetric portion} 
\label{sec:final}
To complete the arctic curve, we resort to geometries similar to that of previous sub-sections for the ``normal'' and ``shear'' portions, 
but where the role of the $x$ and $y$ directions have been exchanged.
As already discussed, the vertex weights are not invariant under this symmetry but their modified 
values are simply obtained by changing $\mu$ into $-\mu$
in \eqref{eq:explweights}. As a consequence, new portions of arctic curve are immediately obtained from the known ones upon the 
simultaneous changes $x\leftrightarrow y$ and $\mu\leftrightarrow -\mu$.
If we start from the ``normal'' portion, we get tangent line parametric equations
\begin{equation*}
0=\tilde{F}(x,y;\xi):=x+\tilde{s}(\tilde{\tau}(\xi)) (y-1)-\tilde{r}(\tilde{\tau}(\xi))
\end{equation*}
with $\tilde{\tau}(\xi)$, $\tilde{r}(\tilde{\tau}(\xi)$ and $\tilde{s}(\tilde{\tau}(\xi))$ given by 
\begin{equation}
\begin{split}
\tilde{\tau}(\xi)&=\frac{\sin(\lambda+\eta)\sin\left(\frac{\lambda+3\eta-\mu}{2}\right)\sin(\xi+\lambda-\eta)\sin\left(
\xi+\frac{\lambda-\eta-\mu}{2}\right)}{\sin(\lambda-\eta)\sin\left(\frac{\lambda-\eta-\mu}{2}\right)\sin(\xi+\lambda+\eta)\sin\left(
\xi+\frac{\lambda+3\eta-\mu}{2}\right)}
\\
\tilde{r}(\tilde{\tau}(\xi))
&=\left(\cot(\xi+\lambda-\eta)-\cot(\xi)+\alpha\cot(\alpha\, \xi)
-\alpha\cot(\alpha(\xi+\lambda-\eta))\right)\\
&\ \ \times
\frac{\sin(\xi+\lambda+\eta)\sin(\xi+\lambda-\eta)\sin\left(
\xi+\frac{\lambda-\eta-\mu}{2}\right)\sin\left(
\xi+\frac{\lambda+3\eta-\mu}{2}\right)}{\sin(2\eta)\left(\sin(\xi+\lambda+\eta)\sin(\xi+\lambda-\eta)+\sin\left(
\xi+\frac{\lambda-\eta-\mu}{2}\right)\sin\left(
\xi+\frac{\lambda+3\eta-\mu}{2}\right)\right)}
\\
\end{split}
\label{eq:tildetaur}
\end{equation}
and 
\begin{equation*}
\tilde{s}(\tilde{\tau}(\xi)=
\frac{\sin(\xi+\lambda+\eta)\sin(\xi+\lambda-\eta)\left(\sin(\xi)\sin(\xi+2\eta)+\sin\left(\xi+\frac{\lambda-\eta-\mu}{2}\right)
\sin\left(\xi+\frac{\lambda+3\eta-\mu}{2}\right)\right)}{\sin(\xi)\sin(\xi+2\eta)\left(\sin(\xi+\lambda+\eta)\sin(\xi+\lambda-\eta)+\sin\left(\xi+\frac{\lambda-\eta-\mu}{2}\right)
\sin\left(\xi+\frac{\lambda+3\eta-\mu}{2}\right)\right)}\ .
\end{equation*}
Here $\xi$ runs again over $\left[0,\pi-\lambda-\eta\right]$. These tangent lines form the same family as those leading to the ``normal'' portion, due to the
identity 
\begin{equation*}
\begin{split}
&\tilde{F}(x,y;\pi-\lambda-\eta-\xi)=\\
&\ \frac{\sin(\xi)\sin(\xi+2\eta)\left(\sin(\xi+\lambda-\eta)\sin(\xi+\lambda+\eta)+\sin\left(\xi+\frac{\lambda-\eta+\mu}{2}\right)
\sin\left(\xi+\frac{\lambda+3\eta+\mu}{2}\right)\right)}{\sin(\xi+\lambda-\eta)\sin(\xi+\lambda+\eta)\left(\sin(\xi)\sin(\xi+2\eta)+\sin\left(\xi+\frac{\lambda-\eta+\mu}{2}\right)
\sin\left(\xi+\frac{\lambda+3\eta+\mu}{2}\right)\right)}
F(x,y;\xi)\ .\\
\end{split}
\end{equation*} 
We therefore recover the \emph{same} ``normal'' portion of the arctic curve. This could be expected since this portion has tangents intersecting the positive $x$ and $y$ axes,
and therefore may be attained by the tangent method using escape paths with displaced endpoints on the positive $x$ axis or displaced starting points
on the positive $y$ axis.
\begin{figure}
\begin{center}
\includegraphics[width=13cm]{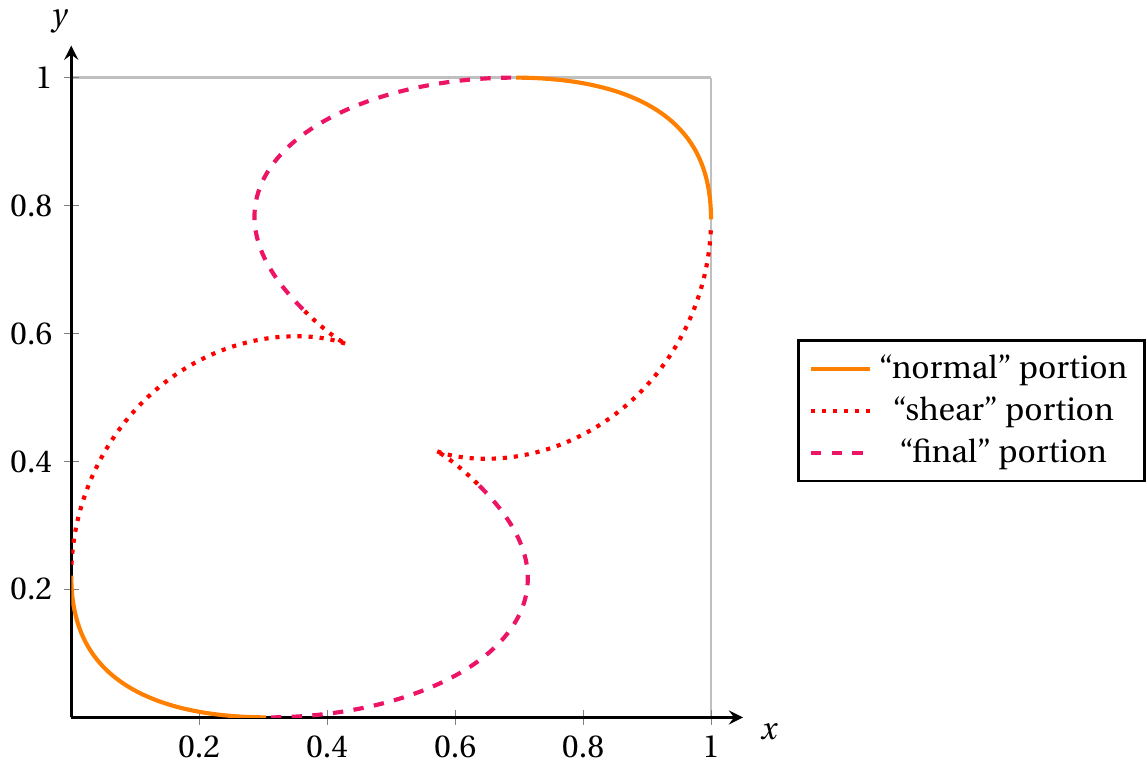} 
\end{center}
\caption{\small The arctic curve for $\eta=\pi/12$, $\lambda=10 \pi/12$ and $\mu=5\pi/12$. It is formed of a ``normal'' portion (solid curve tangent to the $x=1$ and $y=1$ lines) , a ``shear'' portion (dotted curve tangent to the $x=1$ line) 
and a final'' portion (dashed curve tangent to the $y=1$ line)
and their three symmetric portions under $180^\circ$ rotation.}
\label{fig:ACpisurdouze}
\end{figure}

More interestingly, if we start instead from the ``shear'' portion, we get a \emph{new} portion of arctic curve, hereafter called the ``final'' portion.
The corresponding tangent lines have parametric equation
\begin{equation*}
0=\tilde{\bar{F}}(x,y;\xi):=x+\tilde{\bar{s}}(\tilde{\tau}(\xi)) (y-1)-\tilde{r}(\tilde{\tau}(\xi))
\end{equation*}
with $\tilde{\tau}(\xi)$ and $\tilde{r}(\tilde{\tau}(\xi))$ given by \eqref{eq:tildetaur} and 
\begin{equation*}
\tilde{\bar{s}}(\tilde{\tau}(\xi))=\frac{\sin(\xi+\lambda+\eta)\sin(\xi+\lambda-\eta)\sin\left(
2\xi+\frac{\lambda-\eta-\mu}{2}\right)\sin\left(
\frac{\lambda+3\eta-\mu}{2}\right)}{\sin(2\eta-\xi)\sin(\xi)\left(\sin(\xi+\lambda+\eta)\sin(\xi+\lambda-\eta)+\sin\left(
\xi+\frac{\lambda-\eta-\mu}{2}\right)\sin\left(
\xi+\frac{\lambda+3\eta-\mu}{2}\right)\right)}\ .
\end{equation*}
Here $\xi$ now runs over $\left[-\frac{\lambda-\eta-\mu}{2},0\right]$.
The corresponding portion of arctic curve has the parametric expression $({x}(\xi),{y}(\xi))$ with
\begin{equation*}
{x}(\xi)=\tilde{r}(\tilde{\tau}(\xi))- \tilde{\bar{s}}(\tilde{\tau}(\xi)) 
\frac{\partial_\xi \tilde{r}(\tilde{\tau}(\xi))}{\partial_\xi \tilde{\bar{s}}(\tilde{\tau}(\xi))}\ ,\qquad
{y}(\xi)=1+\frac{\partial_\xi \tilde{r}(\tilde{\tau}(\xi))}{\partial_\xi \tilde{\bar{s}}(\tilde{\tau}(\xi))} \ ,
\qquad \xi \in \left[-\frac{\lambda-\eta-\mu}{2},0\right]\ .
\end{equation*}
This completes the proof of the third branch of arctic curve in Theorem~\ref{thm:acgeneral} with the identifications  $(x_f(\xi),y_f(\xi))=(x(\xi),y(\xi))$,
$R_f(\xi)=\tilde{r}(\tilde{\tau}(\xi))$ and $S_s(\xi)=\tilde{\bar{s}}(\tilde{\tau}(\xi))$.
From the symmetry DWBC1/2 under rotation by $180^\circ$, we deduce again that this ``final'' portion of arctic curve $({x}(\xi),{y}(\xi))_{\xi\in \left[-\frac{\lambda-\eta-\mu}{2},0\right]}$
has a symmetric portion $(1-{x}(\xi),1-{y}(\xi))_{\xi\in \left[-\frac{\lambda-\eta-\mu}{2},0\right]}$.
The six portions of arctic curve computed so far, namely the ``normal'', the ``shear'' and the ``final'' portions together with their symmetric portions by $180^\circ$ rotation, 
constitute the entire arctic curve. An example of such arctic curve is displayed in Figure~\ref{fig:ACpisurdouze}.

A last remark is in order: the junction between the ``shear'' and ``final'' portions takes place at a point where the common tangent coincides with
the second diagonal $x+y=1$. It is indeed easily checked that  $\bar{s}\left(\tau\left(-\frac{\lambda-\eta+\mu}{2}\right)\right)
=\tilde{\bar{s}}\left(\tilde{\tau}\left(-\frac{\lambda-\eta-\mu}{2}\right)\right)=1$, while $r\left(\tau\left(-\frac{\lambda-\eta+\mu}{2}\right)\right)
=\tilde{r}\left(\tilde{\tau}\left(-\frac{\lambda-\eta-\mu}{2}\right)\right)=0$. 

\subsection{Phases and plots}
\label{sec:phases}
\begin{figure}
\begin{center}
\includegraphics[width=13cm]{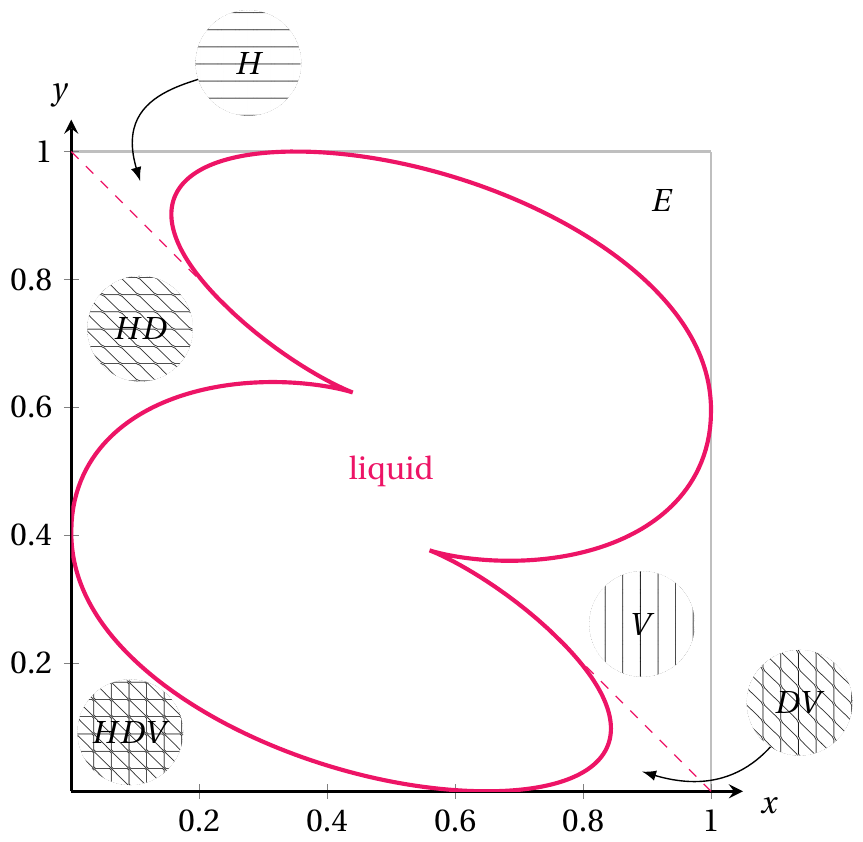} 
\end{center}
\caption{\small Generic phases of the integrable 20V model with DWBC1 or 2. The six frozen phases are labelled by their occupied edges, with $H$, $D$ and $V$
standing for horizontal, diagonal and vertical, while $E$ stands for ``empty''. The two dashed segments are portions of the second diagonal separating
different types of frozen phases while the red curve is the arctic curve encompassing the liquid phase.}
\label{fig:phases}
\end{figure}

The arctic curve is the limit between a liquid phase (inside the curve) and a number of frozen phases (outside of the curve).
Let us describe these frozen phases in the case of generic values of $\eta$, $\lambda$ and $\mu$. As displayed in Figure~\ref{fig:phases}, 
there are six different frozen regions around the arctic curve, each made of a single type of vertex. The phase denoted by $E$ is made 
of empty vertices only (top row of Figure~\ref{fig:generalweights} with weight $\omega_0$) while the symmetric region under $180^\circ$ rotation,
denoted by $HDV$ corresponds to a phase with fully occupied vertices (bottom row of Figure~\ref{fig:generalweights} with weight $\omega_0$).
Similarly, the phase denoted by $H$ (resp. $V$) is made 
of vertices crossed by a single horizontal (resp. vertical) path (top row of Figure~\ref{fig:generalweights} with weight $\omega_6$, resp. $\omega_1$) 
while the symmetric region under $180^\circ$ rotation,
denoted by $DV$ (resp. $HD$) corresponds to a phase with the complementary vertex (bottom row of Figure~\ref{fig:generalweights} with weight $\omega_6$,
resp. $\omega_1$).

\begin{figure}
\begin{center}
\includegraphics[width=12cm]{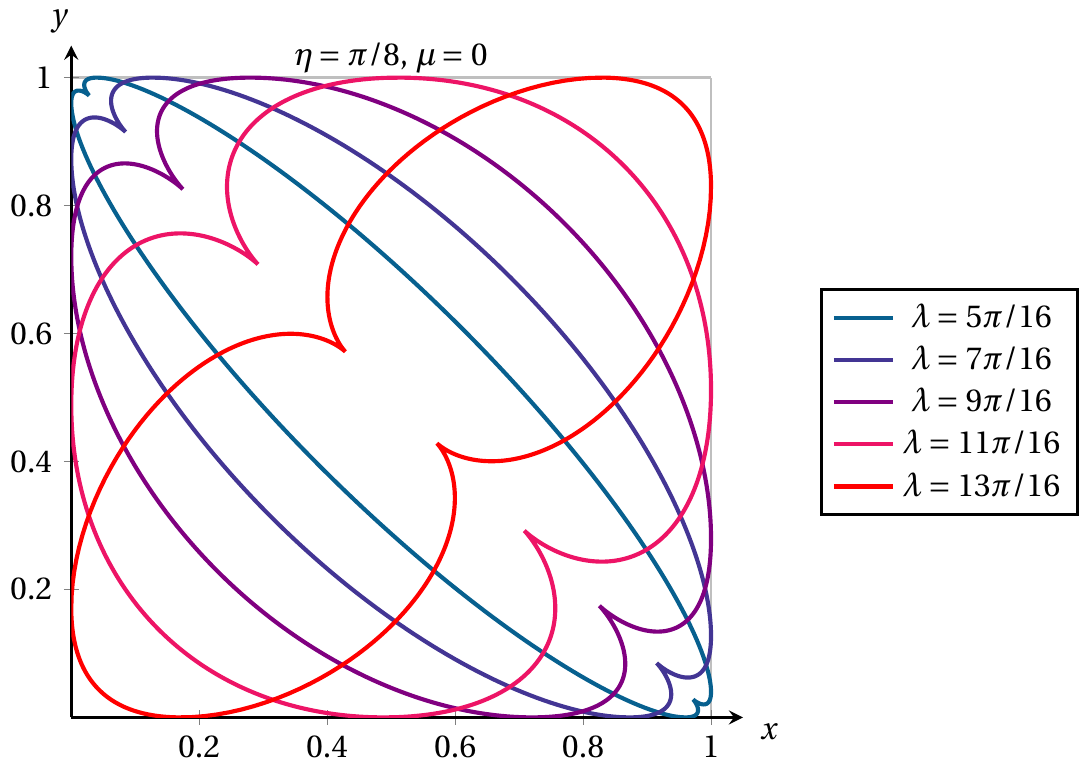} 
\end{center}
\caption{\small The arctic curve for $\eta=\pi/8$, $\mu=0$ and $\lambda=(5,7,9,11,13)\times\pi/16$.}
\label{fig:plot1}
\end{figure}
\begin{figure}
\begin{center}
\includegraphics[width=11.8cm]{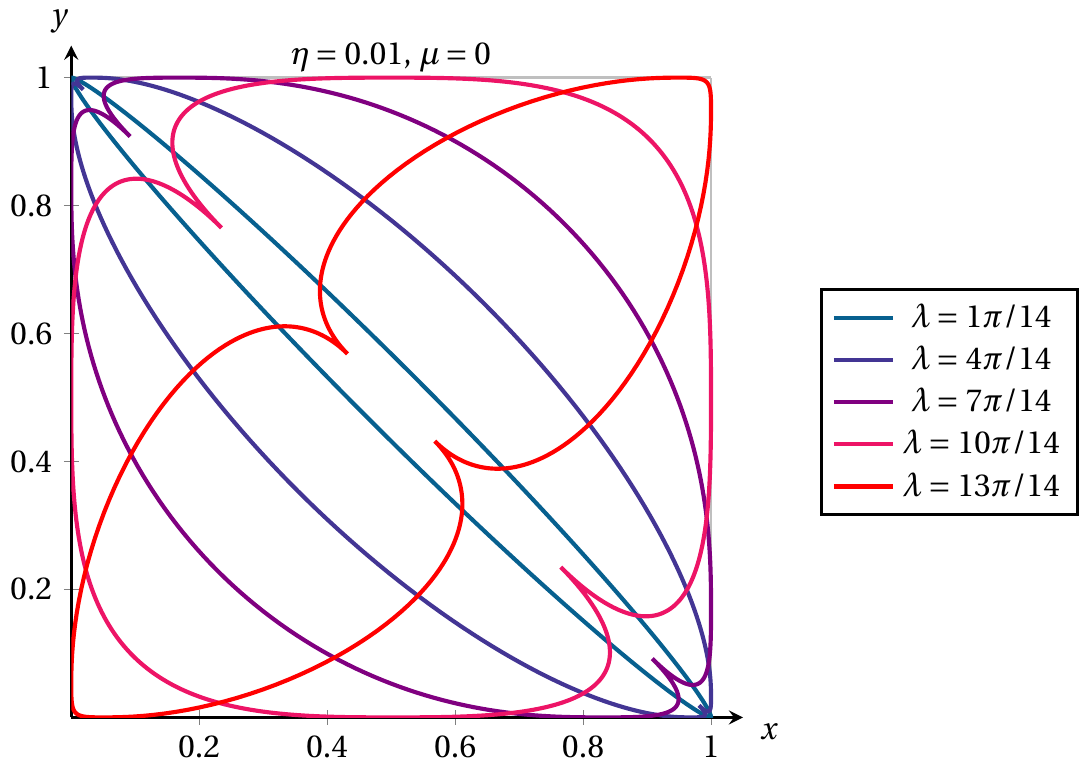} 
\end{center}
\caption{\small The arctic curve for $\eta\to 0$ (here $\eta=0.01$), $\mu=0$ and $\lambda=(1,4,7,10,13)\times\pi/14$.}
\label{fig:plot2}
\end{figure}
\begin{figure}
\begin{center}
\includegraphics[width=11.8cm]{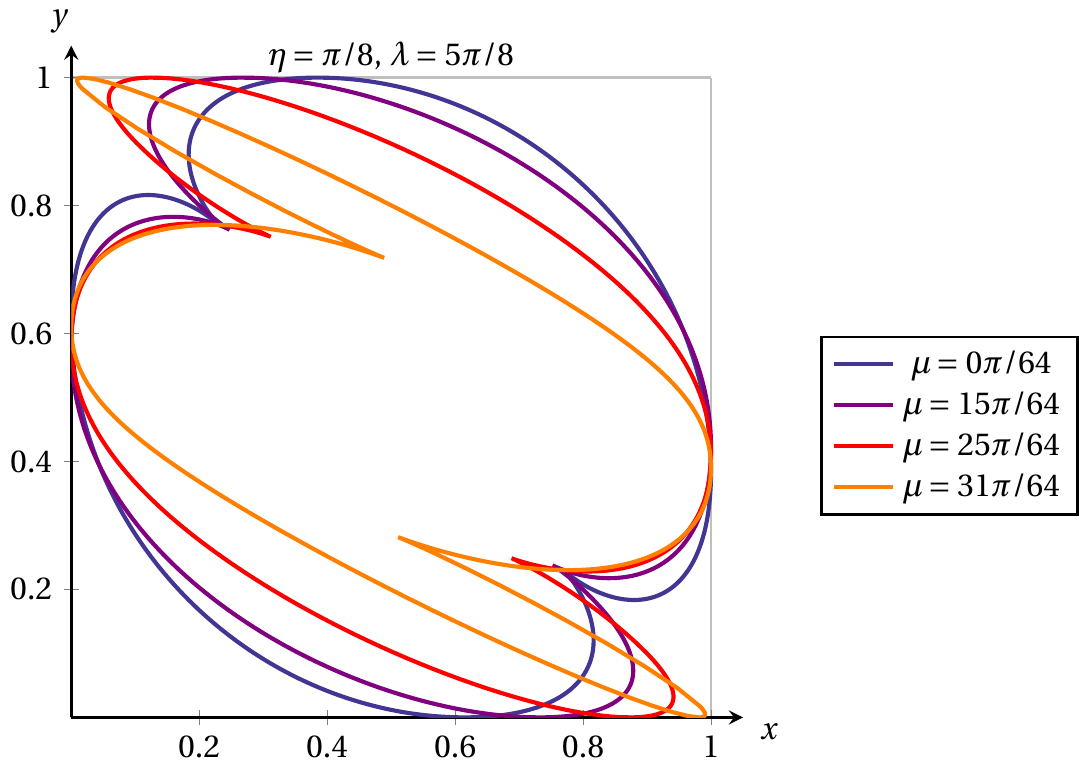} 
\end{center}
\caption{\small The arctic curve for $\eta=\pi/8$, $\lambda=5\pi/8$ and $\mu=(0,15,25,31)\times\pi/64$.}
\label{fig:plot3}
\end{figure}
\begin{figure}
\begin{center}
\includegraphics[width=16cm]{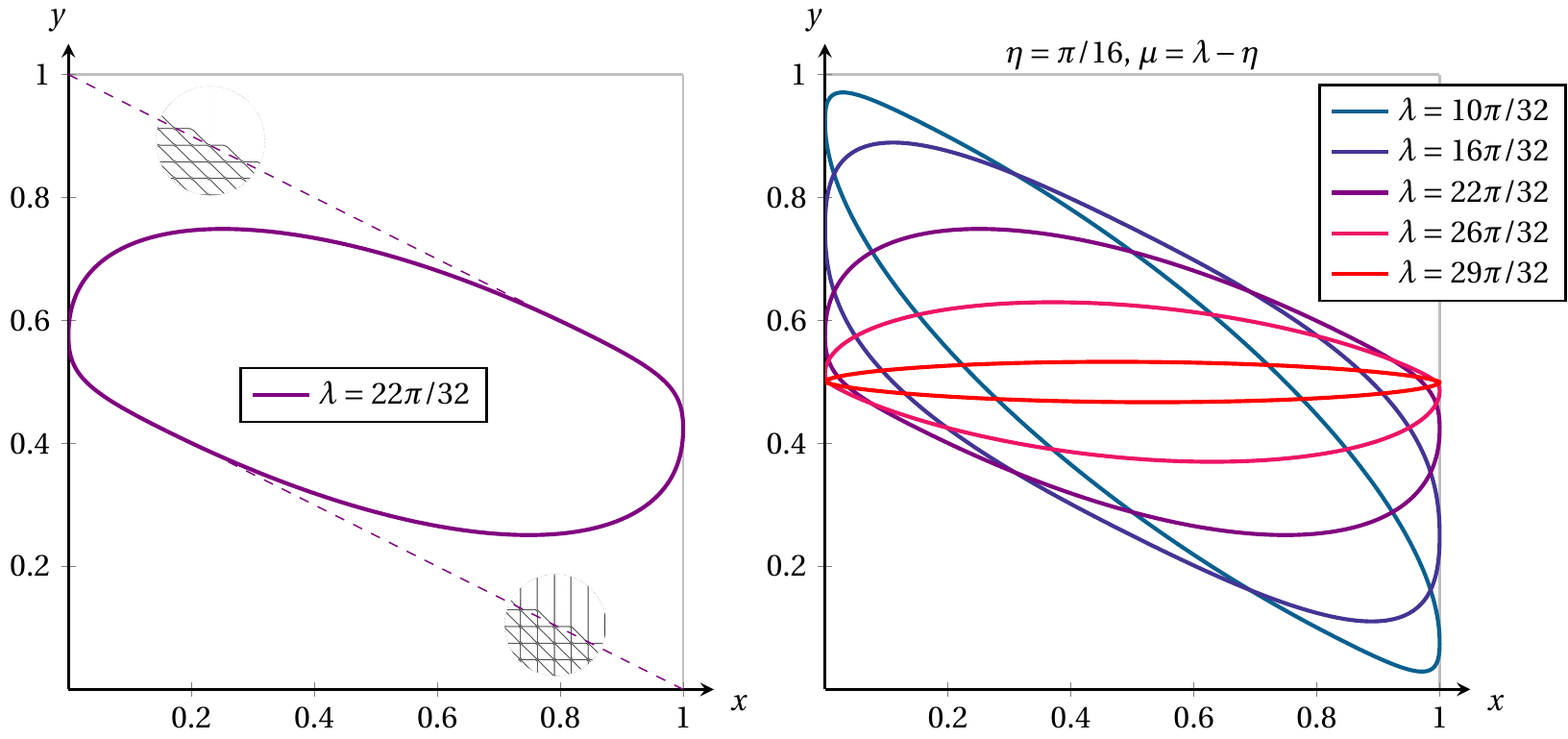} 
\end{center}
\caption{\small The arctic curve for $\eta=\pi/16$, $\mu=\lambda-\eta$ and $\lambda=(10,16,22,26,29)\times\pi/32$ (right).
At $\mu=\lambda-\eta$, a direct transition takes place between $HD$ and $E$ (resp. between $HDV$ and $V$) along a
segment of slope $-1/2$ dictated by the local path geometry, as shown (left), here for $\lambda=22\pi/32$.}
\label{fig:plot6}
\end{figure}
\begin{figure}
\begin{center}
\includegraphics[width=16cm]{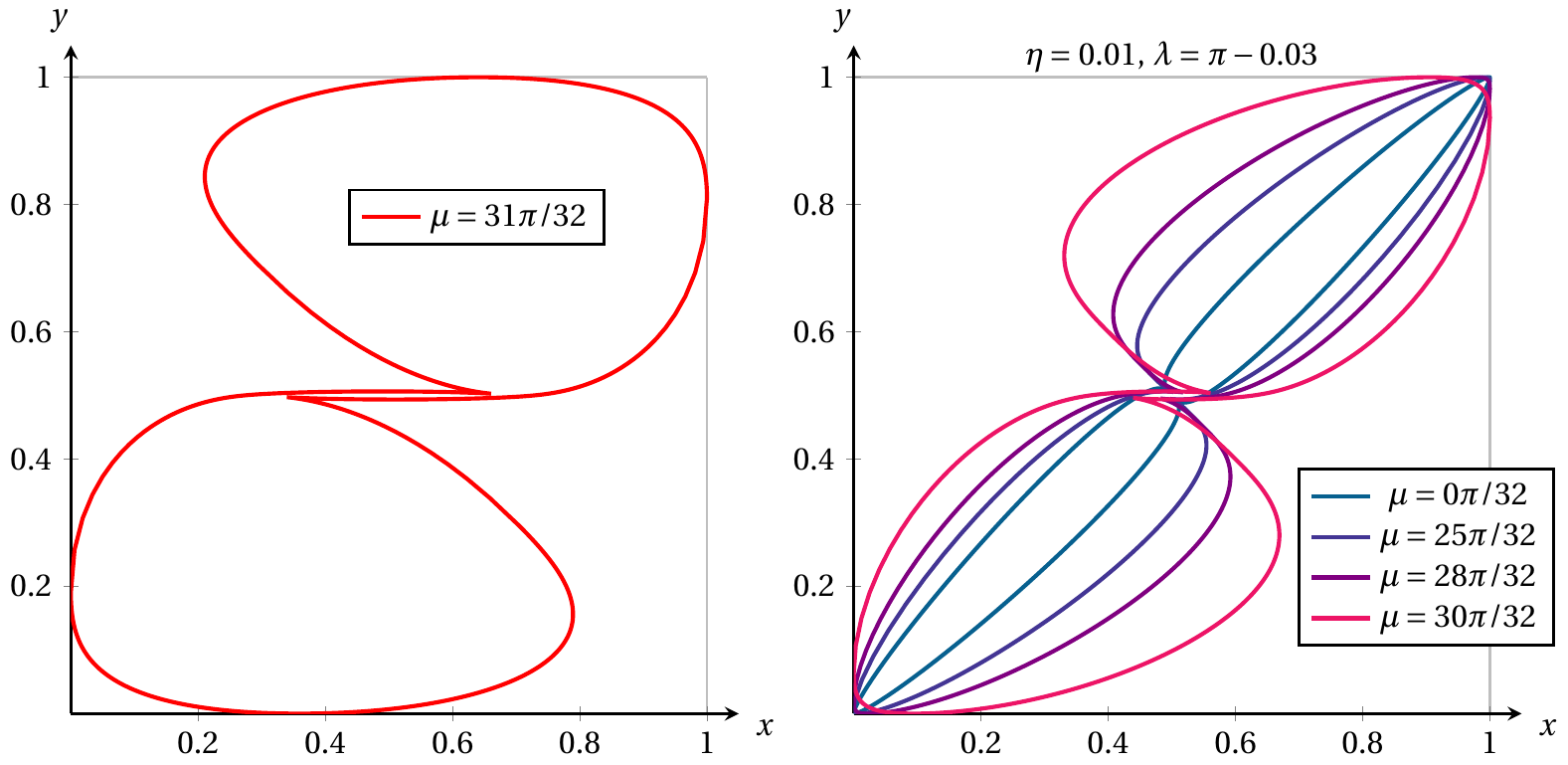} 
\end{center}
\caption{\small The arctic curve for $\eta\to 0$ (here $\eta=0.01$), $\lambda+\eta\to \pi$ (here $\lambda=\pi-0.03$) and $\mu=(0,25,28,30)\times\pi/32$
(right) and $\mu=31\pi/32$ (left).}
\label{fig:plot4}
\end{figure}
Let us now present a number of plots to illustrate the evolution of the arctic curve with varying $\eta$, $\lambda$ and $\mu$.
The arctic curves are all invariant under rotation by $180^\circ$.
As we already saw, the parameter $\mu$ (which varies in the range $]-\lambda+\eta,\lambda-\eta[$ controls the asymmetry of the 
arctic curve under the $x\leftrightarrow y$ transformation. Figures~\ref{fig:plot1} and \ref{fig:plot2} display arctic curves for $\mu=0$, 
hence symmetric under $x\leftrightarrow y$. The first figure is for $\eta=\pi/8$ with a varying $\lambda$ in the range $]\pi/8,7\pi/8[$,
while the second figure is for a value of $\eta \to 0$ (in practice $\eta=0.01$) with a varying $\lambda$ in the range $]0,\pi[$.
In both cases, the limiting curve $\lambda\to\eta$ degenerates into the second diagonal $x+y=1$, as easily understood from the 
values $\omega_{2,3,5,6}=0$ which imply that only one configuration survives, with all edges occupied below the second diagonal
and none above. Similarly, in the limit $\lambda\to \pi-\eta$, the arctic curve degenerates into the first diagonal $y=x$, as easily understood
from the value $\omega_0=0$ which forces a unique configuration in which the occupied edges are: all horizontal edges above the first diagonal,
all vertical edges below the first diagonal, and all diagonal edges below the second diagonal. 

Figure~\ref{fig:plot3} shows the increasing asymmetry of the arctic curve under $x\leftrightarrow y$ for increasing $\mu$ in the range
$]0,\pi/2[$ for fixed values $\eta=\pi/8$ and $\lambda=5\pi/8$. When the parameter $\mu$ tends to its maximal value $\lambda-\eta$ (here equal to $\pi/2$), 
the arctic curve displays two outgrowths which become narrower and narrower and eventually degenerate into 
two segments so that the arctic curve is formed of a single convex curve (see Figure~\ref{fig:plot6} left). The two limiting segments are 
tangent to this curve and have a slope $-1/2$ independently of 
$\lambda$ and $\eta$. This phenomenon has a simple explanation: for $\mu=\lambda-\eta$, we have $\omega_6=0$
so that the $H$ and $DV$ frozen phases cannot exist anymore. A direct transition then takes place between the frozen 
phases $HD$ and $E$ (resp. between $HDV$ and $V$) and the slope $-1/2$ of the transition line is directly dictated by the path geometry,
as illustrated in Figure~\ref{fig:plot6} left).   
The evolution of the $\mu=\lambda-\eta$ convex arctic curve with fixed $\eta=\pi/16$ and varying $\lambda$ is represented in Figure~\ref{fig:plot6} right (with $\lambda\in ]\pi/16,15\pi/16[$). A similar phenomenon occurs when $\mu= \eta-\lambda$ ($\omega_1= 0$), creating 
arctic curve configurations which are symmetric to those just described under the exchange $x\leftrightarrow y$.

Figure \ref{fig:plot4} displays the evolution with $\mu$ of the arctic
curve for a very small $\eta$ and a value of $\lambda$ close to $\pi$ (recall the $\lambda<\pi-\eta$, here we took $\eta=0.01$ and $\lambda=\pi-0.03$).  
For large value of $\mu$ (recall that $\mu<\lambda-\eta$, hence $\mu<\pi-0.04$), for instance $\mu=31\pi/32$, the arctic curve is formed of two symmetric 
convex pieces connected by a narrow isthmus. 

More precisely, a well-defined limit can be reached by setting
\begin{equation}
\eta=\epsilon\ \Lambda_1\ , \quad \lambda=\pi-\epsilon(\Lambda_1+\Lambda_2)\ , \quad \mu=\pi-\epsilon(2\Lambda_1+\Lambda_2+\Lambda_3)
 \label{eq:Lambdalimit}
 \end{equation}
and sending $\epsilon\to 0$, with fixed $\Lambda_{1,2,3}>0$ so that the parameters remain in the admissible range \eqref{eq:admissible}.
Up to an overall $\epsilon^3$, all the weights $\omega_i$ remain finite and are positive homogeneous polynomials of degree $3$ in the $\Lambda_i$'s.
In this limit, the arctic curve of Theorem~\ref{thm:acgeneral} tends to an algebraic curve made of two symmetric pieces: the first piece lies in the
upper half of the rescaled square domain and is tangent to the lines $y=1$, $x=1$ and $y=1/2$. The second piece is its image under 
$180^\circ$ and lies in the lower half of the square. As for the isthmus observed in Figure~\ref{fig:plot4}, it degenerates into a horizontal segment joining the two tangency points 
along the $y=1/2$ line. This segment corresponds to a new, direct transition between the $HD$ and $V$ frozen regions and its horizontality is
dictated by the path geometry.  

\begin{figure}
\begin{center}
\includegraphics[width=13cm]{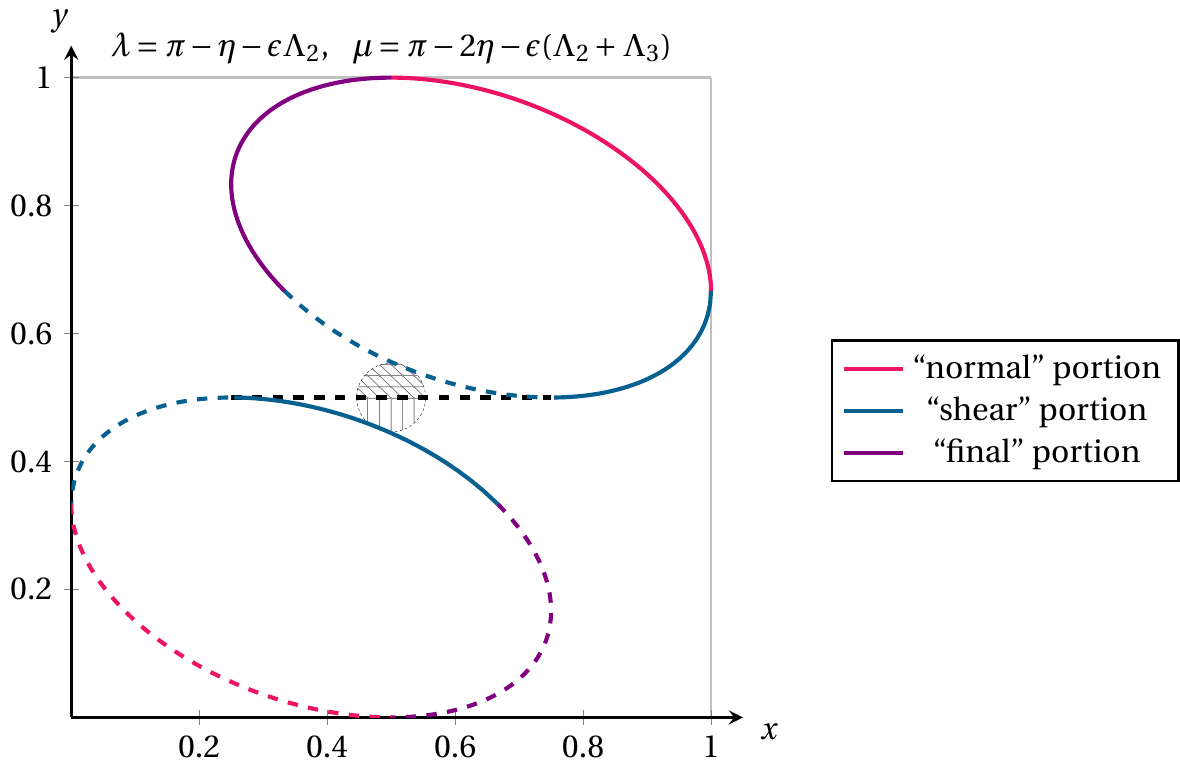} 
\end{center}
\caption{\small The arctic curve for $\lambda=\pi-\eta-\epsilon\, \Lambda_2$ and $\mu=\pi-\eta-\epsilon(\Lambda_2+\Lambda_3)$ 
with $\epsilon \to 0$ and some arbitrarily fixed $\eta$ (here for $\Lambda_3/\Lambda_2=2$). The limiting shape is independent of $\eta$ and
formed of two symmetric ellipses tangent to the $y=1/2$ line. Note that the
``shear'' portion in Theorem~\ref{thm:acgeneral} gives rise to two portions in this limit, one in each ellipse. The black dashed segment is the locus of the transition between the $HD$ and $V$ frozen regions.}
\label{fig:plot9}
\end{figure}

It is interesting to follow more precisely the limit  \eqref{eq:Lambdalimit} of each of the six portions of arctic curve.
To get the limit of the ``normal'' portion, we must simultaneously set $\xi= \epsilon \Upsilon$ so that $\Upsilon$ varies in the range
$[0,\Lambda_2]$. Similarly, the limit of the ``final'' portion is obtained by setting  $\xi= \epsilon \Upsilon$ with $\Upsilon$ varying in the range
$\left[-\frac{\Lambda_3}{2},0\right]$. These two portions then lead to two adjacent connected portions of the first piece of the algebraic curve (see Figure~\ref{fig:plot9}
for an illustration) while their symmetric portions contribute to the second piece. The case of the ``shear'' portion is more subtle as
the original range $\left[-\frac{\lambda-\eta+\mu}{2},0\right]$ gives rise by rescaling to two limiting intervals for the variable $\xi$: the vicinity of $\xi=0$
is probed by setting $\xi= \epsilon \Upsilon$ with  $\Upsilon$ in the range $]-\infty,0]$ while the vicinity of $-\frac{\lambda-\eta+\mu}{2}$ is
probed by setting $\xi= -\pi+\epsilon \Upsilon$ with $\Upsilon$ in the range $\left[2\Lambda_1+\Lambda_2+\frac{\Lambda_3}{2},+\infty\right[$. This eventually 
gives rise to two separate portions of arctic curve, one contributing to the upper piece and the other of the lower one
(see Figure~\ref{fig:plot9}). Since the ``shear'' portion was originally connected, 
the two separate limiting portions are in fact connected by the horizontal segment encountered above, at the transition between the $HD$ and $V$ regions. 
This segment however is no longer part of the arctic  curve as it is not incident to the liquid phase. The symmetric of the ``shear'' portion
finally builds the missing parts of the upper and lower pieces. 
 
The same phenomenon of splitting of the arctic curve is in fact observed by keeping a finite value of $\eta$ whenever $\lambda$ and $\mu$ tend to their maximal admissible values, namely $\lambda=\pi-\eta-\epsilon\, \Lambda_2$ and $\mu= \pi-2\eta
-\epsilon(\Lambda_2+\Lambda_3)$ with $\epsilon\to 0$, $\Lambda_{2,3}$ finite and positive. This leads to a limiting arctic curve depending on $\Lambda_{2,3}$
but \emph{independent of $\ \eta$} (assuming that $\eta$ itself is kept fixed and does not scale with $\epsilon$) corresponding to limiting weights $\omega_{0,4,5,6}\to 0$ and 
(projectively) $\omega_{1,2,3}\to 1$.
This $\eta$-independent limit may be reached  in the above setting \eqref{eq:Lambdalimit} by sending $\Lambda_1\to \infty$. Remarkably, the corresponding limiting arctic curve
is made of two ellipses (exchanged by $180^\circ$ rotation) with respective equations
\begin{equation*}
\begin{split}
&\big(\Lambda_3(x-y)+2\Lambda_2(x+2y-2)\big)^2 + 8 \Lambda_2\Lambda_3 (1-y)(1-2y)=0\ ,\\
&\big(\Lambda_3(x-y)+2\Lambda_2(x+2y-1)\big)^2 - 8 \Lambda_2\Lambda_3 y(1-2y)=0\ .\\
\end{split}
\end{equation*}
In particular, these curves depend on the ratio $\Lambda_3/\Lambda_2$, hence on the precise way in which the weights $\omega_i$ reach their
limiting values $0$ or $1$. The corresponding  arctic curve for $\Lambda_3/\Lambda_2=2$ is displayed in Figure~\ref{fig:plot9}.
  
\section{Simulations}
\label{sec:simul} 

\subsection{Method}
\label{sec:nummethod} 
Numerical studies of the arctic curve phenomenon in the 6V model with DWBC (or ``domain-wall-like'' boundary conditions) are numerous \cite{Syljuasen2004,Allison2005,Cugliandolo2015,Lyberg2017a,Keesman2017, Lyberg2017b}. In this section we will adapt a Markov-chain Monte Carlo method due to Allison and Reshetikhin discussed in \cite{Allison2005, Lyberg2017a,Lyberg2017b}. This algorithm exploits the bijection with osculating paths to design a local-move Markov-chain whose stationary distribution is that associated with the weights \eqref{eq:explweights}. For the sake of definiteness we will consider the 20V model with DWBC1, but the discussion is easily extended to other fixed boundary conditions, including DWBC2.

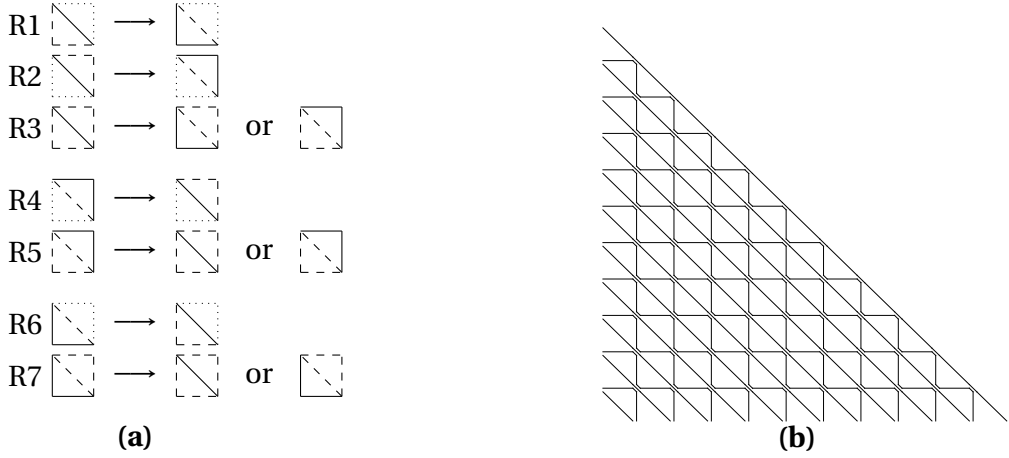
\begin{figure}

\begin{center}
\begin{tikzpicture}[scale=0.55]
\draw(18,0) node{\includegraphics[scale=0.3]{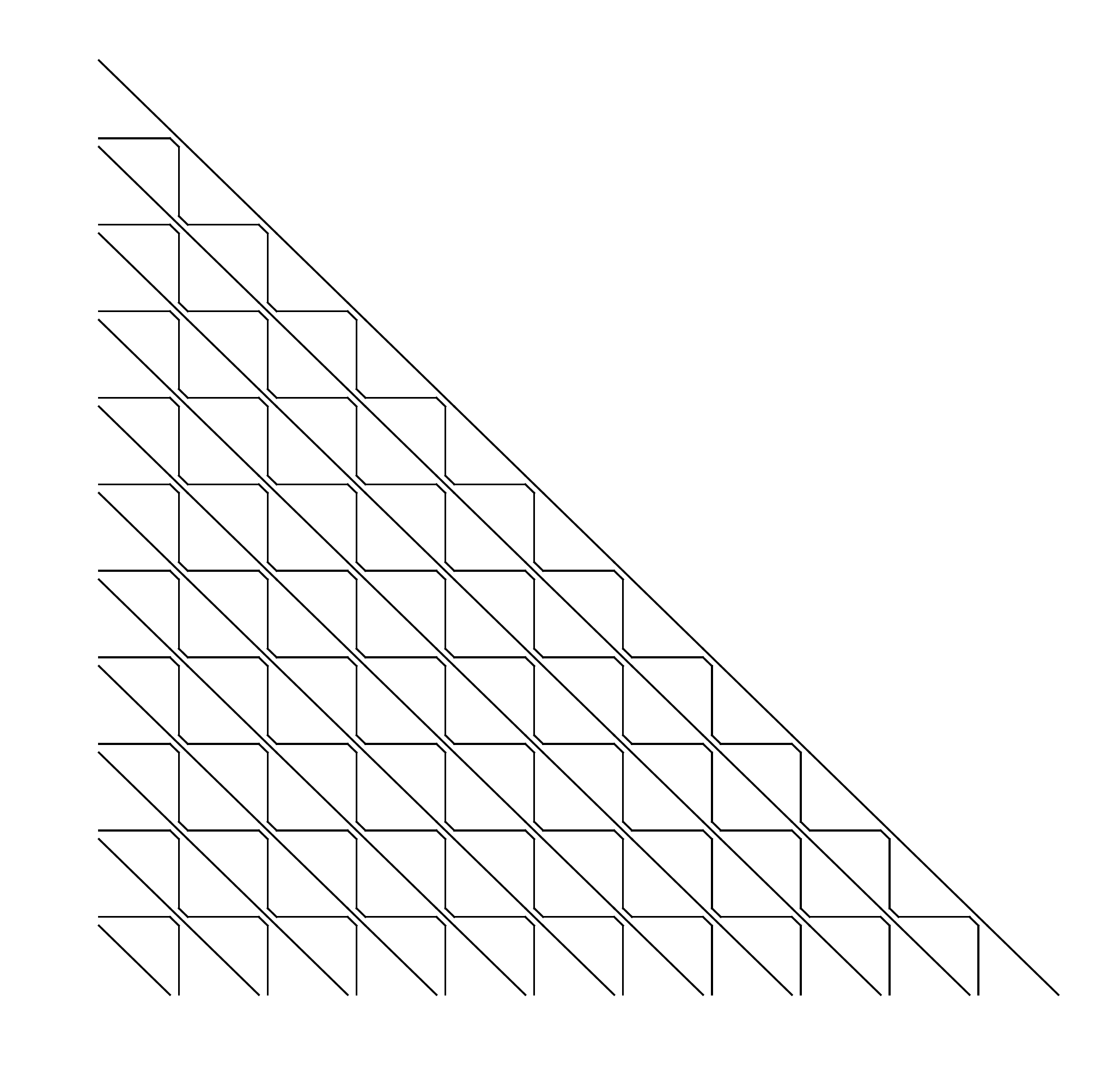}} ;
\draw (2,-5) node{\textbf{(a)}};
\draw (18,-5) node{\textbf{(b)}};

\begin{scope}[xshift=0cm,yshift=4.5cm]
\draw (0,0.5) node[left]{R1};
\draw[dashed] (0,0)--++(1,0);
\draw[dashed] (0,0)--++(0,1);
\draw (1,0)--(0,1);
\draw[dotted] (1,0)--++(0,1);
\draw[dotted] (0,1)--++(1,0);
\draw (2,0.5) node{$\longrightarrow$};
\begin{scope}[xshift=3cm]
\draw (0,0)--++(1,0);
\draw (0,0)--++(0,1);
\draw[dashed] (1,0)--(0,1);
\draw[dotted] (1,0)--++(0,1);
\draw[dotted] (0,1)--++(1,0);
\draw (2,0.5) node{};
\end{scope}
\end{scope}

\begin{scope}[xshift=0cm,yshift=3.25cm]
\draw (0,0.5) node[left]{R2};
\draw[dotted] (0,0)--++(1,0);
\draw[dotted] (0,0)--++(0,1);
\draw (1,0)--(0,1);
\draw[dashed] (1,0)--++(0,1);
\draw[dashed] (0,1)--++(1,0);
\draw (2,0.5) node{$\longrightarrow$};
\begin{scope}[xshift=3cm]
\draw[dotted] (0,0)--++(1,0);
\draw[dotted] (0,0)--++(0,1);
\draw[dashed] (1,0)--(0,1);
\draw (1,0)--++(0,1);
\draw (0,1)--++(1,0);
\draw (2,0.5) node{};
\end{scope}
\end{scope}

\begin{scope}[xshift=0cm,yshift=2cm]
\draw (0,0.5) node[left]{R3};
\draw[dashed] (0,0)--++(1,0);
\draw[dashed] (0,0)--++(0,1);
\draw (1,0)--(0,1);
\draw[dashed] (1,0)--++(0,1);
\draw[dashed] (0,1)--++(1,0);
\draw (2,0.5) node{$\longrightarrow$};
\begin{scope}[xshift=3cm]
\draw (0,0)--++(1,0);
\draw (0,0)--++(0,1);
\draw[dashed] (1,0)--(0,1);
\draw[dashed] (1,0)--++(0,1);
\draw[dashed] (0,1)--++(1,0);
\draw (2,0.5) node{or};
\begin{scope}[xshift=3cm]
\draw[dashed] (0,0)--++(1,0);
\draw[dashed] (0,0)--++(0,1);
\draw[dashed] (1,0)--(0,1);
\draw (1,0)--++(0,1);
\draw (0,1)--++(1,0);
\end{scope}
\end{scope}
\end{scope}

\begin{scope}[xshift=0cm,yshift=0.25cm]
\draw (0,0.5) node[left]{R4};
\draw[dotted] (0,0)--++(1,0);
\draw[dotted] (0,0)--++(0,1);
\draw[dashed] (1,0)--(0,1);
\draw (1,0)--++(0,1);
\draw (0,1)--++(1,0);
\draw (2,0.5) node{$\longrightarrow$};
\begin{scope}[xshift=3cm]
\draw[dotted] (0,0)--++(1,0);
\draw[dotted] (0,0)--++(0,1);
\draw (1,0)--(0,1);
\draw[dashed] (1,0)--++(0,1);
\draw[dashed] (0,1)--++(1,0);
\draw (2,0.5) node{};
\end{scope}
\end{scope}

\begin{scope}[xshift=0cm,yshift=-1cm]
\draw (0,0.5) node[left]{R5};
\draw[dashed] (0,0)--++(0,1);
\draw[dashed] (0,0)--++(1,0);
\draw[dashed] (1,0)--(0,1);
\draw (1,0)--++(0,1);
\draw (0,1)--++(1,0);
\draw (2,0.5) node{$\longrightarrow$};
\begin{scope}[xshift=3cm]
\draw[dashed] (0,0)--++(1,0);
\draw[dashed] (0,0)--++(0,1);
\draw (1,0)--(0,1);
\draw[dashed] (1,0)--++(0,1);
\draw[dashed] (0,1)--++(1,0);
\draw (2,0.5) node{or};
\begin{scope}[xshift=3cm]
\draw[dashed] (0,0)--++(0,1);
\draw[dashed] (0,0)--++(1,0);
\draw[dashed] (1,0)--(0,1);
\draw (1,0)--++(0,1);
\draw (0,1)--++(1,0);
\end{scope}
\end{scope}
\end{scope}

\begin{scope}[xshift=0cm,yshift=-2.75cm]
\draw (0,0.5) node[left]{R6};
\draw (0,0)--++(1,0);
\draw (0,0)--++(0,1);
\draw[dashed] (1,0)--(0,1);
\draw[dotted] (1,0)--++(0,1);
\draw[dotted] (0,1)--++(1,0);
\draw (2,0.5) node{$\longrightarrow$};
\begin{scope}[xshift=3cm]
\draw[dashed] (0,0)--++(1,0);
\draw[dashed] (0,0)--++(0,1);
\draw (1,0)--(0,1);
\draw[dotted] (1,0)--++(0,1);
\draw[dotted] (0,1)--++(1,0);
\draw (2,0.5) node{};
\end{scope}
\end{scope}

\begin{scope}[xshift=0cm,yshift=-4cm]
\draw (0,0.5) node[left]{R7};
\draw (0,0)--++(0,1);
\draw (0,0)--++(1,0);
\draw[dashed] (1,0)--(0,1);
\draw[dashed] (1,0)--++(0,1);
\draw[dashed] (0,1)--++(1,0);
\draw (2,0.5) node{$\longrightarrow$};
\begin{scope}[xshift=3cm]
\draw[dashed] (0,0)--++(1,0);
\draw[dashed] (0,0)--++(0,1);
\draw (1,0)--(0,1);
\draw[dashed] (1,0)--++(0,1);
\draw[dashed] (0,1)--++(1,0);
\draw (2,0.5) node{or};
\begin{scope}[xshift=3cm]
\draw (0,0)--++(0,1);
\draw (0,0)--++(1,0);
\draw[dashed] (1,0)--(0,1);
\draw[dashed] (1,0)--++(0,1);
\draw[dashed] (0,1)--++(1,0);
\end{scope}
\end{scope}
\end{scope}
\end{tikzpicture}
\end{center}
\caption{\textbf{(a)} The rules R$1-7$ used to perform the elementary moves of our Markov-chain. Plain lines indicate occupied edges, dashed lines 
empty edges while at least one of the dotted line is occupied. When two outputs are drawn, one is chosen with probability $1/2$. \textbf{(b)}  The ``diagonal'' configuration used as initial state of our Markov-chain. It is made of a triangular $HDV$ region and a triangular $E$ region.}
\label{fig_initial_diago}
\end{figure}

Let us first consider the model with uniform distribution, i.e. with all the $\omega_i$'s equal ($\eta=\pi/8$, $\lambda=5\eta$ and $\mu=0$). The Markov-chain starts from an allowed configuration (for example, the ``diagonal'' configuration displayed in Figure \ref{fig_initial_diago}-(b)) and preserves the non-crossing property at each step. At any given iteration, either the configuration remains unchanged or some elementary move is performed on a plaquette. The algorihm goes as follows: start by selecting a plaquette at random, and, if the plaquette has at least a section of path connecting its Northwest to its Southeast corner,  randomly choose which section of path to update (\begin{tikzpicture} \draw[dotted](0,0) rectangle(0.3,0.3);\draw(0.3,0)--(0,0.3); \end{tikzpicture}, \begin{tikzpicture} \draw[dotted](0,0) rectangle(0.3,0.3);\draw(0.3,0)--(0,0)--(0,0.3); \end{tikzpicture} or \begin{tikzpicture} \draw[dotted](0,0) rectangle(0.3,0.3);\draw(0.3,0)--(0.3,0.3)--(0,0.3); \end{tikzpicture} ). The four possible moves are \begin{tikzpicture} \draw[dotted](0,0) rectangle(0.3,0.3);\draw(0.3,0)--(0,0.3); \end{tikzpicture} $\rightarrow$ \begin{tikzpicture} \draw[dotted](0,0) rectangle(0.3,0.3);\draw(0.3,0)--(0,0)--(0,0.3); \end{tikzpicture}, \begin{tikzpicture} \draw[dotted](0,0) rectangle(0.3,0.3);\draw(0.3,0)--(0,0)--(0,0.3); \end{tikzpicture} $\rightarrow$ \begin{tikzpicture} \draw[dotted](0,0) rectangle(0.3,0.3);\draw(0.3,0)--(0,0.3); \end{tikzpicture}, \begin{tikzpicture} \draw[dotted](0,0) rectangle(0.3,0.3);\draw(0.3,0)--(0,0.3); \end{tikzpicture} $\rightarrow$  \begin{tikzpicture} \draw[dotted](0,0) rectangle(0.3,0.3);\draw(0.3,0)--(0.3,0.3)--(0,0.3); \end{tikzpicture}  and  \begin{tikzpicture} \draw[dotted](0,0) rectangle(0.3,0.3);\draw(0.3,0)--(0.3,0.3)--(0,0.3); \end{tikzpicture}$ \rightarrow$ \begin{tikzpicture} \draw[dotted](0,0) rectangle(0.3,0.3);\draw(0.3,0)--(0,0.3); \end{tikzpicture}, if allowed by the local environment. If no move 
is possible for this selected section of path, remain in the same configuration for this iteration of the chain. Otherwise, perform a move 
according to the rules R$i$, $i=1,2,\cdots,7$ presented in Figure \ref{fig_initial_diago}-(a). Repeat the process until the number of iterations is 
``sufficient'' (see the discussion below).

Since every configuration of the model can be obtained from any other configuration by finitely many elementary moves, the Markov-chain is ergodic. One can also verify that the transition probability $p(\mathcal{C}\to \mathcal{C}')$ from a configuration $\mathcal{C}$ to a configuration $\mathcal{C}'$ is symmetric: $p(\mathcal{C}'\to \mathcal{C})=p(\mathcal{C}\to \mathcal{C}')$. In particular, the two possibilities for the output in rules R$5$ and R$7$ are precisely designed so as
to ensure this property. Hence the detailed balance condition is satisfied for the uniform distribution, which implies that the stationary distribution of this Markov-chain is uniform, as required.

\begin{figure}
\includegraphics[scale=0.83]{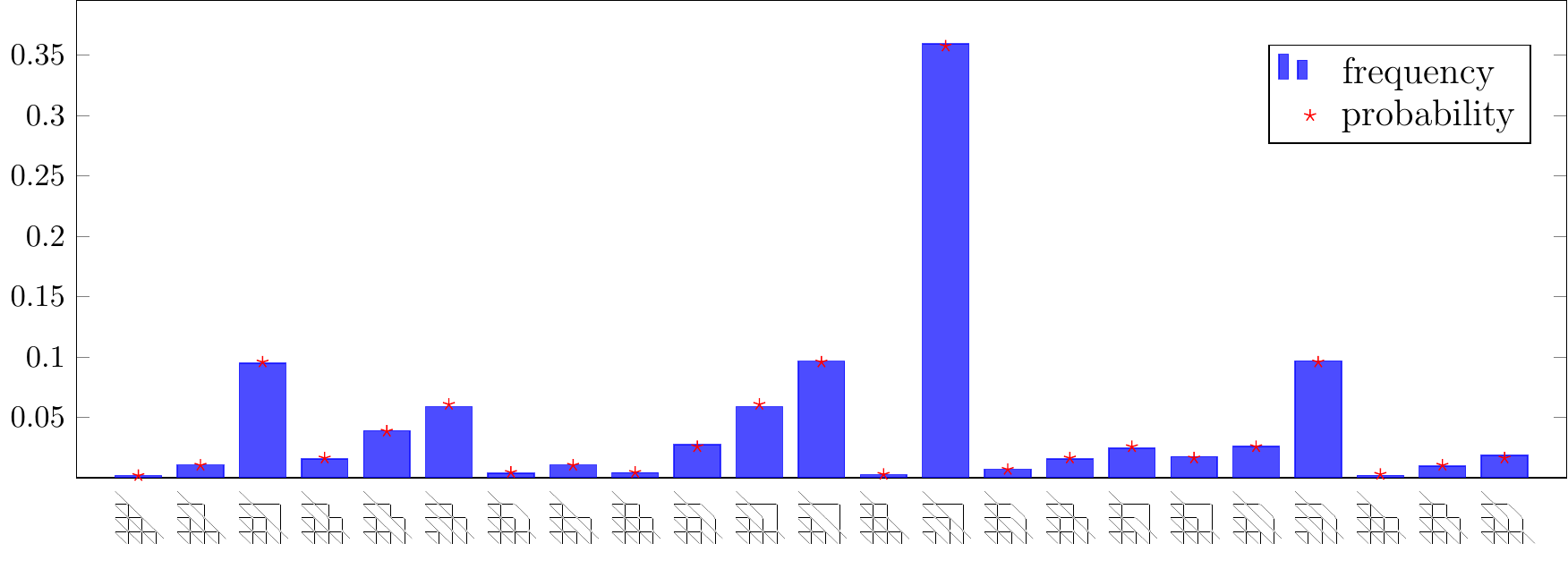}
\caption{The observed frequencies of the 23 possible configurations of the 20V model with DWBC1 at $n=3$, compared with the corresponding theoretical probabilities, here for $\eta=\pi/12$, $\lambda=10\pi/12$ and $\mu=\lambda-5\eta$.  The frequencies were measured 
from $23000$ generated configurations.}
\label{fig_verif_etapiover12_lamb10piover12}
\end{figure}

The uniform version of the algorithm is therefore such that, at every iteration, we either stay in the same configuration $\mathcal{C}$ (the move is rejected) or we perform a move and obtain a new configuration $\mathcal{C}'$. Generalizing the algorithm to the non-uniform case
of arbitrary weights \eqref{eq:explweights} can then be achieved by further rejecting some of the moves in a way that depends on the 
weight of the putative new configuration $\mathcal{C}'$. Assume that the move is attempted on a plaquette whose center is located at $(i,j)$ and call $W_{i,j}(\mathcal{C}')$ the product of the weights of the four nodes around this plaquette in the configuration $\mathcal{C}'$. The move is then accepted (in a similar manner as in \cite{Allison2005, Lyberg2017a,Lyberg2017b})
with a probability equal to: 
\begin{equation}
P=\frac{W_{i,j}(\mathcal{C}')}{W_0},
\label{eq_prob_lyberg}
\end{equation}
where the normalisation $W_0=\left(\max\limits_{k=0,\dots,6}\omega_k\right)^4$ 
ensures that $P\leq 1$. This Markov-chain is ergodic and satisfies the detailed balance condition for the probability distribution induced
by the relative weights $\omega_i$. The ability of this algorithm to generate configurations with the correct frequency was tested for a small size ($n=3$,
hence $23$ configurations) and for several choices of $\eta$, $\lambda$ and $\mu$ (see Figure~\ref{fig_verif_etapiover12_lamb10piover12} for an example).

The Markov-chain converges to the desired distribution in the limit of an infinite number of iterations. In practice, we need a criterion to decide
when to stop the simulation: the main criterion we used, also invoked in \cite{Lyberg2017a,Lyberg2017b}, is the stabilization of the arctic curve,
namely that no qualitative change in this curve is observed. In the cases where the convergence is slow, our estimation is checked against that
obtained by running another Markov-chain starting from a completely different configuration, as is done in \cite{Allison2005}, and we make sure that the results are comparable. 

Notice that instead of \eqref{eq_prob_lyberg} one can alternatively use the metropolis probability 
\begin{equation}
P=\min\left(1,\frac{W_{i,j}(\mathcal{C}')}{W_{i,j}(\mathcal{C})}\right).
\label{eq_prob_metropolis}
\end{equation}
For generic $\omega_i'$s, this choice lowers the rejection of moves and hence increases the thermalization speed. Unless stated otherwise we used \eqref{eq_prob_lyberg}.

\subsection{Results}
\label{sec:numresults}
\begin{figure}[h!]
\begin{tikzpicture}[scale=0.90]
\draw (0,0) node{\includegraphics[scale=0.171]{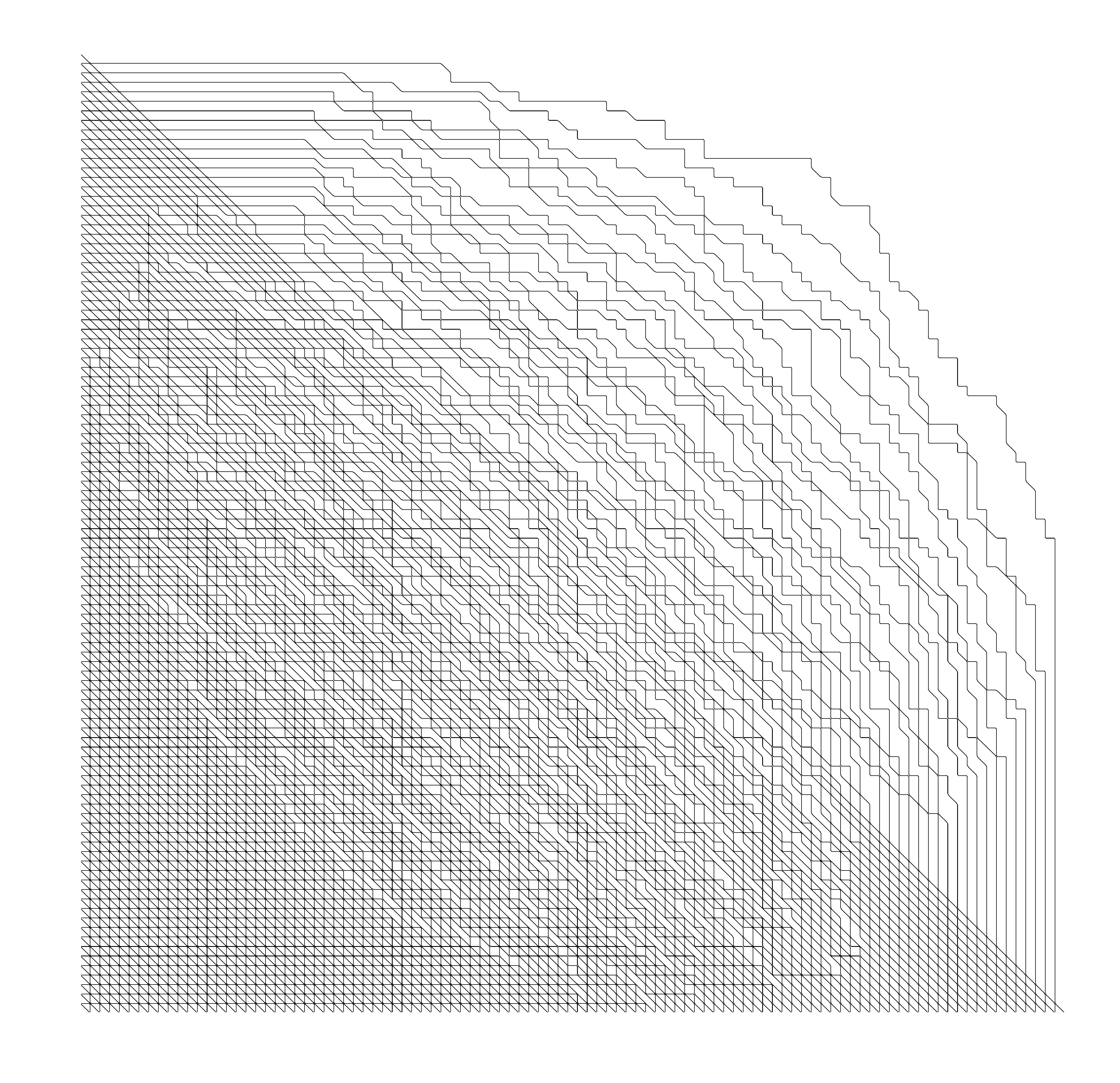}};
\draw (8.5,0) node{\includegraphics[scale=0.183825]{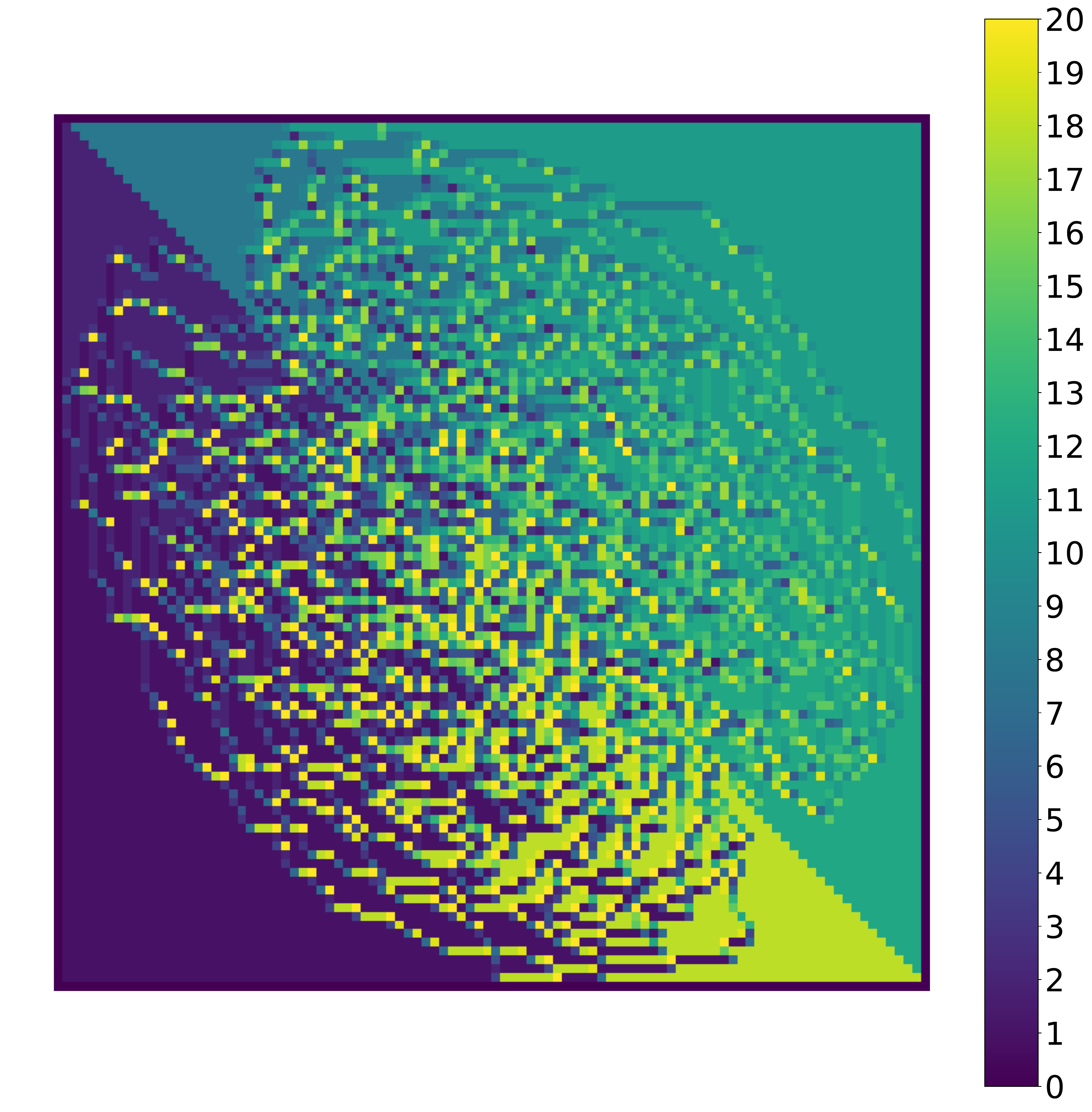}};
\end{tikzpicture}
\caption{A typical configuration of the 20V model with DWB1 for $n=100$ and with uniform distribution in the osculating path representation 
(left) and in a colored vertex coding (right). Here vertices are colored according to their label running from $1$ to $20$ in the order of 
Figure \ref{fig:twentyV}.}
\label{fig_n100_single_configuration}
\end{figure}

\begin{figure}[h!]
\begin{center}
\includegraphics[scale=0.168]{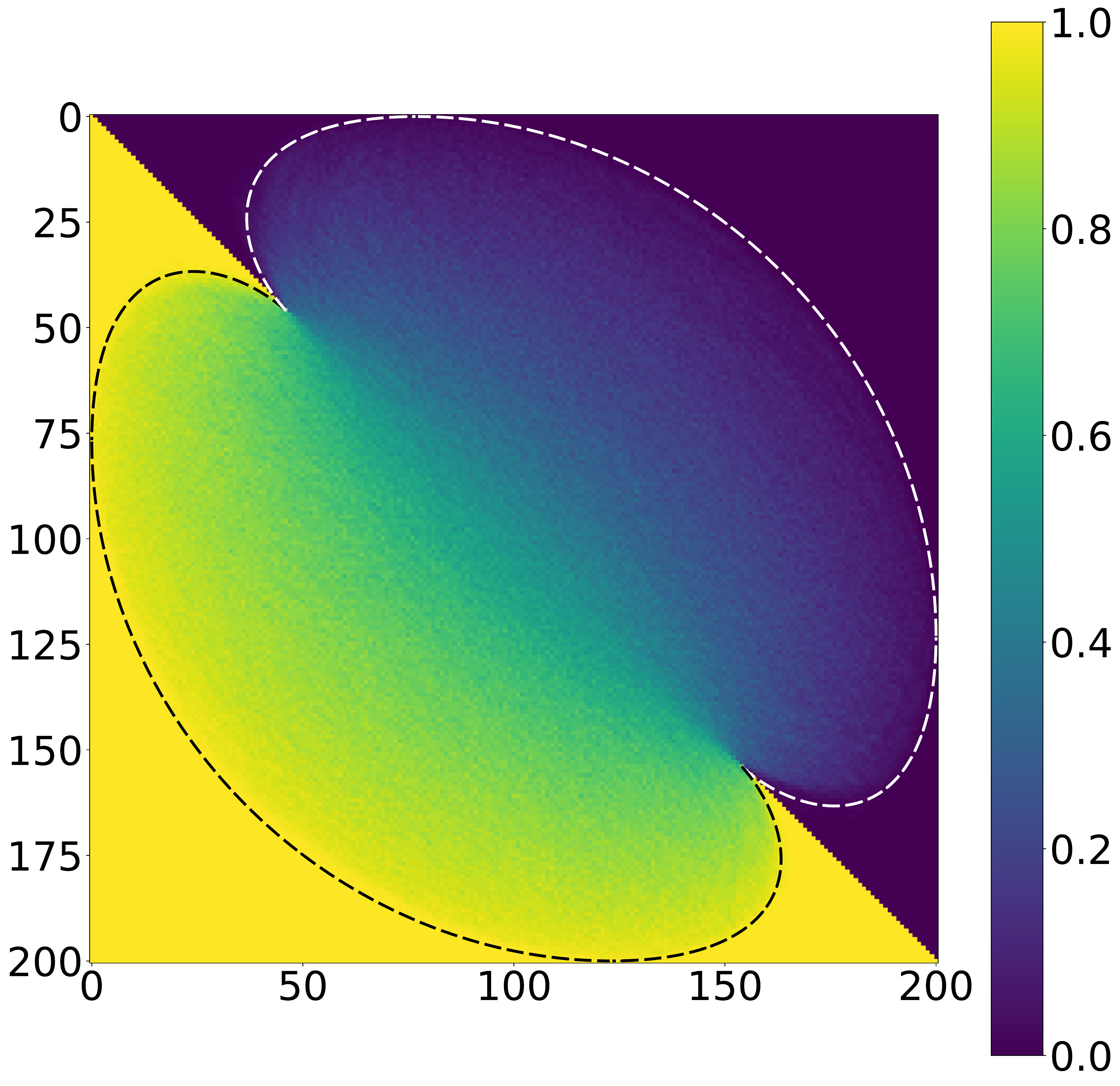}
\end{center}
\caption{Local density of diagonal steps for the 20V model with DWBC1 with $n=200$. The dashed curve is plotted from the analytical expression of the arctic curve (with
a scale factor of $n$).}
\label{fig:cafe}
\end{figure}

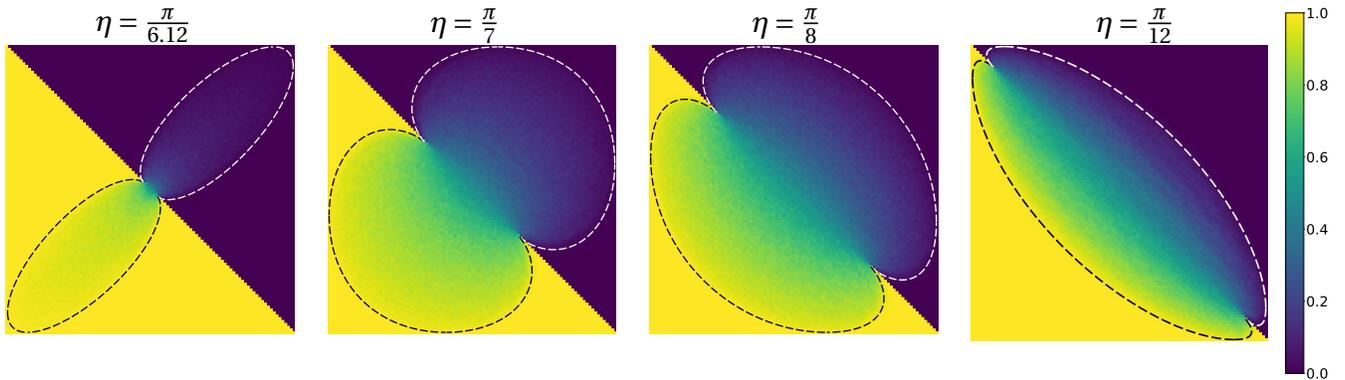
\begin{figure}[h!]
\hspace*{-1cm}
\begin{tikzpicture}[scale=0.9]
\draw (0,0) node{\includegraphics[scale=0.1]{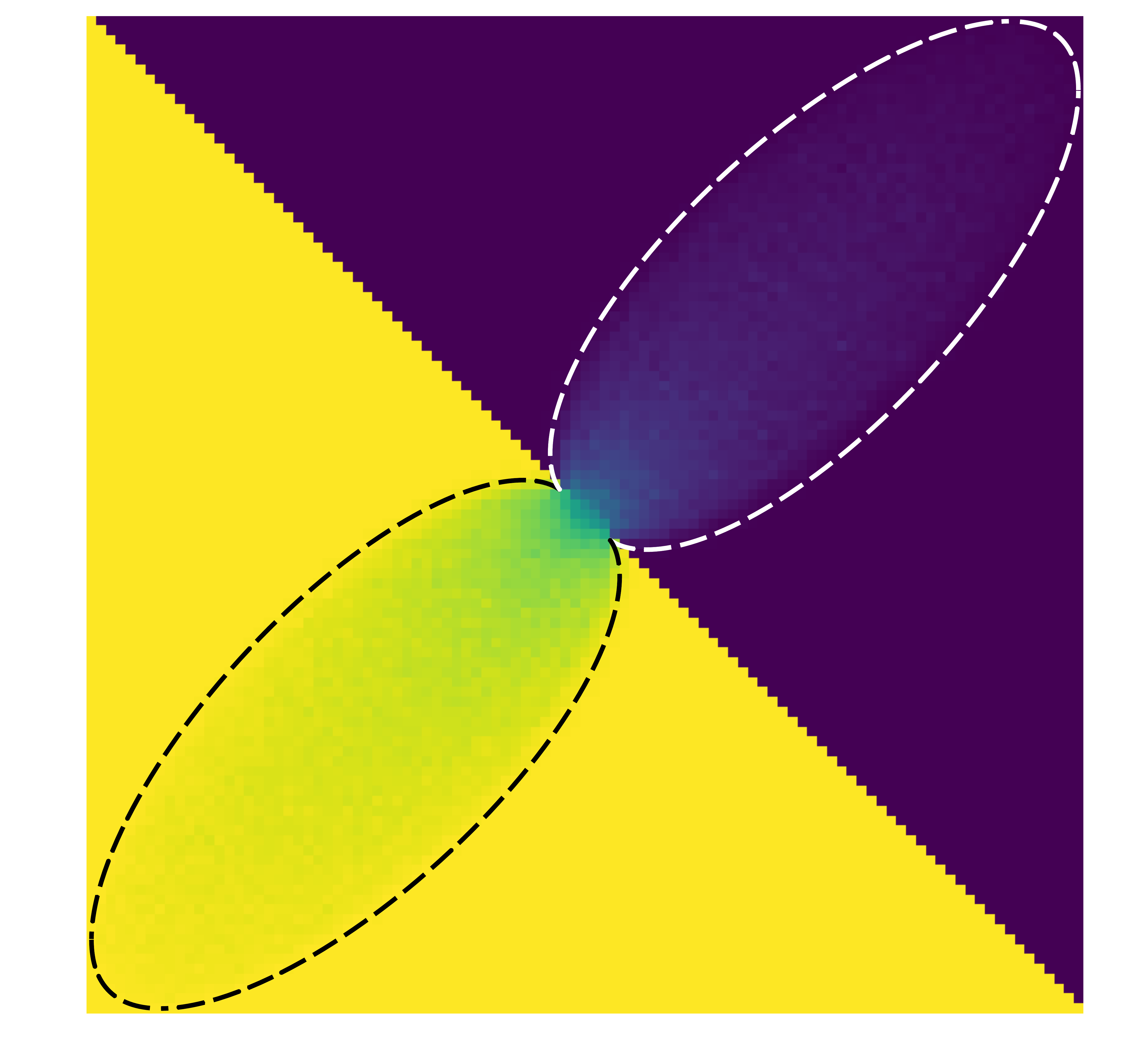}};
\draw (0,2.5)  node{$\eta=\frac{\pi}{6.12}$};
\draw (4.75,0) node{\includegraphics[scale=0.1]{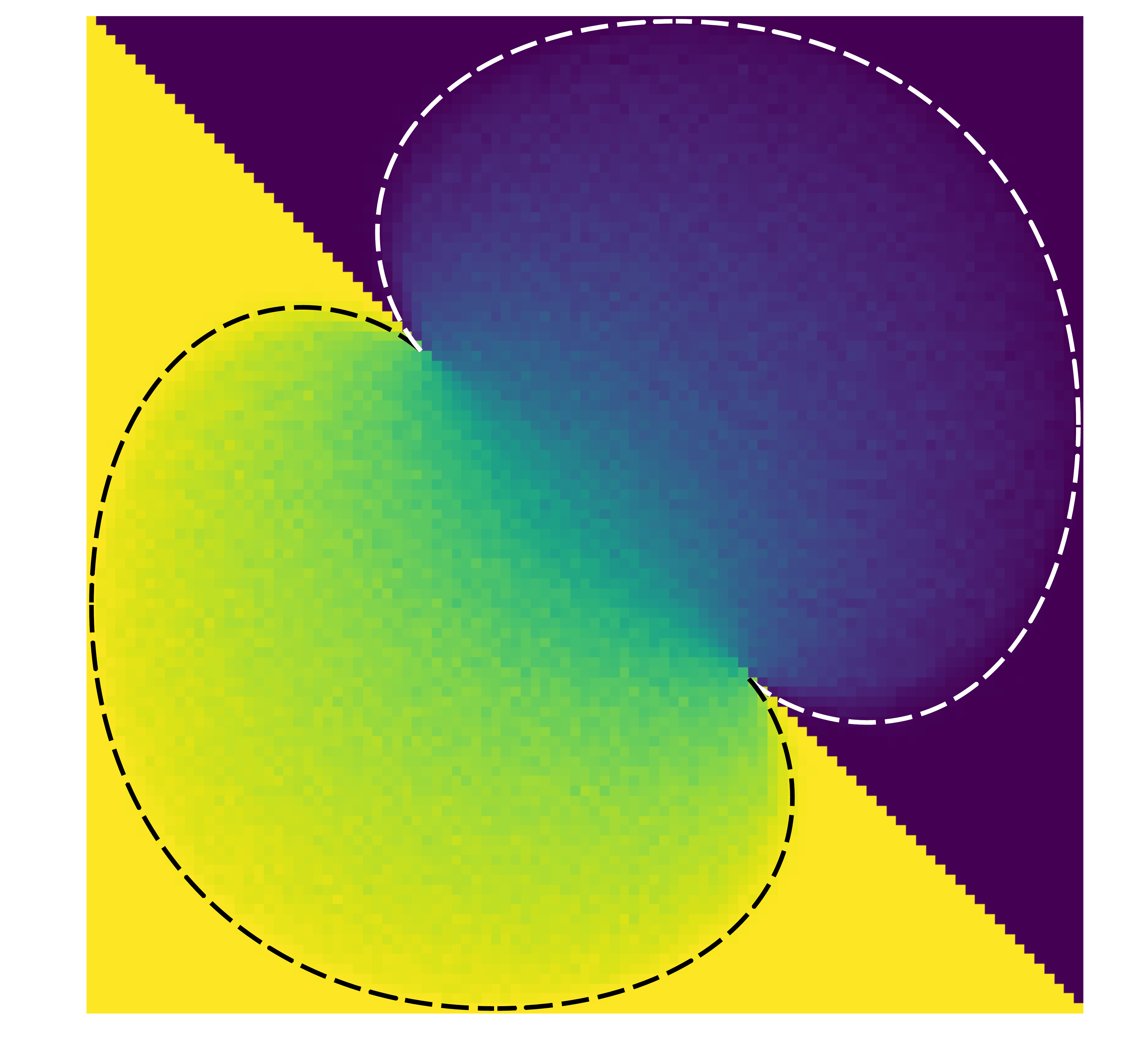}};
\draw (4.75,2.5)  node{$\eta=\frac{\pi}{7}$};
\draw (9.5,0) node{\includegraphics[scale=0.1]{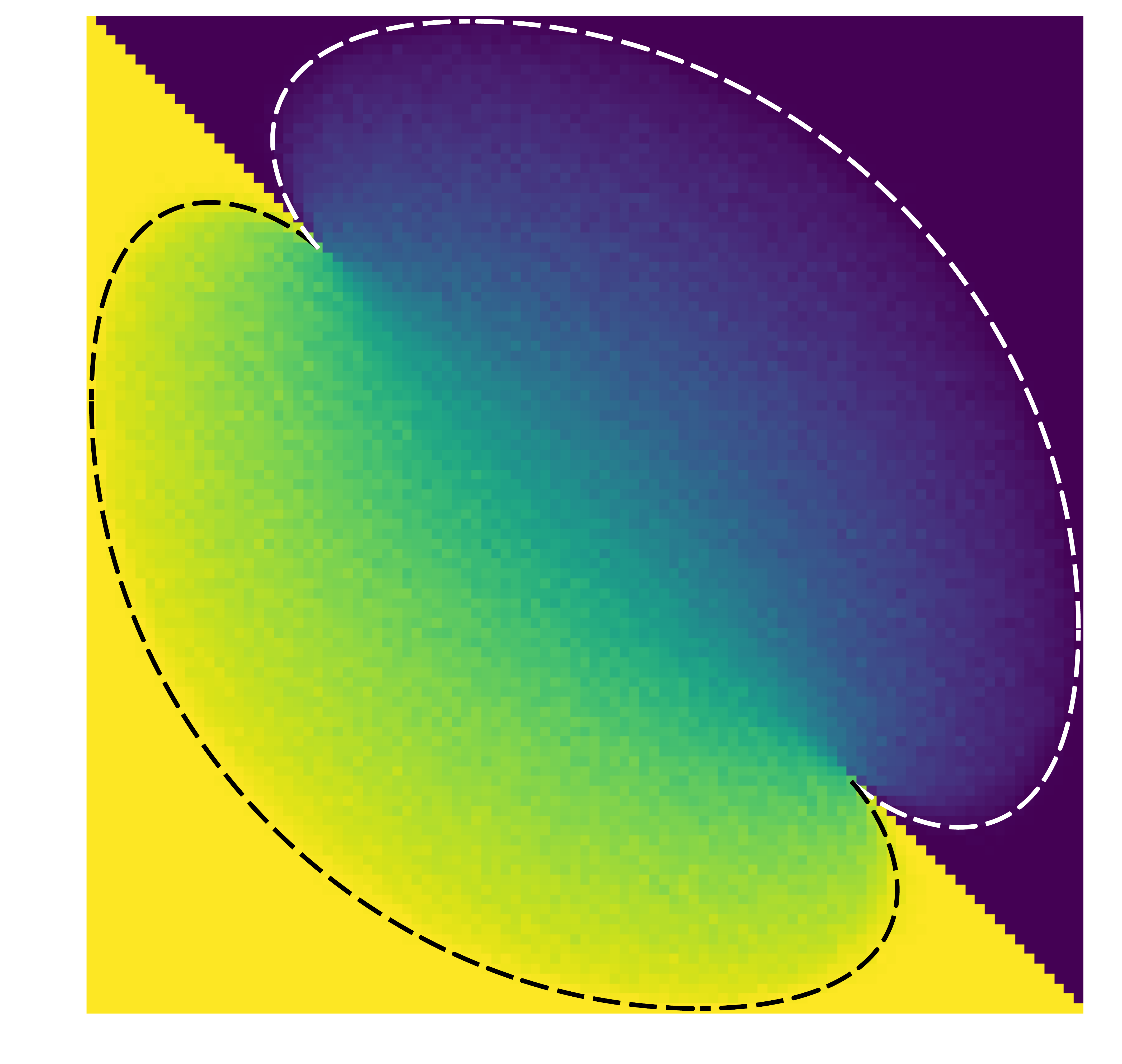}};
\draw (9.5,2.5)  node{$\eta=\frac{\pi}{8}$};
\draw (14.65,0) node{\includegraphics[scale=0.125]{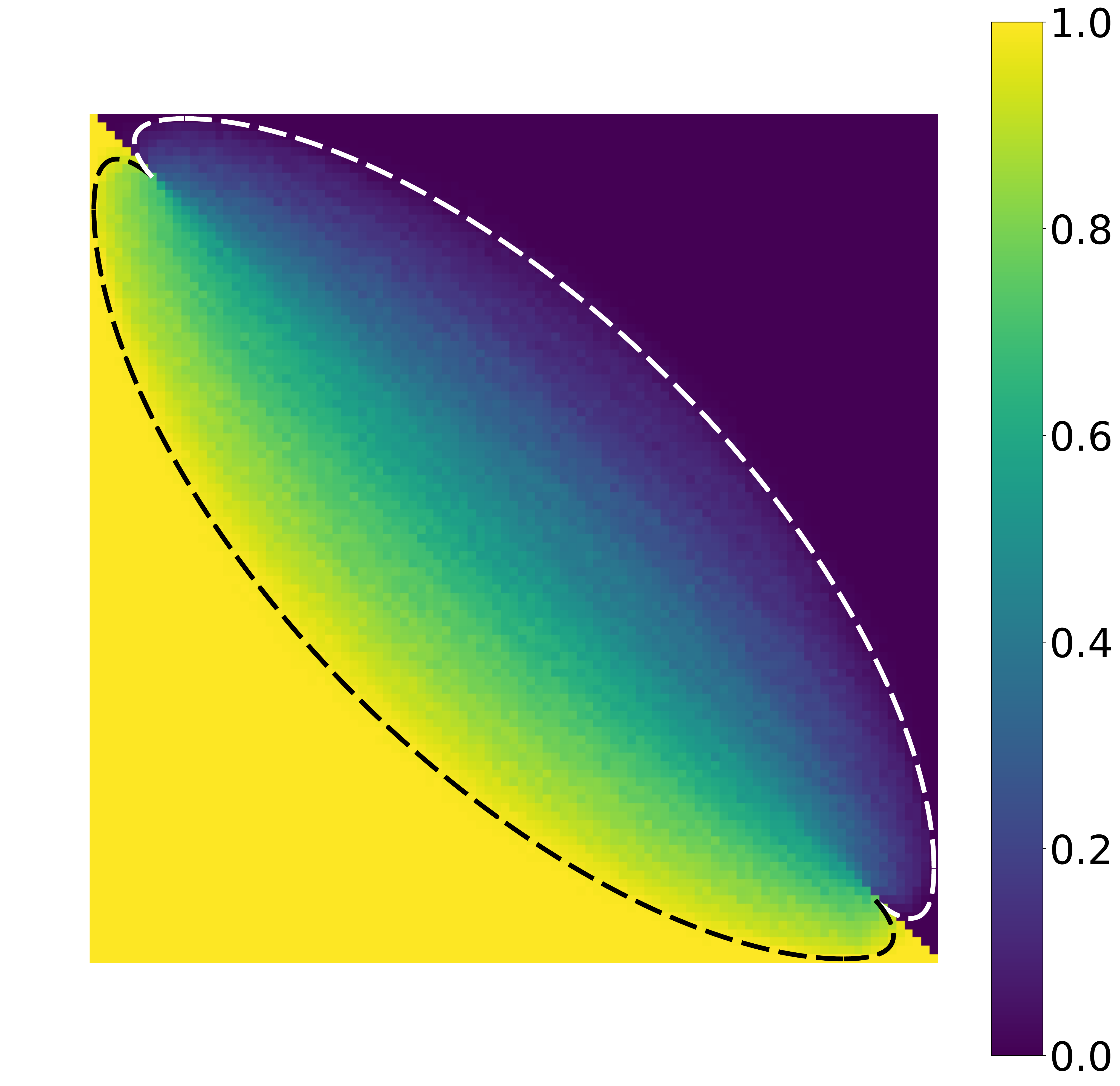}};
\draw (14.65,2.5)  node{$\eta=\frac{\pi}{12}$};
\end{tikzpicture}
\caption{The local density of diagonal steps of the 20V model with DWBC1 for $\lambda=5\eta$, $\mu=0$ and $n=100$, for
several values of $\eta\in]\pi/6,\pi/12]$. The order parameter is 
averaged over $1000$ configurations. The dashed curve is the analytic prediction for the arctic curve.}
\label{fig_n100_average_lambdafixed}
\end{figure}

\begin{figure}[h!]
\hspace*{-1cm}
\begin{tikzpicture}[scale=.9]
\draw (0,2.5)  node{$\lambda=4 \eta$};
\draw (0,0) node{\includegraphics[scale=0.1]
{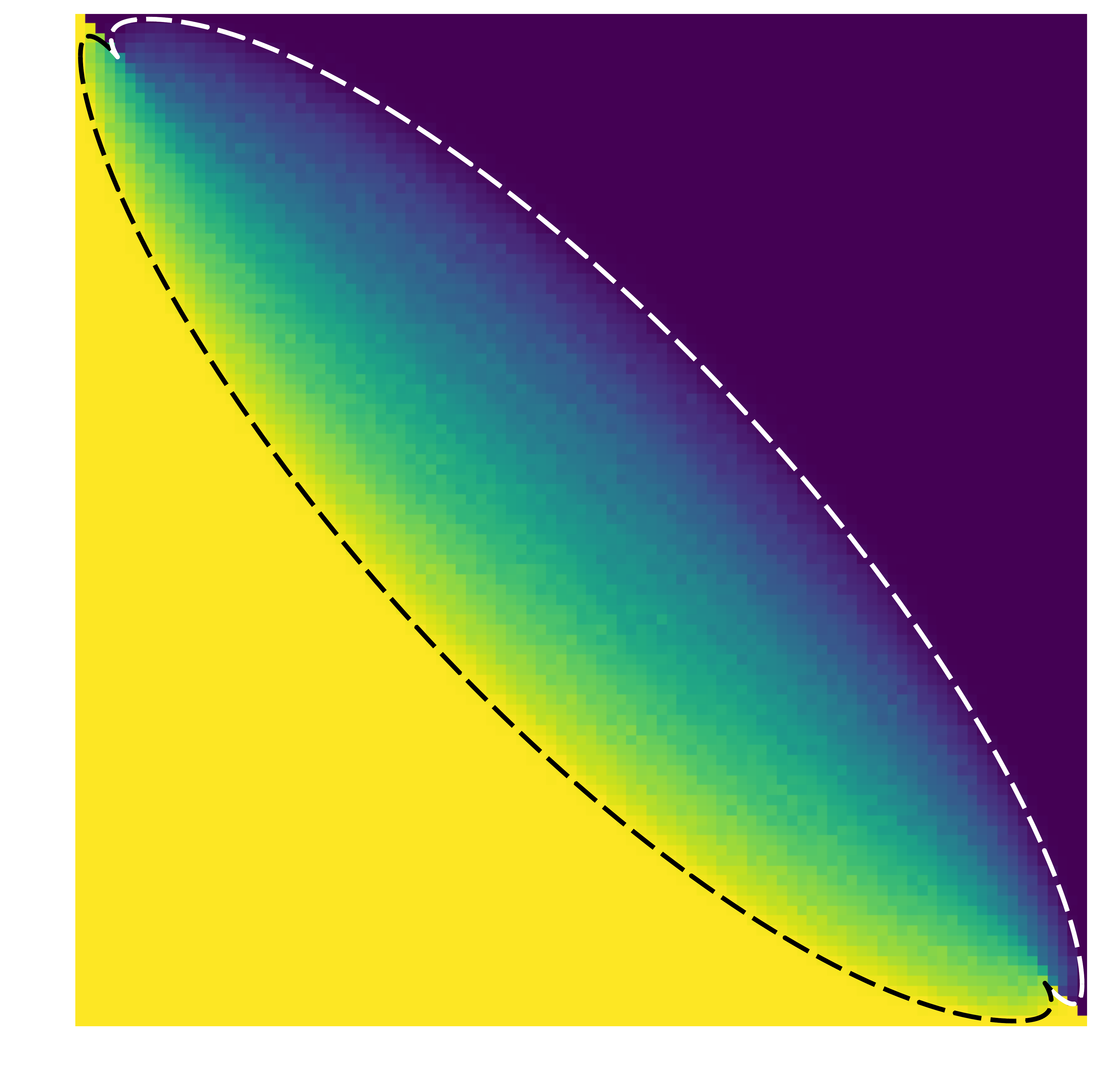}};
\draw (4.75,2.5)  node{$\lambda=6 \eta$};
\draw (4.75,0) node{\includegraphics[scale=0.1]
{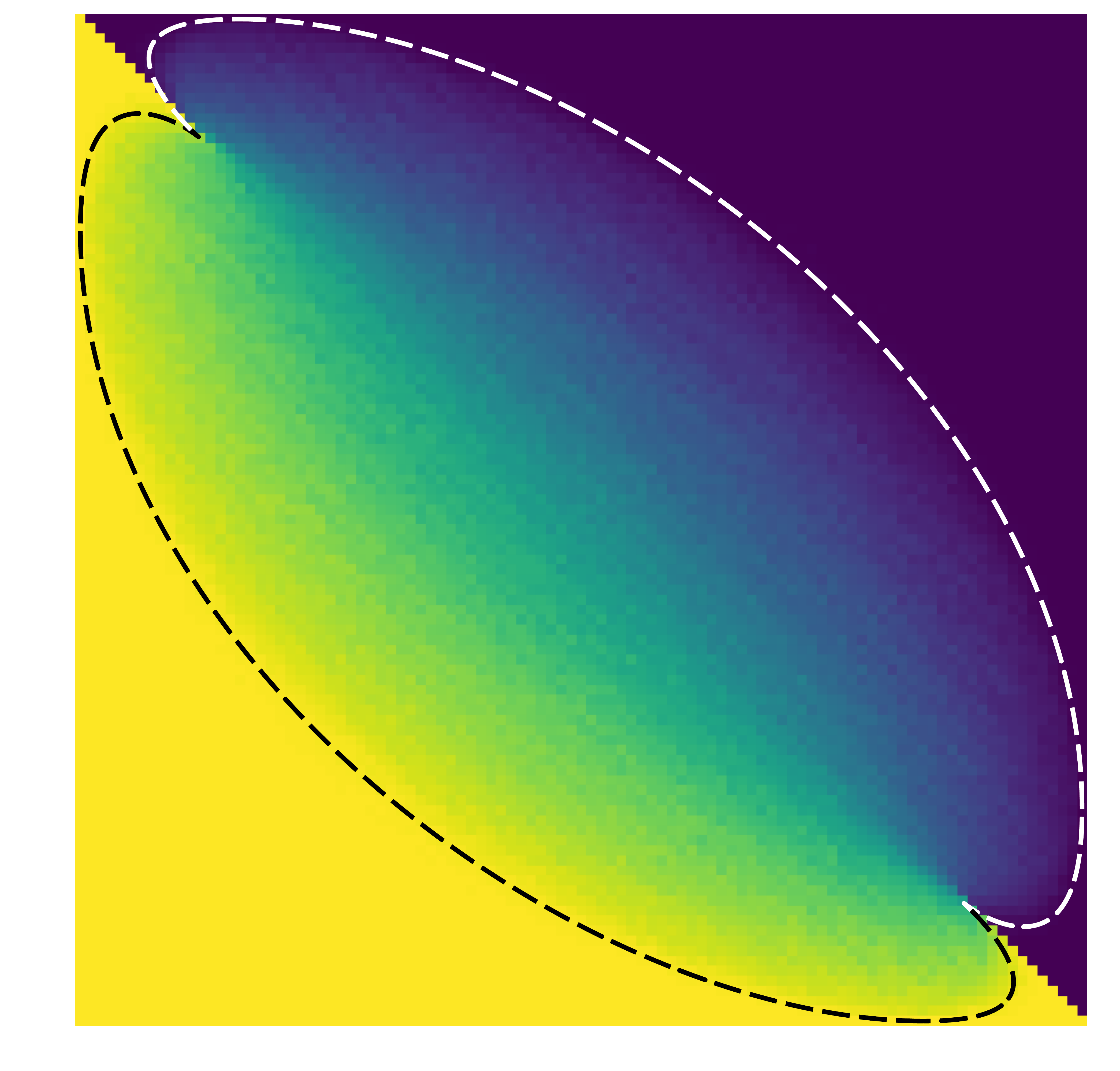}};
\draw (9.5,2.5)  node{$\lambda=8 \eta$};
\draw (9.5,0) node{\includegraphics[scale=0.1]
{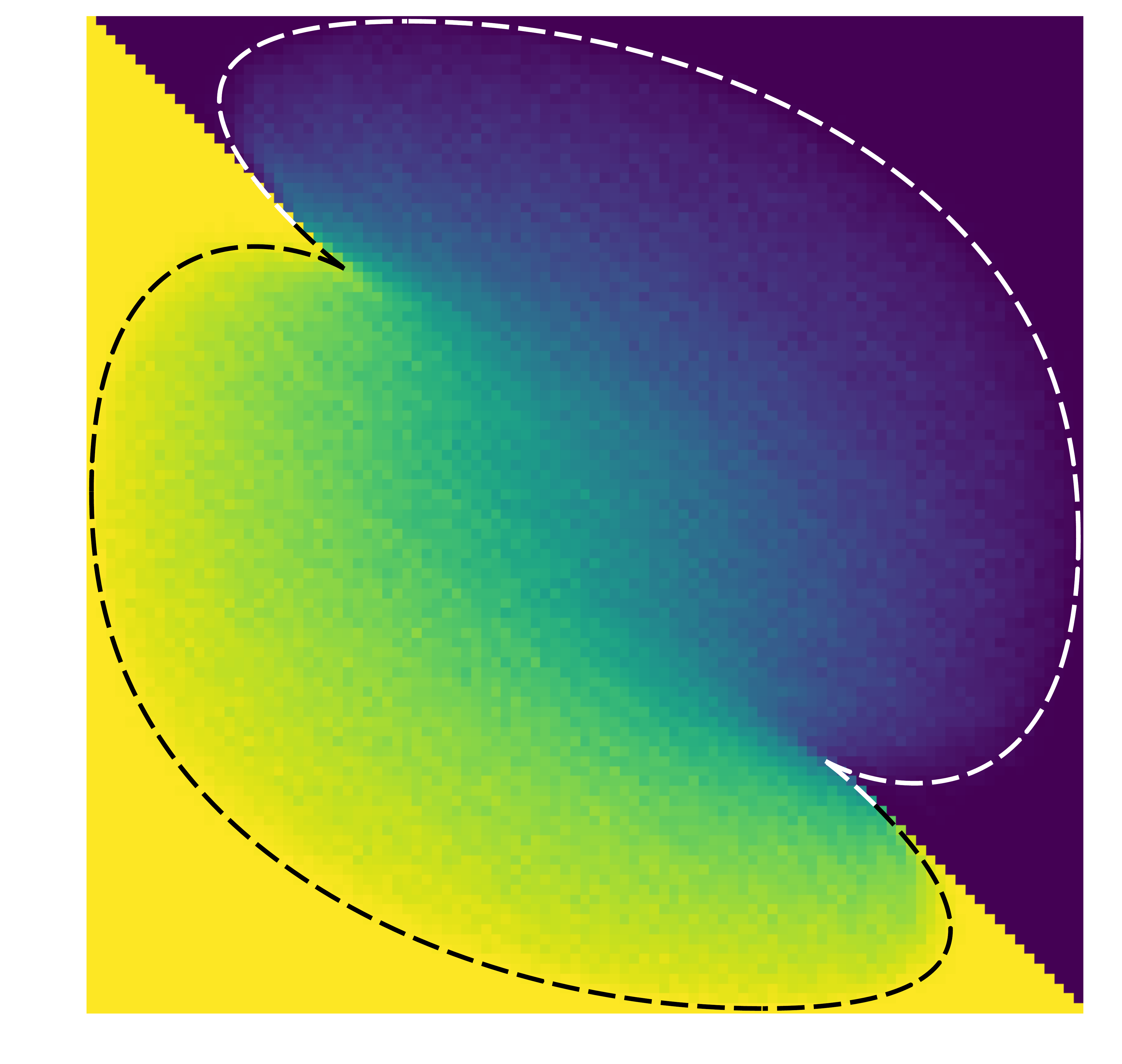}};
\draw (14.65,0) node{\includegraphics[scale=0.125]
{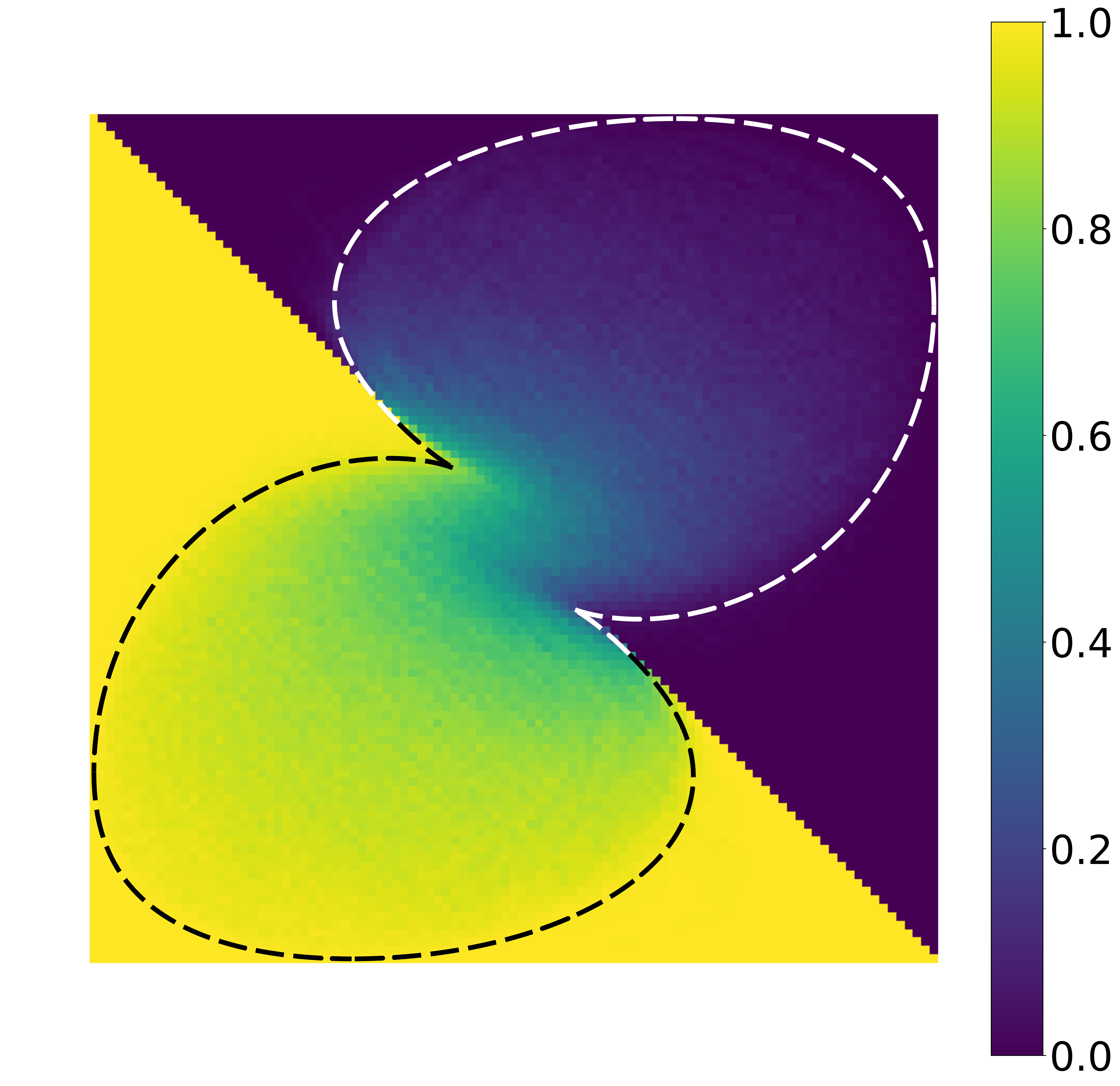}};
\draw (14.65,2.5)  node{$\lambda=10 \eta$};
\end{tikzpicture}
\caption{The local density of diagonal steps of the 20V model with DWBC1 for $\eta=\frac{\pi}{12}$, $\mu=\lambda-5\eta$ and $n=100$,
for several values of $\lambda\in [4\eta,10\eta]$. The order parameter is averaged over $1000$ configurations. The dashed curve is the analytic prediction for the arctic curve.}
\label{fig_n100_average_etafixed}
\end{figure}
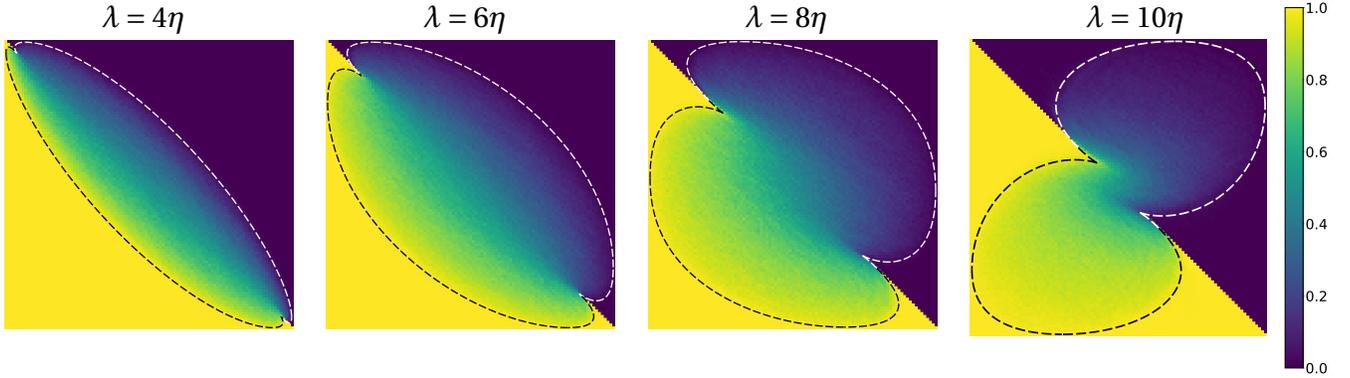

\begin{figure}[h!]
\hspace*{-1cm}
\begin{tikzpicture}[scale=.9]
\draw (0,0) node{\includegraphics[scale=0.13]
{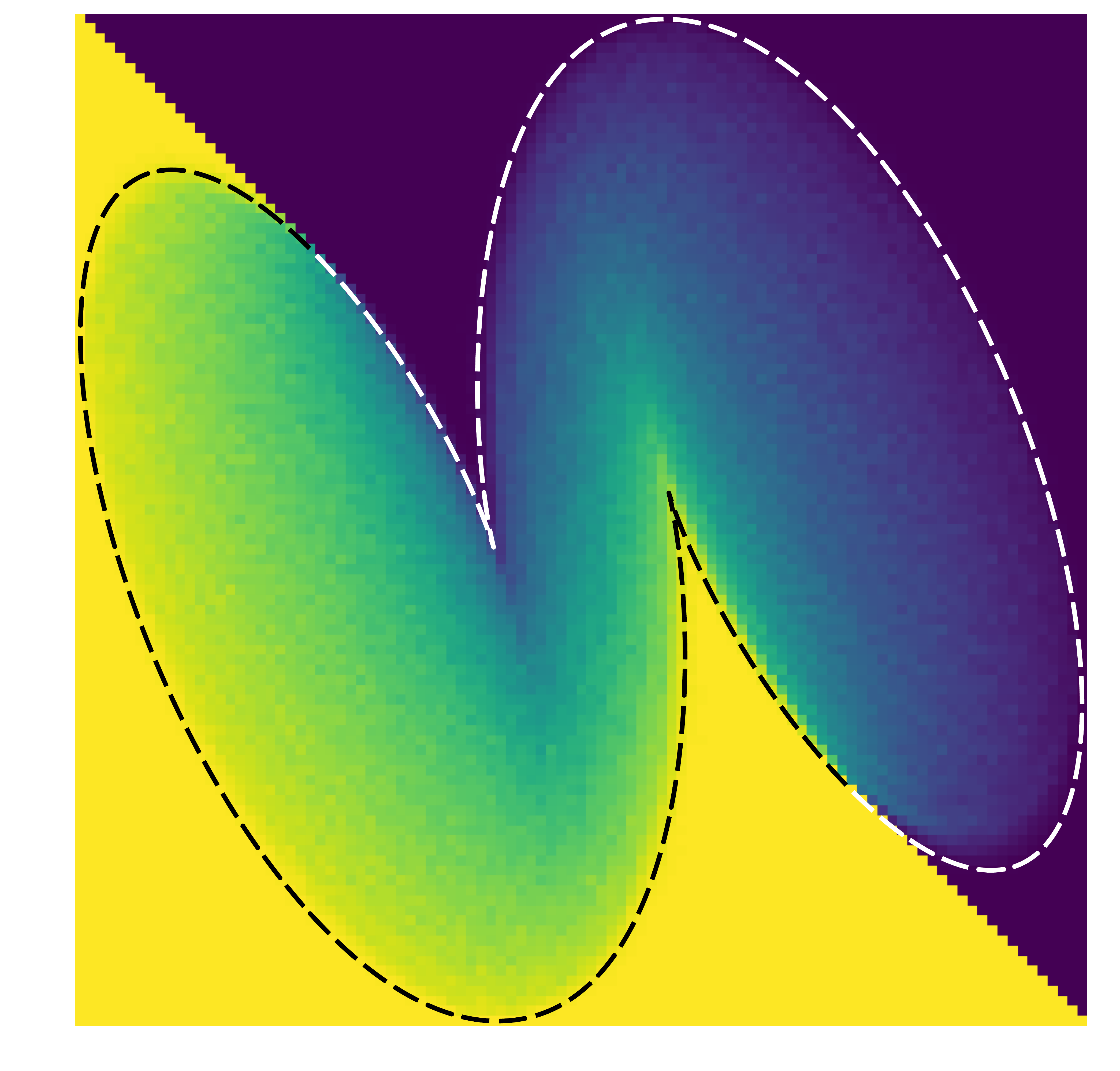}};
\draw (8,0) node{\includegraphics[scale=0.155]{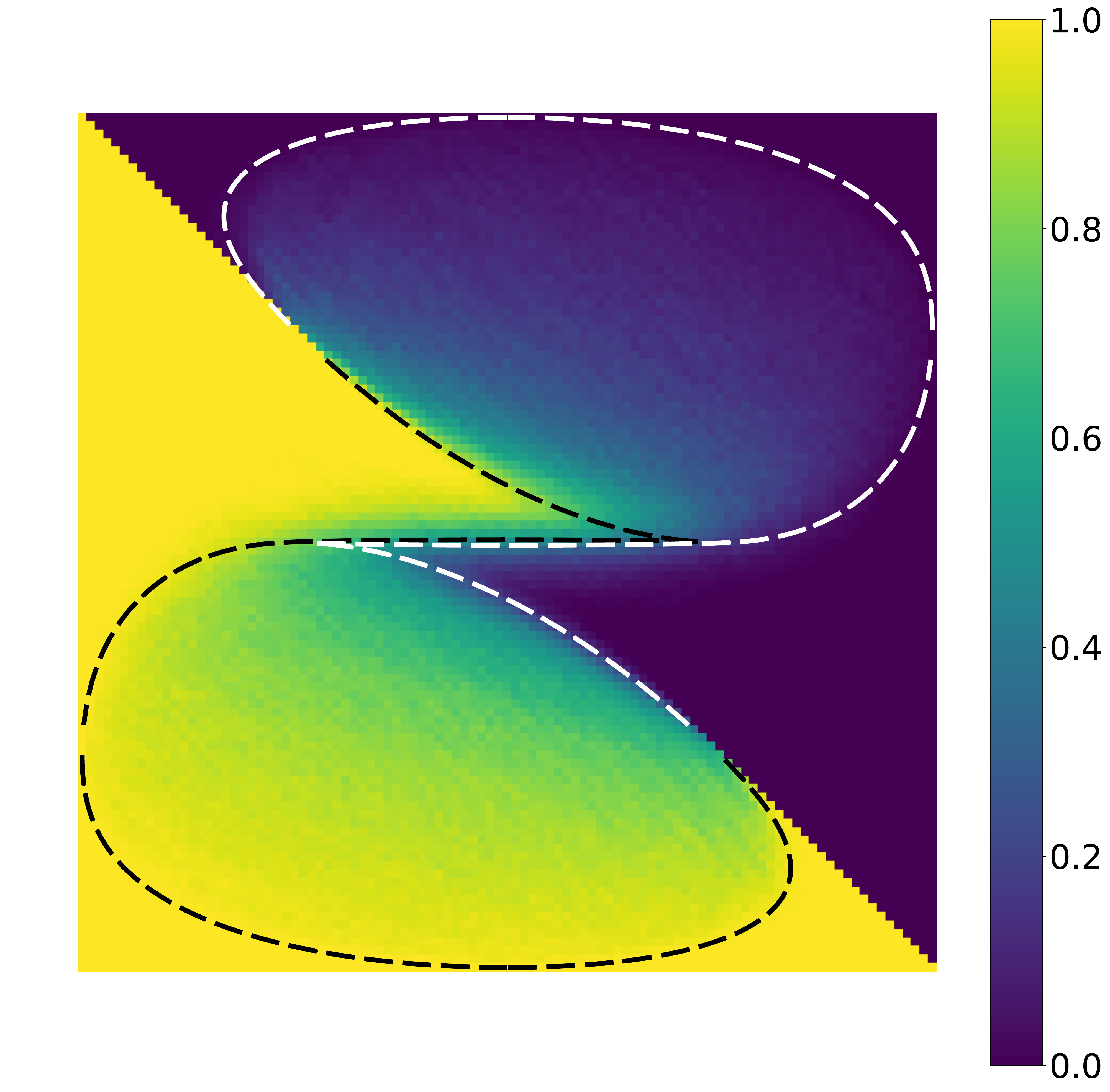}};
\end{tikzpicture}
\caption{The local density of diagonal steps of the 20V model with DWBC1 for $n=100$ in the case $\mu \neq \lambda-5\eta$. On the left $\eta=\pi/6$, $\lambda=9\pi/12$ and $\mu=-\pi/2$. On the right $\eta=1/200$, $\lambda=\pi-3\eta$ and $\mu=\lambda-5\eta$. The metropolis version of the algorithm was used.}
\label{fig_n100_average_etatozero}
\end{figure}

\begin{figure}[h!]
\includegraphics[scale=1.2]{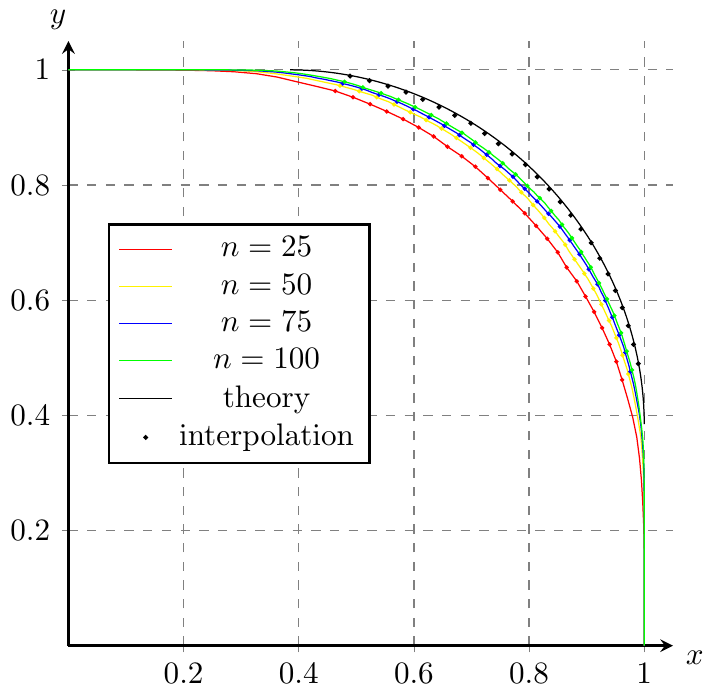}
\caption{Outermost path position averaged over $500$ configurations of the twenty-vertex model with DWBC1 for $n=25,50,75,100$, drawn in the rescaled domain. We estimate the curve reached asymptotically when $n \to \infty$, in the coordinates $u=(y+x)/2$ and $v=(y-x)/2$. For a fixed $v$ let us call $u_n(v)$ the corresponding $u$ on the average path $n$. We assume the scaling $u_n(v)=u(v)-n^{-2/3} \text{corr}(v)$, and extract an estimation of the arctic curve $u(v)$. As a consistency check one can alternatively estimate the scaling exponent $\alpha$ defined by $u(v)-u_n(v)=n^\alpha \text{corr}(v)$ by using for $u(v)$ the theoretical prediction. It is found that on average $\alpha=-0.627$ with a standard deviation of $\sigma=0.103$.}
\label{fig_interpolation}
\end{figure}

Figure \ref{fig_n100_single_configuration} displays a configuration with a stabilized arctic curve generated by our algorithm for the uniform distribution,
both in the path formulation (left) and vertex formulation (right).  Rather than displaying the precise 20V environments of each node, we choose to use 
as order parameter the local density of diagonal steps. Indeed this average density is $1$ in the frozen phases $HD$, $DV$ and
$HDV$ of Figure \ref{fig:phases}, and $0$ in $H$, $E$ and $V$, while it varies continuously in the liquid region. Figure~\ref{fig:cafe} displays the value
of this order parameter in the uniform case $\eta=\pi/8$, $\lambda=5\pi/8$ and $\mu=0$ and shows that it is indeed a good indicator for the position of the arctic curve.
The evolution of the arctic curve with varying parameters is shown in Figures~\ref{fig_n100_average_lambdafixed}, \ref{fig_n100_average_etafixed} and \ref{fig_n100_average_etatozero}. In all cases we also indicate the theoretical prediction (dashed curve): the agreement is quite good, despite what looks like an ``attraction'' of the finite-size arctic curve towards the liquid region. These finite size effects are estimated in Figure \ref{fig_interpolation} by evaluating the average outermost path for different sizes, here in the uniform case. Similar finite size effects were analyzed \cite{johansson2005} in the case of the uniform domino tiling of the Aztec diamond, and found to be governed by a scaling exponent $\alpha=-2/3$. Our results are compatible
with a scaling exponent  $\alpha=-2/3$ as well.

\section{The arctic curve for the Quarter Turn symmetric Holey Aztec Domino Tiling model}
\label{sec:QTHADT}
\begin{figure}
\begin{center}
\includegraphics[width=10cm]{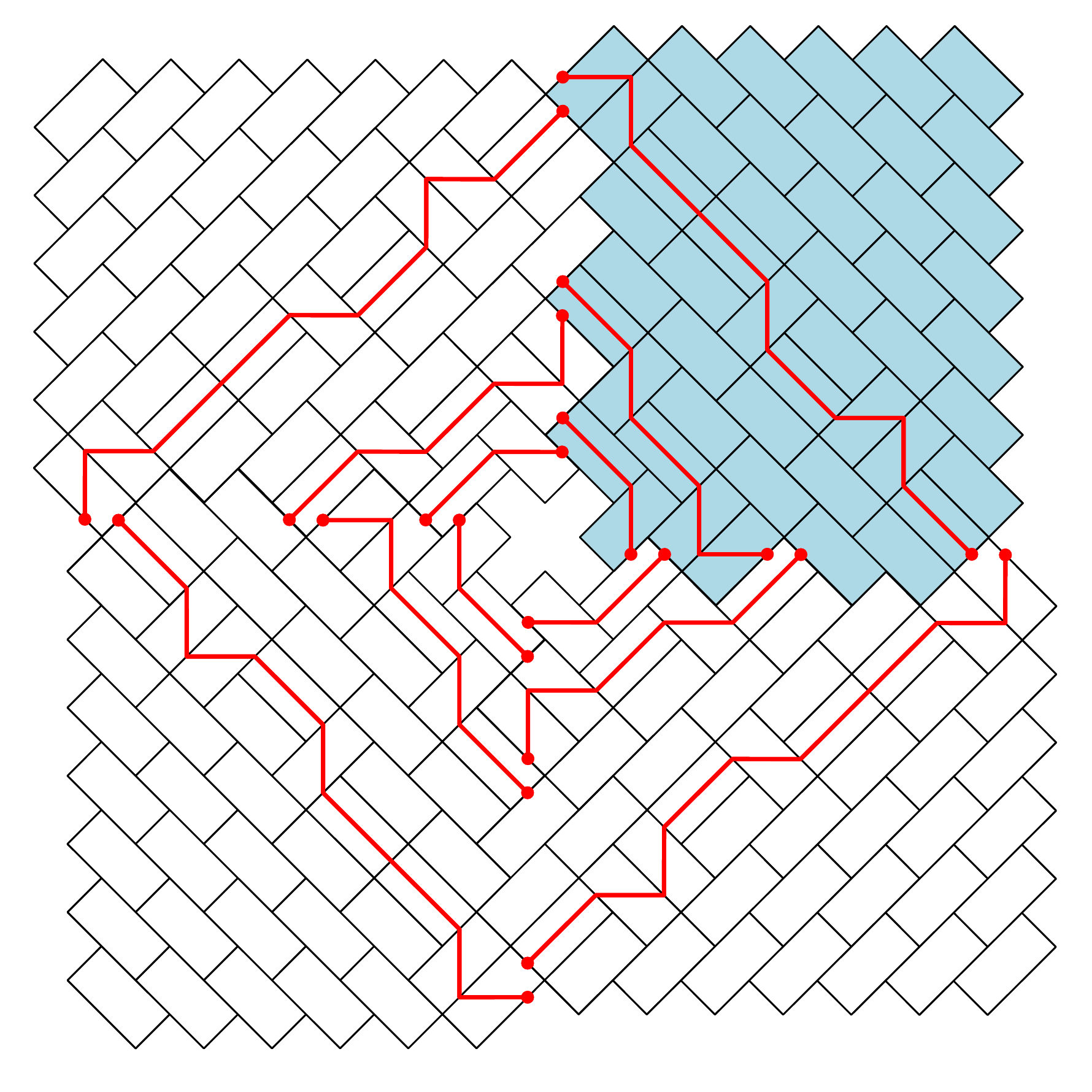}
\end{center}
\caption{\small A sample configuration of the QTHADT model together with the associated non-intersecting Sch\"oder path configuration.
The path configuration of the fundamental domain (upper right quadrant) is repeated by symmetry in the whole domain.}
\label{fig:pavage}
\end{figure}
In \cite{DFG20V}, it was shown that the set of configurations of the 20V model with DWBC1 or 2 on an $n\times n$ square is equinumerous to that of
domino tilings of a quasi-square domain of Aztec-like shape of size $2n \times 2n$ with a cross- shaped hole in the middle
that are invariant under a quarter-turn rotation (i.e.\ a rotation by $90^\circ$) around the center of the cross (see \cite{DFG20V} for a detailed definition).
We shall refer to this model as the Quarter Turn symmetric Holey Aztec Domino Tiling (QTHADT) model, see Figure~\ref{fig:pavage}. A natural question 
is then that of the shape of the arctic curve of this domino tiling problem. Here we give the answer to this question
based again on the tangent method. For simplicity, we limit ourselves to the derivation of a single portion of arctic curve (analogous to the
``normal'' portion in the 20V problem). The remainder of the arctic curve is then obtained by analytic continuation (see discussion below).
As it turns out, our results are validated by numerical simulations, with a perfect agreement.   

By symmetry, any configuration of the QTHADT model is entirely determined by its intersection with the fundamental domain, say
the upper right quadrant. As shown in \cite{DFG20V}, configurations in the fundamental domain may 
be reformulated in terms of non-intersecting Schr\"oder paths, with horizontal, diagonal and vertical steps, with symmetic starting and ending points
along the boundary and subject to particular restrictions (see Figure~\ref{fig:pavage} for an illustration).   
The number of paths is not fixed and may vary between $0$ and $n$, and all path steps receive the same weight $1$. Here we slightly generalize
the model by introducing an extra weight $\gamma$ per diagonal step. This amounts in the tiling language to assign a weight $\gamma$ to a particular type of tile. 

\subsection{Partition function}
\label{sec:PF}
As discussed in \cite{DFG20V} for $\gamma=1$, and straightforwardly generalized to an arbitrary $\gamma$, the partition function of the QTHADT problem is given by $\det(A)$, where 
\begin{equation*}
\begin{split} 
&A_{i,j}=F_A(u,v)\vert_{u^i v^j} \quad (i,j=0,1,...,n-1)\ ,\\
& F_A(u,v)=\frac{1}{1-u v} +\frac{(1+\gamma) u}{(1-u)(1-u-v-\gamma u v)}\ .\\
\end{split}
\end{equation*}
The latter function corresponds to a matrix $A={\mathbb I}+M$, where $M_{i,j}$ enumerates Schr\"oder path configurations from to
$(0,j+1)$ to $(i,0)$ which are ``restricted'' so that their first step cannot be a down step. The generating function 
$F_M(u,v)=\sum _{i,j\geq 0}M_{i,j}u^iv^j$ is obtained as follows: by combining at least one first horizontal step (an arbitrary $k\geq 1$ number of them,
generated by $\frac{u}{1-u}$)
followed by a vertical step (generated by $v$), or an arbitrary number $k\geq 0$ of horizontal steps
(generated by $\frac{1}{1-u}$) followed by a diagonal step (generated by $\gamma u v$), both 
followed by a Schr\"oder path with $\gamma$ weight on diagonal steps (generated by 
$1/(1-u-v-\gamma u v)$), we build all the desired restricted paths. The result is
$$F_M(u,v)=\frac{1}{v}\left(\frac{u}{1-u} v+\frac{1}{1-u} \gamma u v \right)\frac{1}{1-u-v-\gamma u v}\ ,$$
hence the second term in the equation above (as in \cite{DFG20V}, the prefactor $1/v$ accounts for the fact that the height of the starting point is 
$j+1$, not $j$).

\subsection{Refined partition function}
\label{sec:RPF}
The path enumeration may be refined as follows: we introduce an extra multiplicative weight $\tau$ per horizontal step
along the row of maximal height $n$. For paths that start at position $(0,n)$ this changes the weight
as follows: paths are obtained either by combining at least one first horizontal step (an arbitrary $k\geq 1$ number of them,
generated by $\frac{\tau u}{1-\tau u}$) 
followed by a vertical step (generated by $v$), or by combining an arbitrary number $k\geq 0$ of horizontal steps
(generated by $\frac{1}{1-\tau u}$) followed by a diagonal step (generated by $\gamma u v$), both 
followed by a Schr\"oder path starting at height $n-1$ (generated by 
$1/(1-u-v-\gamma u v)$). The net result is a change $A\to A(\tau)={\mathbb I}+M(\tau)$ where $M(\tau)$ differs from $M$ in its $n$-th column 
only, now  generated by
$$\sum_{i=0}^{\infty} M_{i,n-1}(\tau) u^i=\left. \frac{1}{v}\left(\frac{\tau u}{1-\tau u} v+\frac{1}{1-\tau u} \gamma u v \right)\frac{1}{1-u-v-\gamma u v}\right\vert_{v^{n-1}}\!\!\!\!\!\!=
\left. \frac{(\tau+\gamma) u}{(1-\tau u)(1-u-v-\gamma u v)}\right\vert_{v^{n-1}}
$$
so that the complete generating function for $A(\tau)$ is therefore
\begin{eqnarray*}F_{A(\tau)}(u,v)&=&\frac{1}{1-u v} +\frac{(1+\gamma) u}{(1-u)(1-u-v-\gamma u v)} +
\left\{  \frac{\tau+\gamma}{1-\tau u} -\frac{1+\gamma}{1-u}\right\}
\frac{(1+\gamma u)^{n-1}}{(1-u)^{n}}\,u \,  v^{n-1}\\
&=&\frac{1}{1-u v} +\frac{(1+\gamma) u}{(1-u)(1-u-v-\gamma u v)} + 
\frac{(\tau-1)u}{(1-u)(1-\tau u)} \left(\frac{1+\gamma u}{1-u}\right)^n\, v^{n-1}\ .
\end{eqnarray*}

\subsection{Comparision with the 6V model refined partition function}
\label{sec:QTHADT6V}
The refined partition function for the 6V model with DWBC, with weights $a,b,c$ and an extra weight $\sigma$
per horizontal step in the top row (in the osculating path formulation) is
$\det(B(\sigma))$, where the matrix $B(\sigma)=(B_{i,j}(\sigma))_{0\leq i,j\leq n-1}$ is generated by \cite{BDFPZ1}
$$ F_{B(\sigma)}(u,v)=\frac{1}{1-u v}+ \frac{x u}{(1-u)(1\!-\!y u\!-\!v\!-\!(x\!-\!y)u v)} + \frac{(\sigma-1) x u}{(1-u)(1-(y+x(\sigma-1))u)}
\left( \frac{1+(x-y)u}{1-y u}\right)^n  v^{n-1} $$
with
$$ x=\left(\frac{b}{a}\right)^2,\qquad y=\left(\frac{c}{a}\right)^2 .$$
Comparing with the expression found in the previous section, we are led to identify $y=1$ and $x=1+\gamma$.
In the usual parametrization $a=\sin(\lambda+\eta)$, $b=\sin(\lambda-\eta)$ and $c=\sin(2\eta)$, this corresponds
to taking $\lambda=\pi -3 \eta$, so that $a=c$ and $b/a=2\cos(2\eta)$, leading
to
 $$\gamma=1+2\cos(4\eta)\ .$$
The generating functions $F_{A(\tau)}(u,v)$ and $F_{B(\sigma)}(u,v)$ are therefore identified upon taking
$$ \tau= y+x(\sigma-1)=1+(1+\gamma)(\sigma-1)=1+4\cos^2(2\eta) (\sigma-1) .$$

\subsection{Tangent method for the QTHADT with weight $\gamma$}
\label{sec:QTHADTtg}
The use of the tangent method to determine the arctic curve of the QTHADT is similar to that for the 20V-model 
in the geometry of Section~\ref{sec:final} used for the alternative derivation of the ``normal'' portion. A subtle difference arises
in the definition of the position $L$ of the escape point: if there exists a path with original starting point $(0,n)$, this starting point
is moved to $(0,n+M)$ as usual, leading to a non-trivial value of $L$, while if it does not exist, we set $L=0$ by convention independently
of $M$ (we add in this case a trivial vertical segment with weight $1$ from $(0,n)$ to $(0,n+M)$).
The subsequent extremization with respect to $L$ shows that the most likely value of $L$ is non-zero, hence the
configurations with no path starting at $(0,n)$ are negligible and the arctic curve is therefore well probed in our approach. 
Using \eqref{eq:fsigma} for the present value $\lambda=\pi-3\eta$, we get
\begin{eqnarray*}
\sigma&=&\sigma(\xi)=2\cos(2\eta)\,  \frac{\sin(4\eta-\xi)}{\sin(2\eta-\xi)}\\
f(\xi)&=&{\rm Log}\left(\frac{\sin(2\al \eta) \sin(\al \xi) \sin(4\eta-\xi)}{\al \sin(4\eta)\sin(\xi)\sin(\al(2\eta-\xi)} \right)
\end{eqnarray*}
with $\al=\pi/(\pi -2\eta)$,
leading to
$$ \tau(\xi)=\frac{\sin(2\eta+\xi)}{\sin(2\eta -\xi)}\ . $$
This defines implicitly $f$ as a function of $\tau$.

Setting $M=nm$ and $L=n\ell$ and using renormalized coordinates (divided by $n$),
the equation of the line through the endpoint $(0,1+m)$ and the escape point $(\ell,1)$ is $\frac{\ell}{m}(y-1)+x -\ell=0$.

As for the 20V-model, we may get the most likely value of $\ell$ for fixed $m$ by extremizing the appropriate action $S_0(\ell,\tau)+S(\ell,m,\phi)$, over $\ell,\tau,\phi$, where
\begin{eqnarray*}
S_0(\ell,\tau)&=&f(\xi(\tau))-\ell \, {\rm Log}(\tau)\\
S_1(\ell,m,\phi)&=&(\ell+m-\phi) {\rm Log} (\ell+m-\phi)-(\ell-\phi){\rm Log}(\ell-\phi)\\
&&\qquad -(m-\phi){\rm Log}(m-\phi)-\phi\, {\rm Log}(\phi)+\phi \, {\rm Log}(\gamma) \ .
\end{eqnarray*}
Here, the interpretation of $\phi$ is that $n \phi$ is the total number of diagonal steps.   
This gives
$$\ell=\tau \frac{df}{d\tau}=\frac{\tau(\xi)}{\tau'(\xi)} f'(\xi)=:\rho(\xi),\qquad  \gamma \frac{(\ell-\phi)(m-\phi)}{\phi\, (\ell+m-\phi)}=1,\qquad  \frac{\ell+m-\phi}{\ell-\phi}=\tau$$
and finally
\begin{equation*}
\frac{\ell}{m}=1+\frac{(1+\gamma)\tau(\xi)}{(\tau(\xi)-1)(\tau(\xi)+\gamma)}=:\beta(\xi)
\end{equation*}
so that the equation for the family of tangent lines reads $\beta(\xi)(y-1)+x-\rho(x)=0$.
We end up with the following parametrization of the arctic curve:
\begin{thm}
\label{thm:acqt}
The arctic curve of the QTHADT model in its fundamental domain, as predicted by the tangent method, is given by
\begin{equation}
x(\xi)=\rho(\xi)- \beta(\xi)\frac{\rho'(\xi)}{\beta'(\xi)},\quad y(\xi)=1+\frac{\rho'(\xi)}{\beta'(\xi)}
\ , \qquad \xi\in[0,2\eta]\ ,
\label{eq:ACholey} 
\end{equation}
where
\begin{eqnarray*}
\rho(\xi)&=& \bigg(\al \big(\cot(\al \xi)+\cot(\al (2 \eta-\xi)\big)-\cot(\xi)-\cot(4 \eta-\xi) \bigg) \frac{\sin(\xi+2\eta)\sin(2\eta-\xi)}{\sin(4\eta)}\ ,\\
\beta(\xi)&=&\frac{\sin(2\eta+\xi)\sin(2\eta-\xi)}{\sin(\xi)\sin(4\eta-\xi)}\ .
\end{eqnarray*}
\end{thm}
It is easily checked that $x(\xi)=y(2\eta-\xi)$ hence the arctic curve is symmetric under $x\leftrightarrow y$, as expected.
The range $\xi\in[0,2\eta]$ for the current geometry leads only to a portion of the arctic curve from $(z,1)$ to $(1,z)$ 
with $z=x(0)=y(2\eta)$. We expect that the above expression extends to the range $\xi\in[\max(-2\eta,2\eta-\pi/2),\min(4\eta,\pi/2)]$, leading to two additional 
portions from $(0,z')$ to $(z,1)$ and from $(1,z)$ to $(z',0)$, with $z'=x(\min(4\eta,\pi/2))=y(\max(-2\eta,2\eta-\pi/2))$. This completes the description of the arctic curve in the fundamental domain. The full arctic curve is obtained by iterated $90^\circ$ rotation copies of the latter. Note that, for $\gamma\neq 1$, these new copies differ in general from the analytic continuation of the fundamental domain copy. This analytic continuation corresponds in principle to a different tiling problem where the weighting of tiles does not depend
on the underlying quadrant.

\subsection{Simulations}
\label{sec:QTHADTsimul} 

\begin{figure}
\begin{center}
\begin{tikzpicture}
\clip (-2,-2.5) rectangle (9+1.5,2);
\draw (0,0) node{\includegraphics[scale=0.5,angle=-45]{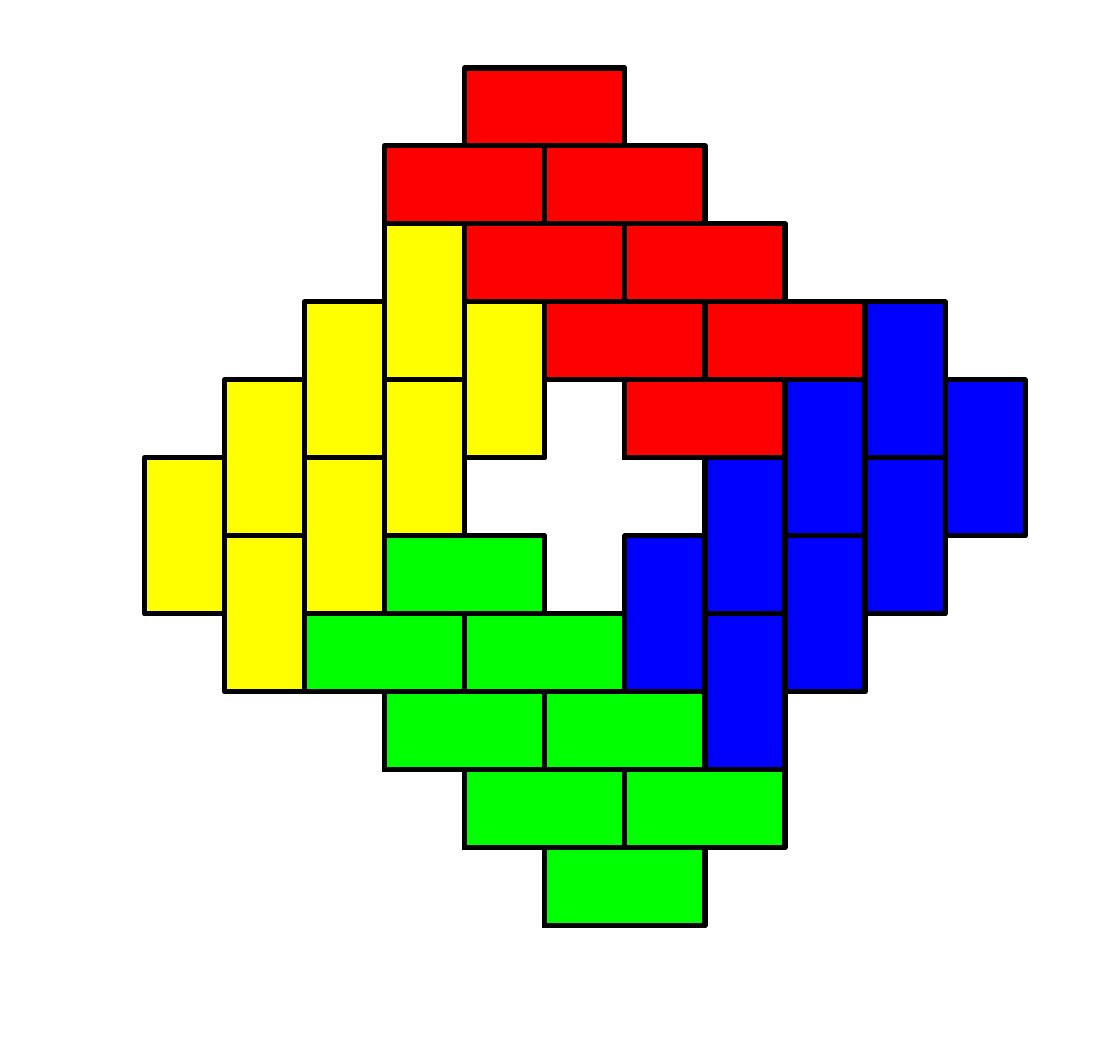}};
\draw (8,0) node{\includegraphics[scale=0.5,angle=-45]{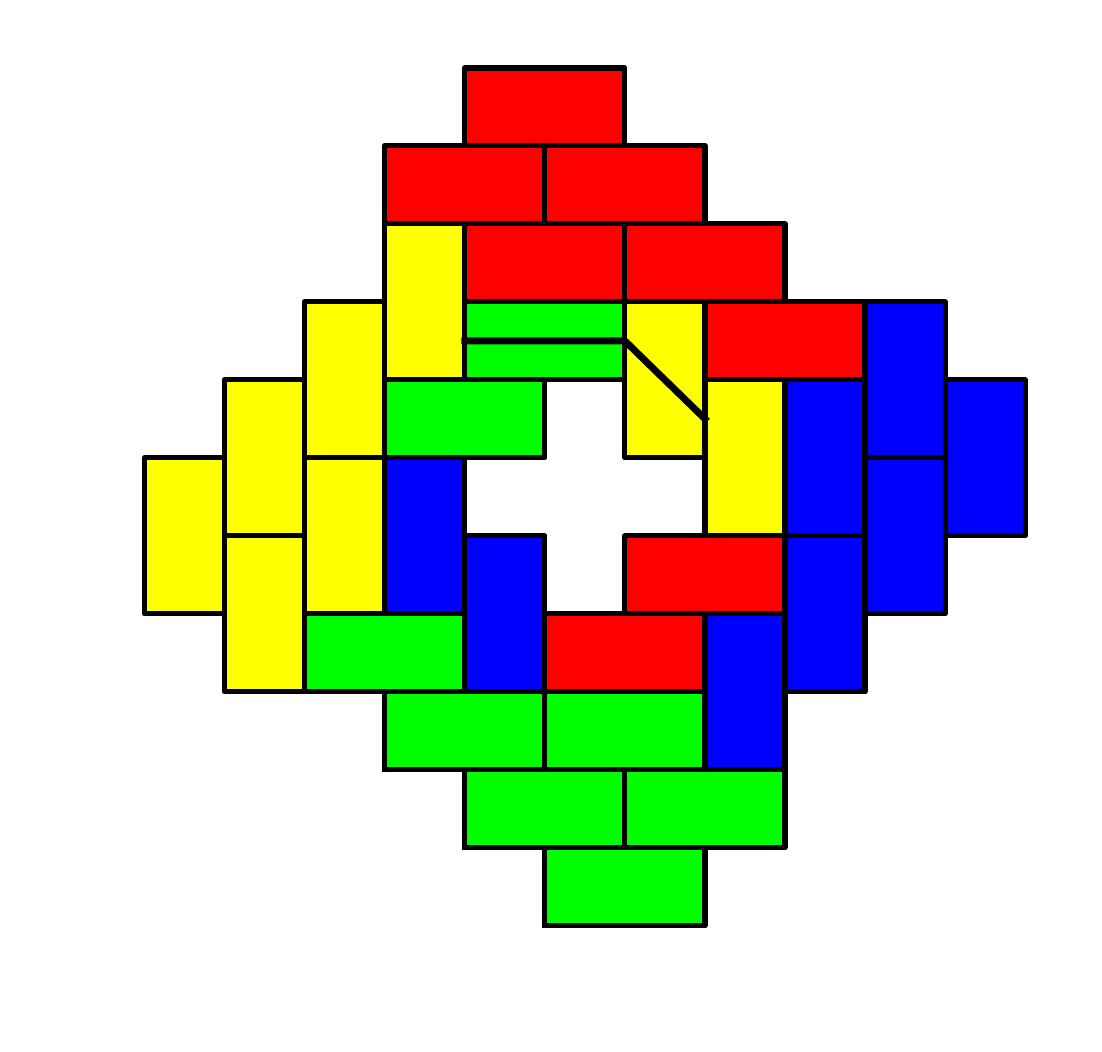}};
\draw[<->,>=latex] (4.12-0.3,0)--++(1,0) ;
\end{tikzpicture}
\caption{The ``cross-move'' performed on the unique configuration of the $\ell=0$ sector (left) for $n=3$. Dominoes are shifted along 
the central cycle around the cross, thus creating a path and sending us to the $\ell=1$ sector (right).  
The created path is drawn in the fundamental domain.}
\label{fig_crossMove}
\end{center}
\end{figure}

\begin{figure}
\includegraphics[scale=0.83]{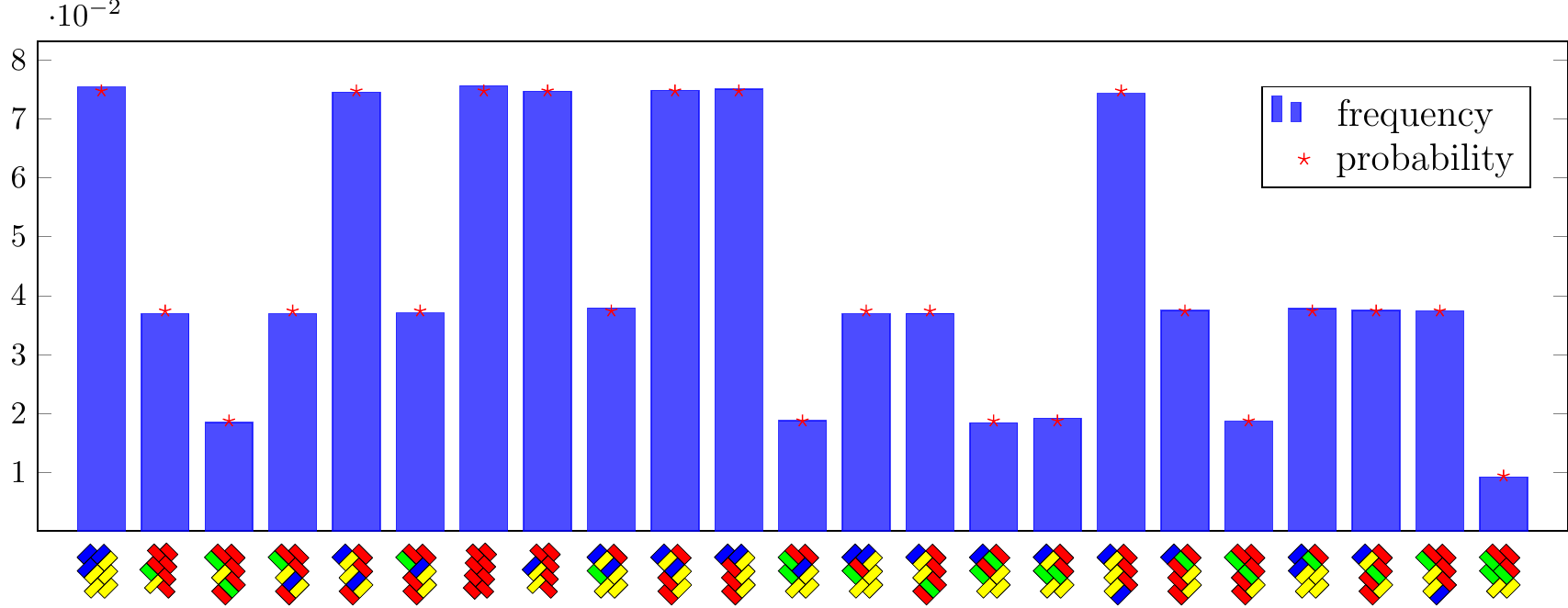}
\caption{The observed frequencies of the 23 possible configurations of the QTHADT
model for $n=3$ and $\gamma=1/2$,
compared with the corresponding theoretical probabilities. The frequencies are measured from the data of $230000$ configurations.}
\label{fig_verif_gamma_onehalf}
\end{figure}

\begin{figure}[h!]
\hspace*{-1cm}
\begin{tikzpicture}
\clip (-2.4,-2) rectangle (16.5,2.7);
\begin{scope}
\clip (-2,-2) rectangle (2,2);
\draw (0,0) node{\includegraphics[scale=0.2,angle=-45]{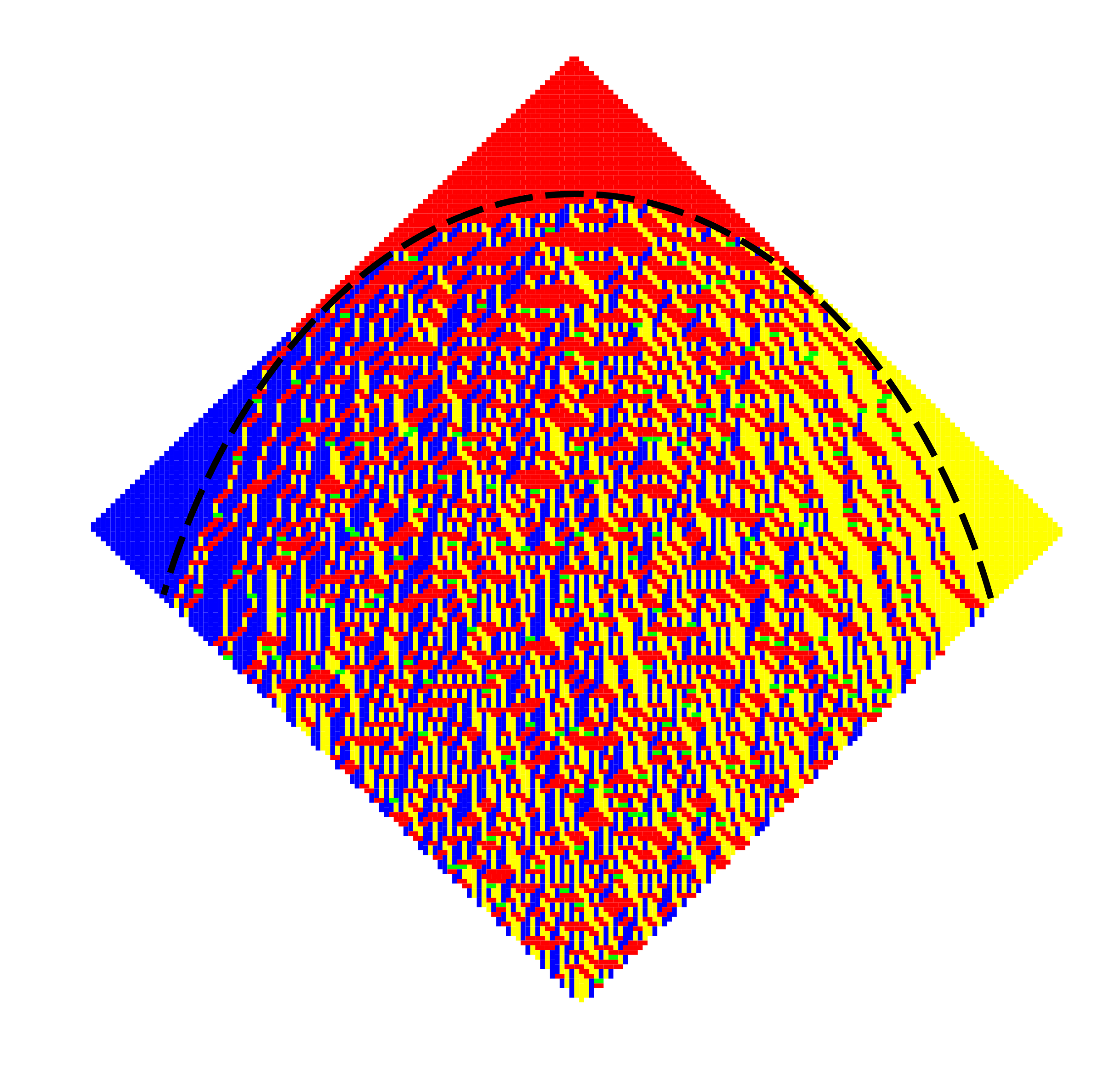}};
\end{scope}

\begin{scope}
\clip (-2+4.75,-2) rectangle (2+4.75,2);
\draw (4.75,0) node{\includegraphics[scale=0.2,angle=-45]{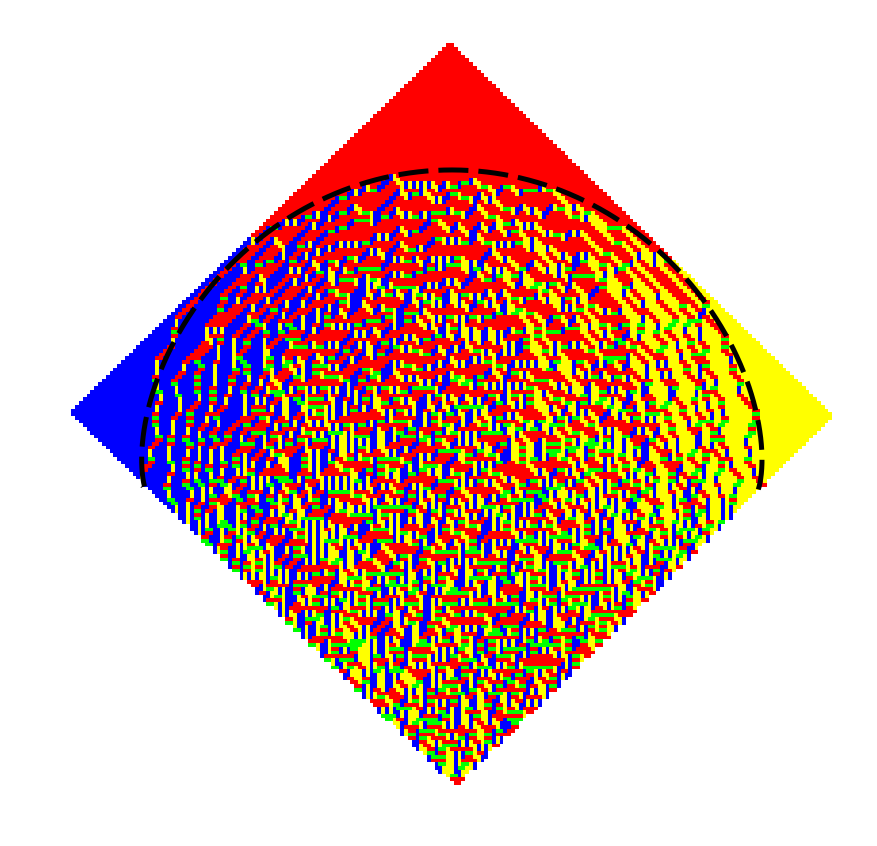}};
\end{scope}

\begin{scope}
\clip (-2+9.5,-2) rectangle (2+9.5,2);
\draw (9.5,0) node{\includegraphics[scale=0.2,angle=-45]{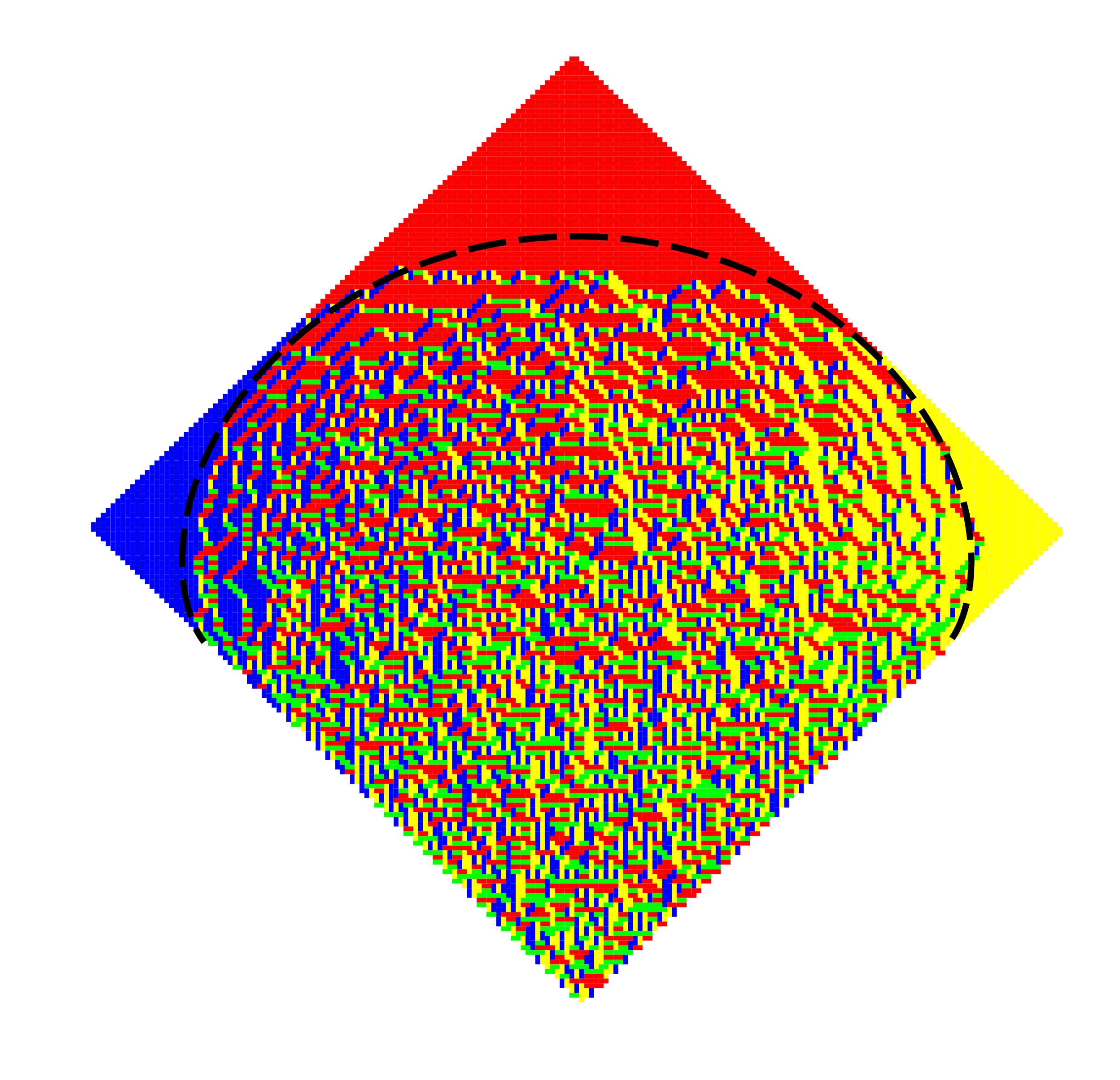}};
\end{scope}

\begin{scope}
\clip (-2+14.25,-2) rectangle (2+14.25,2);
\draw (14.25,0) node{\includegraphics[scale=0.2,angle=-45]{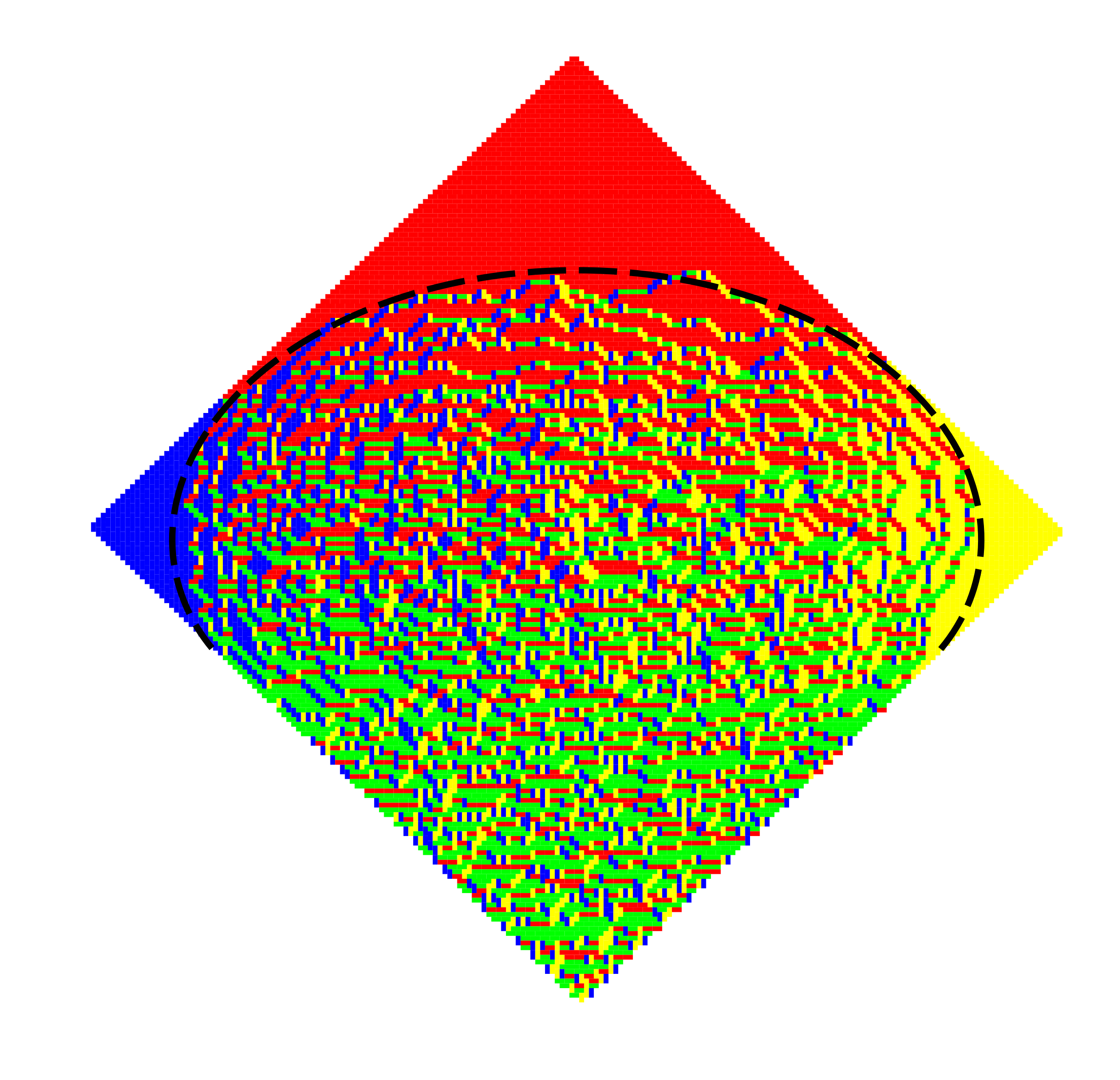}};
\end{scope}

\draw (0,2.3)  node{$\eta=\frac{\pi}{6.12}$};
\draw (4.75,2.3)  node{$\eta=\frac{\pi}{7}$};
\draw (9.5,2.3)  node{$\eta=\frac{\pi}{8}$};
\draw (14.25,2.3)  node{$\eta=\frac{\pi}{12}$};
\end{tikzpicture}
\caption{Typical configurations of the QTHADT 
for $n=100$ for several values of $\gamma=1+2\cos(4\eta)$. We show only the fundamental domain.}
\label{fig_n100_holey_eta}
\end{figure}
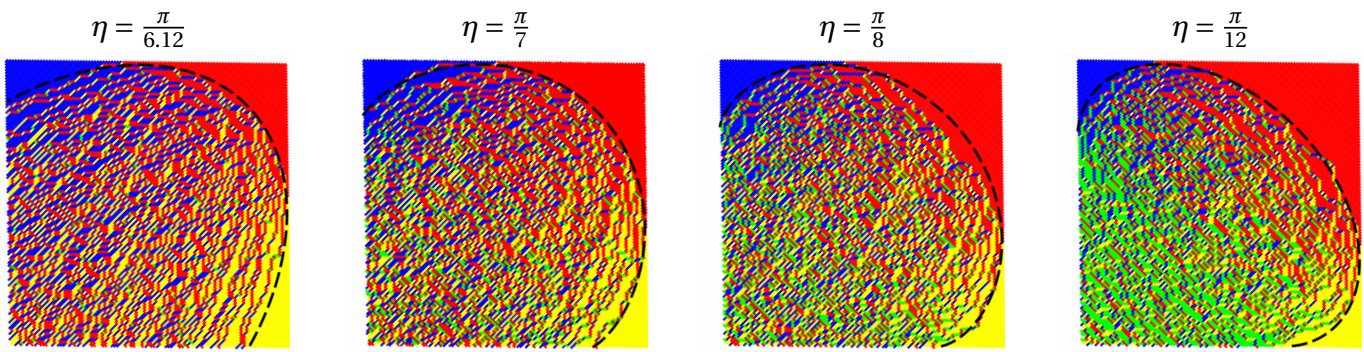

\begin{figure}
\begin{tikzpicture}
\clip (-2.1,-2) rectangle (16,4);
\begin{scope}
\clip (-2,-2) rectangle (2,2);
\draw (0,0) node{\includegraphics[scale=0.2,angle=-45]{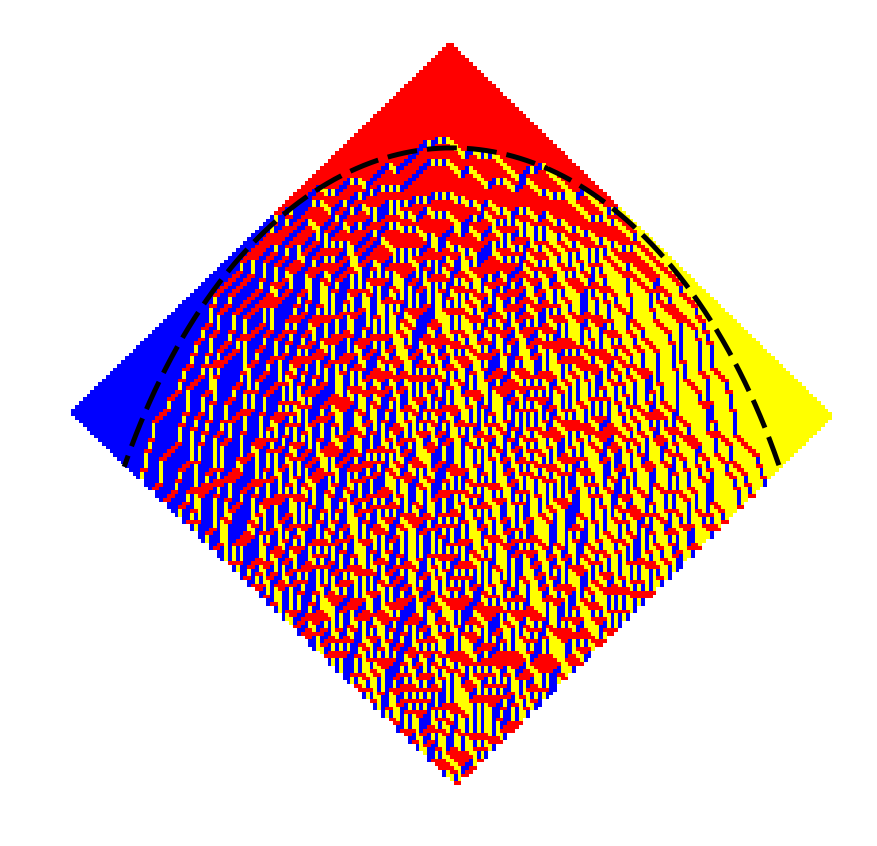}};
\end{scope}
\begin{scope}
\clip (-2+5.5,-2) rectangle (2.5+6,2.5);
\draw (6,0.2) node{\includegraphics[scale=0.23,angle=-45]{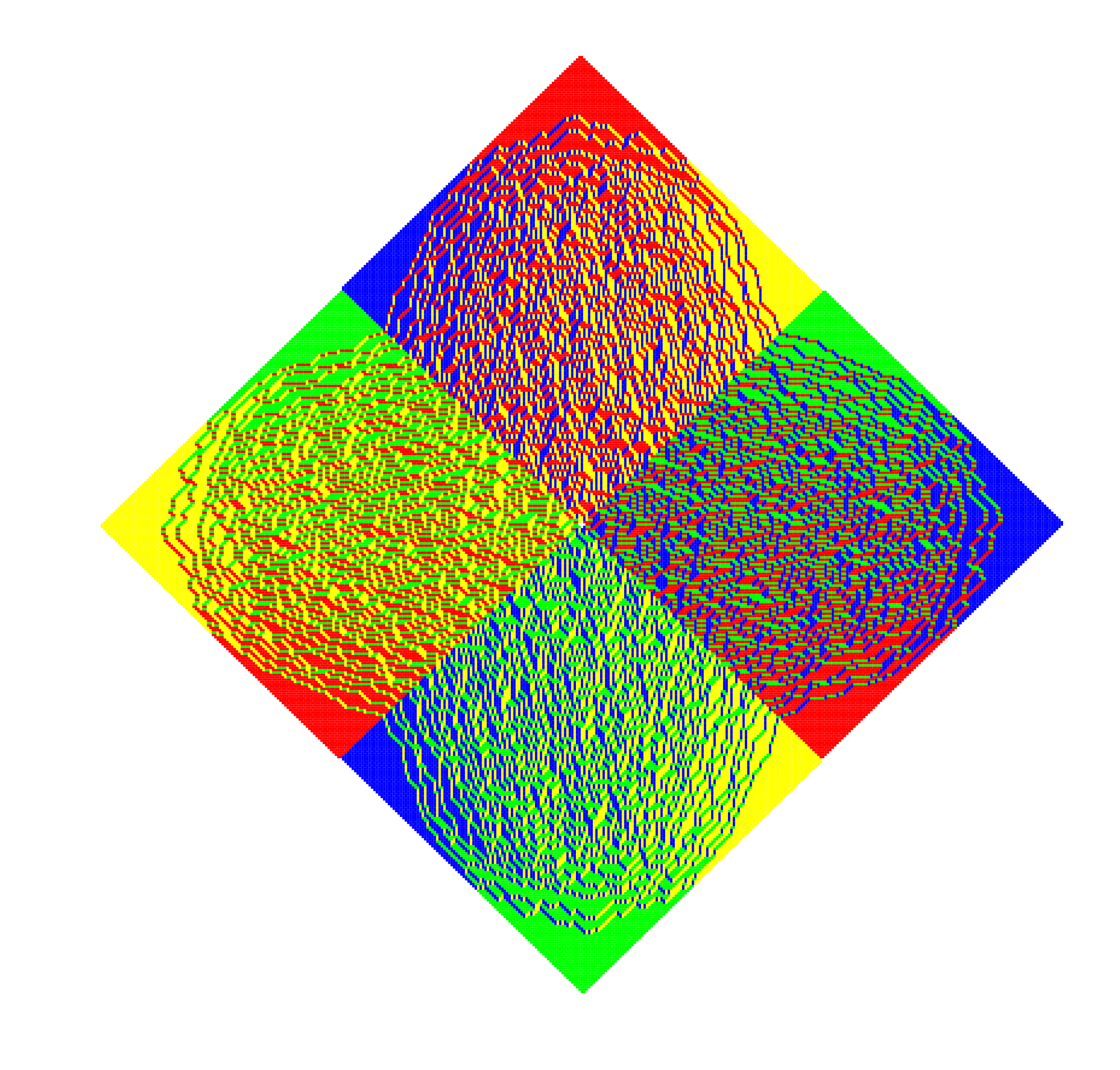}};
\end{scope}
\begin{scope}
\draw (12.5,0.1) node{\includegraphics[scale=0.16]{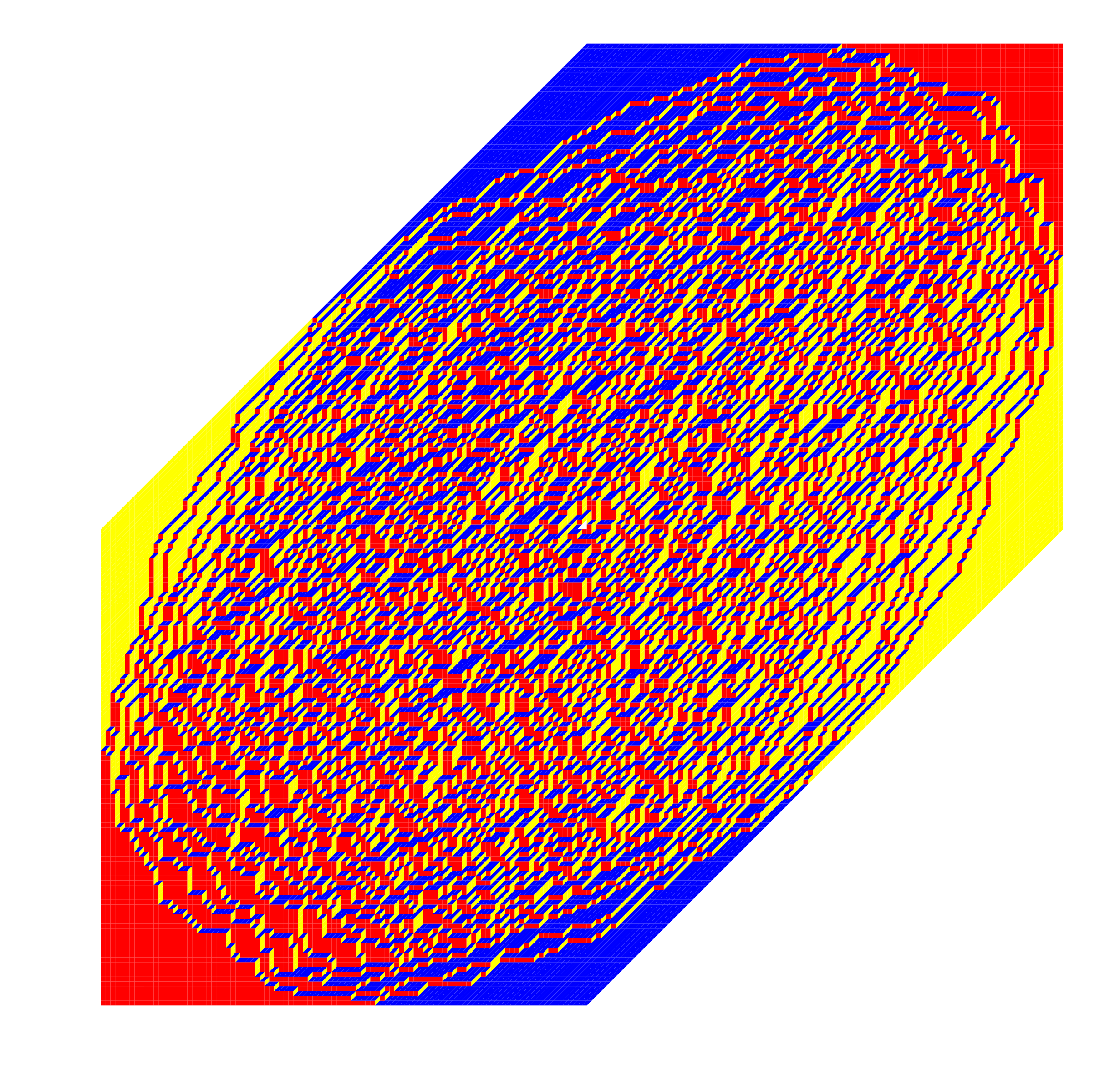}};
\end{scope}
\draw (0,3)  node[text width=5cm,align=center]{Domino tiling of the fundamental domain};
\draw (6.2,3)  node{Whole domain};
\draw (13,3)  node[text width=5cm,align=center]{Rhombus tiling of the Holey Hexagon};
\end{tikzpicture}
\caption{Typical configurations of the QTHADT 
for $n=100$ in the limit $\gamma\to 0$. In this limit, some elementary moves become extremely unlikely, and the convergence may be prohibitively long. To prevent this, one can gradually decrease $\gamma$ during the simulation until a small enough $\gamma$ is reached. The fundamental domain is represented on the left together with the predicted arctic curve. The corresponding symmetric configuration on the whole quasi-square shape domain is displayed in the center. For $\gamma=0$, the domino configurations are in bijection with another tiling problem, namely the cyclically symmetric tiling with rhombic tiles of a Holey hexagon of size $n$. The 3-fold symmetric image of the 4-fold symmetric domino tiling in the center under this
bijection is represented on the right. We note that, with the appropriate color scheme, the two pictures
look in practice very similar and quite indistinguishable in their fundamental domains (upper right square).}
\label{fig_n100_holey_etatozero}
\end{figure}

Like in Section~\ref{sec:simul}, typical random tilings are generated by a Markov process starting from a specific configuration and applying
ergodic moves.
The algorithm used to generate configurations of the QTHADT model with the desired distribution consists of three kinds 
of elementary moves involving pairs of connected dominoes in the fundamental domain (say the first quadrant). Let us first describe the algorithm that generates tilings with the uniform distribution ($\gamma=1$). In the bulk, as well as on the periodic boundaries of the fundamental domain, the elementary moves come in two flavors: \begin{tikzpicture} \draw[rotate=-45] (0,0) rectangle (0.15,0.3); \draw[rotate=-45] (0.15,0) rectangle (0.3,0.3);  \end{tikzpicture} $\rightarrow$ \begin{tikzpicture} \draw[rotate=-45] (0,0) rectangle (0.3,0.15); \draw[rotate=-45] (0,0.15) rectangle (0.3,0.3);  \end{tikzpicture} and \begin{tikzpicture} \draw[rotate=-45] (0,0) rectangle (0.3,0.15); \draw[rotate=-45] (0,0.15) rectangle (0.3,0.3);  \end{tikzpicture} $\rightarrow$ \begin{tikzpicture} \draw[rotate=-45] (0,0) rectangle (0.15,0.3); \draw[rotate=-45] (0.15,0) rectangle (0.3,0.3);  \end{tikzpicture}. 
An additional move is required to ensure ergodicity. Indeed, the two moves above can only deform the paths or change the position of the starting and ending points, but they cannot create or annihilate a path. Hence, by performing these moves only, we conserve the number $\ell$ of paths
so that we stay in a given ``sector'' of the possible tilings with fixed $\ell$. For example, the $\ell=0$ sector consists of a single configuration
with all dominoes in the fundamental domain oriented from Northwest to Southeast (see Figure \ref{fig_crossMove}-left). The third move, referred to as the ``cross-move'', which enables ergodicity, is best described in the complete domain as it involves the cross-shaped hole in its middle. If a tiling contains a cycle of eight dominoes around the hole, the cross-move consists in shifting all dominoes around the cycle by one square (see Figure \ref{fig_crossMove}-right). This has the effect of creating or annihilating a pair of starting and ending points. Note that, in the QTHADT geometry, the eight dominoes reduce, modulo
the quarter turn rotation, to a pair of connected dominoes in the fundamental domain, hence the cross-move is also a flip analogous to the two others. Every step of the Markov-chain goes as follows: 
select at random a position $(i,j)$ in the fundamental domain\footnote{We choose the $(i,j)'$ s on the square lattice on which the corners of the dominoes are.}. If the diamond whose upper vertex ($\bullet$) is at $(i,j)$ is entirely in the bulk, or on the periodic boundary, then perform a regular move \begin{tikzpicture} \clip (0,-0.25) rectangle (0.45,0.3); \draw[rotate=-45] (0,0) rectangle (0.15,0.3); \draw[rotate=-45] (0.15,0) rectangle (0.3,0.3) ; \draw[rotate=-45] (0,0.3) node{$\bullet$};  \end{tikzpicture} $\leftrightarrow$ \begin{tikzpicture}
\clip (0,-0.25) rectangle (0.45,0.3); \draw[rotate=-45] (0,0) rectangle (0.3,0.15); \draw[rotate=-45] (0,0.15) rectangle (0.3,0.3); \draw[rotate=-45] (0,0.3) node{$\bullet$};  \end{tikzpicture} if possible. If $(i,j)$ is adjacent to the cross, perform a cross move if possible. Repeat until the arctic curve has stabilized.

Again the probability  $p(\mathcal{C} \to \mathcal{C}')$ to go from a configuration $\mathcal{C}$ to a configuration $\mathcal{C}'$ is symmetric. Hence detailed balance condition and ergodicity ensure that the stationary distribution is uniform.

As in Section~\ref{sec:simul}, generalizing this Markov-chain to reach a non-uniform probability distribution $\pi(\mathcal{C})$ associated with a weight $\gamma \neq 1$ can be done via a Metropolis algorithm: we accept a move from $\mathcal{C}$ to  $\mathcal{C}'$ with probability $p(\mathcal{C} \to \mathcal{C}')=\min \left( 1,\frac{\pi(\mathcal{C}')}{\pi(\mathcal{C})} \right)$. Ergodicity and detailed balance are again satisfied and the stationary distribution is $\pi(\mathcal{C})$. We checked the validity of our implementation for $n=3$ and $\gamma=1/2$, see Figure \ref{fig_verif_gamma_onehalf}. Figures \ref{fig_n100_holey_eta} and \ref{fig_n100_holey_etatozero} 
display some tilings obtained by this algorithm for various values of $\eta$ as well as the corresponding arctic curve as given by \eqref{eq:ACholey} . 

\section{Discussion/Conclusion}
\label{sec:conclusion}

\subsection{The arctic curve of the 6V model from that of the 20V model}
The 6V model with DWBC may be realized as a particular instance of the 20V model with DWBC1 (resp. DWBC2) by setting $\omega_2=\omega_5=0$. 
Indeed, the condition $\omega_2=\omega_5=0$ forces the diagonal steps to be transmitted through each node so as to arrange into $n$ (resp. $n-1$) complete diagonal lines
below the second diagonal. These lines may be removed and the remaining paths, made
of horizontal and vertical steps only, form configurations of a 6V model with DWBC. The condition $\omega_2=\omega_5=0$
may be reached from the general parametrization \eqref{eq:explweights} by renormalizing the weights $\omega_i$ by an overall factor
into projectively equivalent weights $\omega_i'=4{\rm e}^{{\rm i}\mu}\omega_i$ and taking the limit $\mu\to+{\rm i}\infty$\footnote{Strictly speaking,
this value of $\mu$ exits the allowed domain \eqref{eq:admissible} for positive weights but this is corrected by the 
renormalization $\omega_i\to \omega_i'$.}. This limit corresponds to sending the spectral parameter $t\to 0$.
This results in renormalized 20V weights satisfying $\omega_2'=\omega_5'=0$, as desired, and
\begin{equation*}
\omega_0'=\omega_3=\sin(\lambda+\eta)\ ,\ 
\omega_1'= \sin(\lambda-\eta){\rm e}^{2{\rm i}\eta}\ , \ 
\omega_6'=\sin(\lambda-\eta){\rm e}^{-2{\rm i}\eta}\ , \
\omega_4'=\sin(2\eta)\ .
\end{equation*}
Here we recognize the parametrization $a=\sin(\lambda+\eta)$, $b=\sin(\lambda-\eta)$ and $c=\sin(2\eta)$ of the usual
6V model, apart from a phase factor in $\omega_1'$ and $\omega_6'$. After removing the diagonals of the 20V configurations,
which are all fixed by the boundary condition, all the nodes originally weighted by $\omega_0$ or $\omega_3$ lead to a-type vertices
of the 6V model and receive the correct weight $a$. Similarly, all the nodes originally weighted by $\omega_4$ lead to c-type vertices
of the 6V model and receive the correct weight $c$. The situation for nodes weighted by $\omega_1$ or $\omega_6$ is slightly
more subtle: those under the second diagonal (included for DWBC1, excluded for DWBC2) receiving a weight $\omega_1$ (resp. $\omega_6$) 
lead to 6V nodes of type b 
with two adjacent horizontal (resp. vertical) edges, while those above the second diagonal receiving a weight $\omega_1$ (resp. $\omega_6$)
 lead to 6V nodes of type b with two adjacent vertical (resp. horizontal) edges. These nodes receive a weight $\omega_1'={\rm e}^{2{\rm i}\eta}b$
 (resp. $\omega_6'={\rm e}^{-2{\rm i}\eta}b$). Fortunately, in all 6V DWBC configurations, we have the
 conservation law that the number of b-type nodes with horizontal, resp. vertical edges above any diagonal line parallel to the second diagonal 
 are identical\footnote{This can be seen for instance in the dual language of integer height variables \cite{BL14} at the center of the plaquettes: 
 the height along diagonals varies only at the crossing of a b-type vertex and increases/decreases by $2$ according to its vertical/horizontal nature.
 For DWBC, the difference of height between the two ends of each diagonal line is zero, hence the two types of b-type vertices
 are equinumerous along each diagonal, hence also above each diagonal.}.
 This allows to replace $\omega_1'$ and $\omega_6'$ by $\kappa \omega_1'$ and $\kappa^{-1} \omega_6'$ for any non-zero $\kappa$,
 hence, by choosing $\kappa={\rm e}^{-2{\rm i}\eta}$, to assign the correct weight $b$ to all these nodes. 
 
 As for the arctic curve of the 6V model with DWBC, it is obtained directly from that of the 20V by applying the same $\mu\to+{\rm i}\infty$ 
 limit in the explicit expression of Theorem~\ref{thm:acgeneral}. In practice, only the ``normal'' and ``shear'' portions (and their 
 $180^\circ$ rotation images) are necessary, while the ``final'' portion becomes redundant. More precisely, we get 
\begin{thm} 
\label{thm:ac6Vl}
The arctic curve for the 6V model with DWBC at arbitrary admissible values of the parameters $\eta$ and $\lambda$ ($0<\eta<\lambda<\pi-\eta$) is
made generically of two portions, denoted ``normal'' and ``shear'' with their images under $180^\circ$ rotation.
The two branches have respectively parametric equations:
\begin{equation*}
\begin{split}
\hbox{\rm Normal:}&\qquad x_n(\xi)=1+\frac{\partial_\xi {R}_n(\xi)}{\partial_\xi {S}_n(\xi)}\ ,\qquad y_n(\xi)={R}_n(\xi)- {S}_n(\xi) \frac{\partial_\xi {R}_n(\xi)}{\partial_\xi {S}_n(\xi)}\ , \qquad \xi \in [0,\pi-\lambda-\eta]\\
\hbox{\rm Shear:}&\qquad x_s(\xi)=1+\frac{\partial_\xi {R}_s(\xi)}{\partial_\xi {S}_s(\xi)}\ ,\qquad y_s(\xi)={R}_s(\xi)- {S}_s(\xi) \frac{\partial_\xi {R}_s(\xi)}{\partial_\xi {S}_s(\xi)}\ , \qquad \xi \in [-(\lambda-\eta),0]\\
\end{split}
\end{equation*}
where
\begin{equation*}
\begin{split}
{R}_n(\xi)&={R}_s(\xi)=
\left(\cot(\xi\!+\!\lambda\!-\!\eta)\!-\!\cot(\xi)\!+\!\alpha\cot(\alpha\, \xi)
\!-\!\alpha\cot(\alpha(\xi\!+\!\lambda\!-\!\eta))\right)
\frac{\sin(\xi\!+\!\lambda\!+\!\eta)\sin(\xi\!+\!\lambda\!-\!\eta)}{\sin(2\eta)}\ ,
\\
{S}_n(\xi)&=\frac{\sin(\xi+\lambda+\eta)\sin(\xi+\lambda-\eta)}{\sin(\xi)\sin(\xi+2\eta)}\ ,\quad 
{S}_s(\xi)=\frac{\sin(\xi+\lambda+\eta)\sin(\xi+\lambda-\eta)}{\sin(2\eta-\xi)\sin(\xi))}
\ , \\
\end{split}
\end{equation*}
and $\alpha=\pi/(\pi-2\eta)$.
\end{thm}
This matches the known expressions of \cite{CP2009} for the 6V model with DWBC in its disordered phase.

\subsection{Uniform 20V vs QTHADT and the ASM-DPP correspondence}
\label{sec:ASMDPP}
\begin{figure}
\begin{center}
\includegraphics[width=12cm]{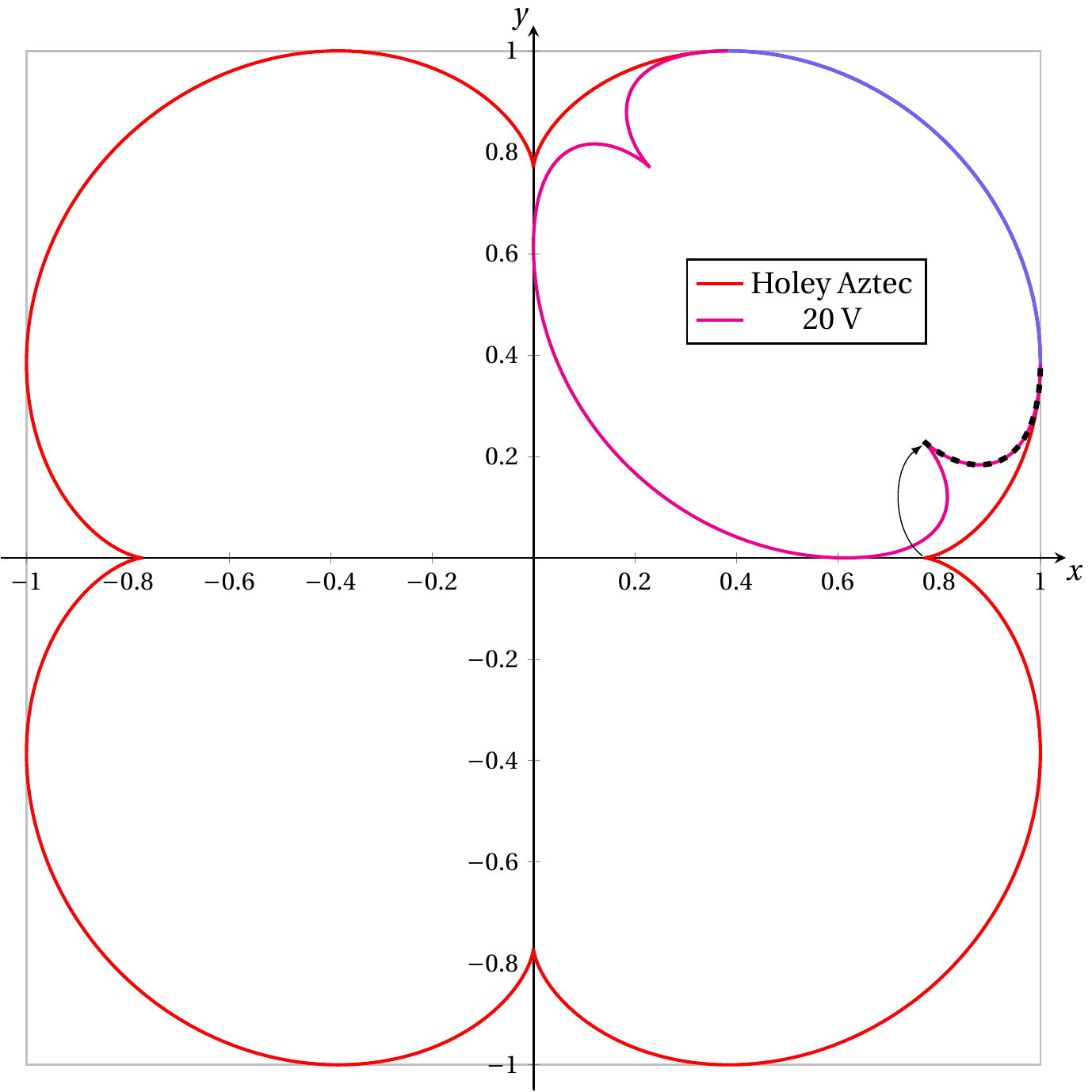} 
\end{center}
\caption{\small The arctic curve of the uniform 20V model with DWBC (in purple) and that of the QTHADT (Holey Aztec) model
with $\gamma=1$ (in red) share a common portion (in violet). For the QTHADT model, the remaining part 
is the analytic continuation of this shared portion, namely the algebraic curve \eqref{eq:ACuniform}. When applied to the correct portion (that joining the tangency point on the right 
boundary to the cusp), the image (dashed) of this analytic continuation
by the shear transformation $y\to 1-x+y$ reproduces the ``shear'' portion of the arctic curve of  the 20V model.}
\label{fig:plot7}
\end{figure}
\begin{figure}
\begin{center}
\includegraphics[width=12cm]{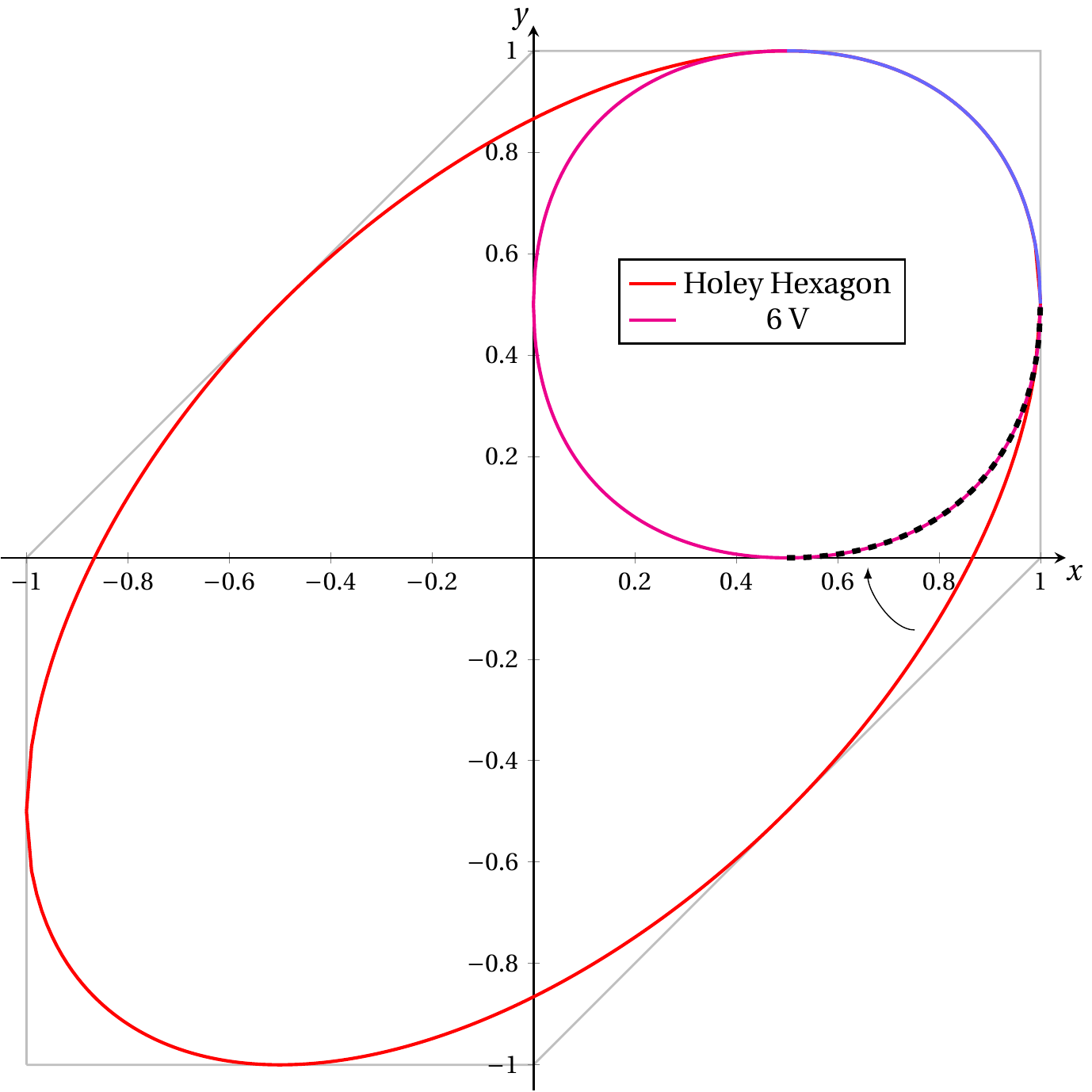} 
\end{center}
\caption{\small The arctic curve of the uniform 6V model with DWBC (in purple) and that of the Holey Hexagon rhombus tiling
(in red) share a common portion (in violet). For the Holey Hexagon rhombus tiling, the remaining part 
is the analytic continuation of this shared portion, forming an ellipse. When applied to the correct portion (that joining the tangency point $(1,1/2)$ on the right 
vertical boundary to the tangency point $(1/2,-1/2)$ on the lower diagonal boundary), the image 
(dashed) of this analytic continuation
by the shear transformation $y\to 1-x+y$ reproduces a portion of the arctic curve of the 6V model. This pattern of correspondences is
in all ways identical to that of Figure~\ref{fig:plot7}.}
\label{fig:plot8}
\end{figure}
Let us discuss a few remarks and open questions.
The first remark concerns the relation between the arctic curve of the 20V model with uniform weights (i.e.\ $\eta=\pi/8$, $\lambda=5\pi/8$ and $\mu=0$
so that all the $\omega_i$ for $i=0,\dots,6$ are equal) and that of the QTHADT model with $\gamma=1$ (i.e.\ $\eta=\pi/8$).
Figure~\ref{fig:plot7} displays the two corresponding arctic curves, as obtained from our expressions above. First, we note that the two curves share a common portion, corresponding to what we called the ``normal'' portion in the 20V model. This property is a direct consequence of the 
refined bijection proved in \cite{DFG20V} (see Theorem 5.2) between the uniform 20V model configurations having, in the path language, their uppermost path hitting the right boundary at height $\ell$ (or equivalently, via the $x\leftrightarrow y$ symmetry, the 
configurations whose uppermost path leaves the upper boundary after $\ell$ steps) and the QTHADT configurations having, 
in the Schr\"oder path language, a path starting at position $(0,n)$ and leaving the upper boundary of the fundamental domain after $\ell$ steps. 
We then note that, for the QTHADT problem at $\gamma=1$, \emph{the entire arctic curve is the analytic continuation of the ``normal'' portion},
i.e.\ it is obtained by extending the original range of the parameter $\xi$ in \eqref{eq:ACholey} from $[0,\pi/4]$ (''normal'' portion) to $[-\pi/4,\pi/2]$ (arctic curve in the fundamental domain), then to $[-11\pi/8,13\pi/8]$, leading to the desired full fourfold symmetric algebraic curve 
\eqref{eq:ACuniform} with its four cusps. Finally, as already discussed in 
Section~\ref{sec:shearbranch}, the ``shear'' portion of the arctic curve of the 20V model is related to this analytic continuation by a simple shear transformation sending the $y=0$ line onto the line $x+y=1$. More precisely, the `shear'' portion of the arctic curve of the 20V model
is itself the image under the shear transformation $y\to 1-x+y$ of the portion of arctic curve of the QTHADT problem between the tangency
point on the right boundary $x=1$ and the cusp at $y=0$ in the fundamental domain ($\xi\in[\pi/4,\pi/2]$). All the other portions of the 20V model arctic curve are obtained by symmetry arguments.

Remarkably, we find exactly the same pattern of correspondences if we compare the arctic curve for cyclically symmetric rhombus tilings of a Holey Hexagon, in bijection with descending plane partitions (DPP) \cite{KrattDPP}
to that of the uniform 6V model (with weights $a=b=c$) with DWBC, in bijection with Alternating Sign Matrices (ASM). The ASM-DPP correspondence 
was proved with its highest level of refinement in \cite{BDFPZ1,BDFPZ2}. In the Holey Hexagon model, the tiled domain is now formed of a fundamental domain with a rhombic shape drawn
on the triangular lattice, and two extra copies of this domain obtained by two $120^\circ$ rotations around a central triangular hole of size $2\times 2\times 2$ so as to 
form a quasi-regular hexagon of shape $n\times (n+2)\times n\times (n+2)\times n\times (n+2)$. The tiles are elementary rhombi covering two adjacent triangles and we demand that the tiling configurations be symmetric under $120^\circ$ rotation. If we redress the fundamental domain into a square, the tiling problem has a Schr\"oder path formulation which corresponds 
precisely to our setting but with paths \emph{without diagonal steps}, i.e.\ to the case $\gamma=0$ ($\eta=\pi/6$). In this redressed geometry,
the arctic curve of the cyclically symmetric Holey Hexagon rhombus tiling is thus obtained via \eqref{eq:ACholey}  with $\eta=\pi/6$, upon taking
$\xi$ in the range $\xi\in[0,\pi/3]$ (``normal'' portion), extended to $\xi \in [-\pi/6,\pi/2]$ (arctic curve in the fundamental domain), then to 
 $\xi \in [-5\pi/6,7\pi/6]$, leading to the complete ellipse $x^2+y^2-x y=3/4$ \cite{CP2010} (see Figure \ref{fig:plot8}). As for the arctic curve of the uniform 6V model, it is made of the very same ``normal'' portion\footnote{The fact that the arctic curve of the Holey Hexagon model and
 that of the uniform 6V model share a common portion is a direct consequence of the refined ASM-DPP correspondence shown in \cite{BDFPZ1}.}, together with three symmetric portions obtained by successive $90^\circ$
 rotations around the center of the fundamental domain. Again, as displayed in Figure \ref{fig:plot8}, the first of these extra portions (that following the
  ``normal'' portion clockwise) 
is the image by the shear transformation $y\to 1-x+y$ of the proper portion of arctic curve of the Holey Hexagon problem, that between
the tangency point $(1,1/2)$ on the right boundary and the tangency point $(1/2,-1/2)$ on the lower diagonal boundary ($\xi\in[\pi/3,2\pi/3]$). All these correspondences follow the same global scheme as that of Figure \ref{fig:plot7} for the 20V/QTHADT relation. These may be occurrences of a more general phenomenon for correspondences between osculating vs non-intersecting path problems, yet to be investigated. 
  
 \subsection{Extension to more general weights}
 Clearly the parametrization \eqref{eq:explweights} for the weights of the 20V model covers only a small subset of the allowed values.
In particular, the restriction to real values of $\eta$ confines the associated 6V model into its so-called disordered phase. 
The result \eqref{eq:fsigma} of \cite{CP2009} for the asymptotics of the one-point function $H^{6V}_n(\sigma)$ was generalized in \cite{CPZ}
for the 6V model with DWBC in its antiferroelectric regime. It should therefore be possible to extend our results to this regime, 
namely, in the parametrization \eqref{eq:explweights}, to the case of imaginary values of $\eta$, $\lambda$ and $\mu$. 
The expressions of \cite{CPZ} involve elliptic functions and this extension might in practice lead to quite involved calculations.

Another question concerns the symmetry of the weights under reversal of the edge orientations. This symmetry was imposed
for convenience throughout the paper and guarantees that the arctic curve of the 20V model is symmetric under $180^\circ$ rotation.
In the case of the 6V model with DWBC, it is easily shown that there are enough sum rules for the numbers of the different types of vertices to
ensure that 
the symmetry of weights under edge orientation reversal can be assumed without loss of generality (see for instance \cite{COSPO}). This is no longer the
case for the 20V model and non symmetric weights may lead to more general arctic curves without the $180^\circ$ rotation symmetry,
a situation yet to be explored.

Another direction of exploration concerns other boundary conditions. In \cite{DFG20V}, another type of boundary conditions for the 20V model, called DWBC3, was introduced and shown to display nice combinatorics. These boundary conditions are expected to give rise to a potentially 
simpler arctic phenomenon with an arctic curve made of a single portion separating the liquid phase from the empty region. 
To derive such a curve, it would be desirable to have more explicit expressions for the model partition function in terms of spectral parameters,
giving access to boundary one-point functions. 
  
Finally, a number of non-intersecting path problems were solved in a more general 
``$Q$-deformed'' framework, which consists in introducing an extra path weight $Q^{\mathcal A}$ involving the area 
$\mathcal{A}$ below the path. In particular, the tangent method was applied successfully to some of these models
to obtain the corresponding $Q$-deformed arctic curve and we may wonder whether this generalization can be carried out in 
our present setting for the QTHADT model. Such a $Q$-deformation was carried out for DPP but not for ASM, hence 
the `$Q$-deformed'' version of the 20V model with DWBC1 or 2 remains a challenge like that of the 6V model with DWBC.

\bigskip
\appendix
\section{The relation between the 6V and 20V refined partition functions in all generality} 
\label{sec:6V20Vgeneralrel}
The aim of this Appendix is to prove the general relations \eqref{eq:genrel}-\eqref{eq:taugofsigma} and \eqref{eq:genrelbis}-\eqref{eq:taugofsigmabis} relating the 
restricted refined partition functions
\begin{equation*}
 Z^{20V_{BC2}\,  \mbox{--}}_n(\tau)=\sum_{\Ell=1}^{n} Z^{20V_{BC2}\, \mbox{--}}_{n;\Ell}\, \tau^{\Ell-1}\ , \qquad
Z^{20V_{BC2}\, {\scriptscriptstyle{\diagdown}}}_n(\tau)=\sum_{\Ell=1}^{n} Z^{20V_{BC2}\, {\scriptscriptstyle{\diagdown}}}_{n;\Ell}\, \tau^{\Ell-1}
\end{equation*}
or their tilde counterparts to the refined partition function 
\begin{equation*}
Z^{6V}_n(\sigma)=\sum_{\Ell=1}^{n} Z^{6V}_{n;\Ell}\, \sigma^{\Ell-1}\ .
\end{equation*}

\begin{figure}
\begin{center}
\includegraphics[width=16cm]{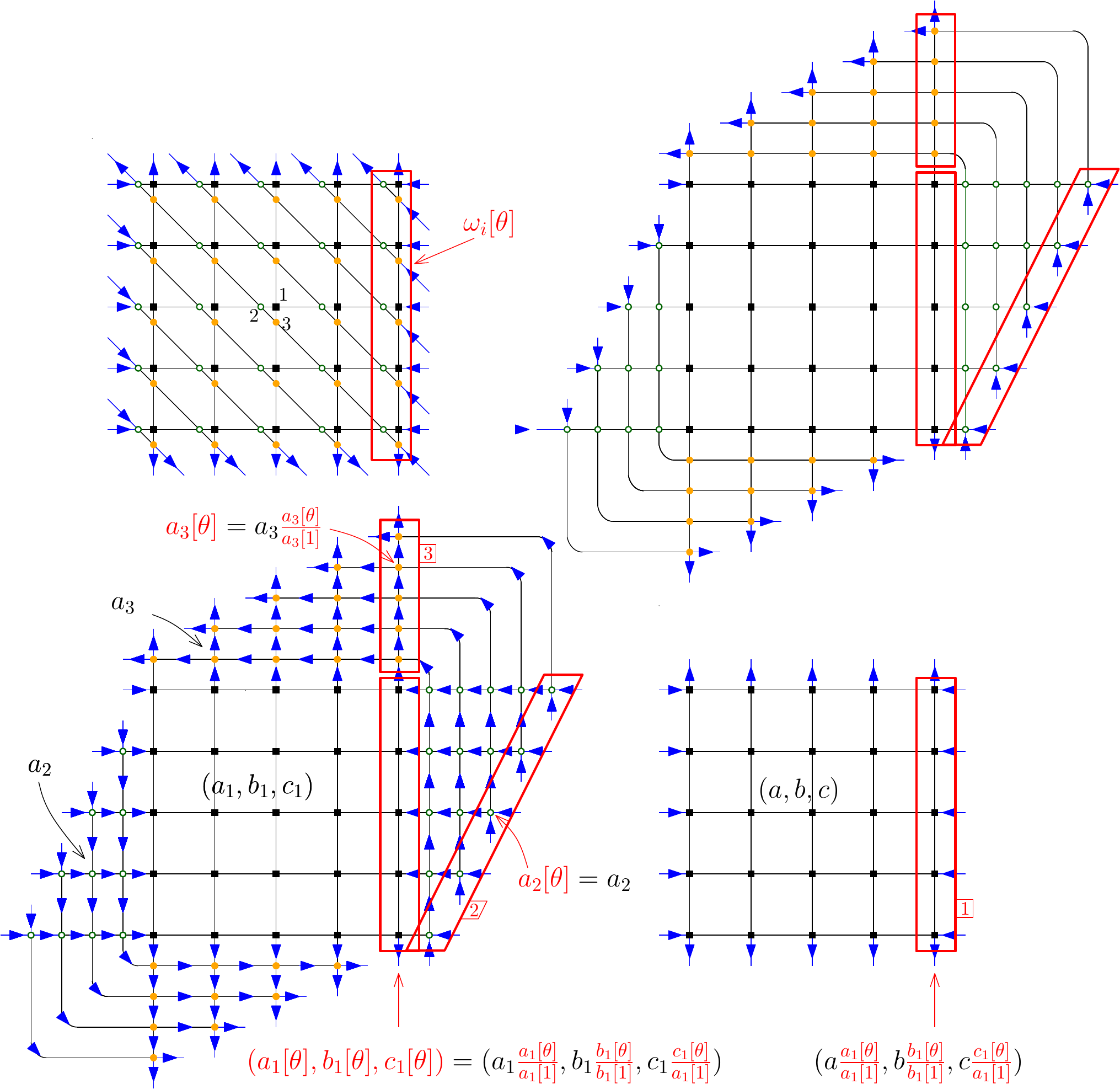}
\end{center}
\caption{\small The unraveling of Figure~\ref{fig:unraveling} in the presence of a spectral parameter $w\theta$
in the last column that modifies the weights for the nodes within the red boxes. In practice, only the weights of sub-lattices $1$ and $3$ are modified
(since those of sub-lattice $2$ do not involve the modified spectral parameter). We thus have a modified weight $a_3[\theta]=a_3\times \frac{a_3[\theta]}{a_3[1]}$
for the $n$ nodes in the upper red box labelled $3$, unmodified weight $a_2[\theta]=a_2$ for the nodes in the red box labeled $2$, and
after the same renormalization as in Figure~\ref{fig:unraveling} to go from $(a_1,b_1,c_1)$ to $(a,b,c)$, modified weights $(a\times \frac{a_1[\theta]}{a_1[1]},
b\times \frac{b_1[\theta]}{b_1[1]},c\times \frac{c_1[\theta]}{c_1[1]})$  in the red box labeled $1$. This leads to the relation \eqref{eq:relwiththeta}.}
\label{fig:unraveling2}
\end{figure}

The relation \eqref{eq:genrel} may be obtained by attaching a different spectral parameter $w\theta$ (instead of $w$) to the last column
$j=n$.
By unravelling the configurations of the 20V model with DWBC2 to a configuration of the 6V model on the sub-lattice $1$,
we have the relation, depicted in Figure~\ref{fig:unraveling2}:
\begin{equation}
Z^{20V_{BC2}}_n[\theta]=\left(\frac{a_2a_3}{t^{1/3}}\right)^{n^2}\ \left(\frac{a_3[\theta]}{a_3[1]}\right)^n\ Z^{6V}_n[\theta]\ .
\label{eq:relwiththeta}
\end{equation}
Here we shall use a different notation with \emph{brackets} (as in $Z^{6V}_n[\theta]$) to indicate that we deal with a model where the 
last column has a modified spectral parameter $w\theta$. It should not be confused with the notation with parentheses (as in $Z^{6V}_n(\sigma)$)
where all the column spectral parameters are left equal to $w$ but where we deal with the refined partition function defined above keeping track of the position
where the uppermost path hits the right boundary. The notation $a_i[\theta]$ (respectively $b_i[\theta]$ and $c_i[\theta]$), $i=1,2,3$, also 
refers to the weights\footnote{As before, when dealing with the 6V model, we use vertex weights normalized to $(a,b,c)$
as in \eqref{eq:abc}. The change $w\to w\theta$ then affects the vertex weights in the last column, equal to
$(a\times \frac{a_1[\theta]}{a_1[1]},
b\times \frac{b_1[\theta]}{b_1[1]},c\times \frac{c_1[\theta]}{c_1[1]})$.}
obtained via the general formula \eqref{eq:intweights} with $w\to w\theta$. Let us now discuss precisely the connection between the two objects
$Z^{6V}_n[\theta]$ and $Z^{6V}_n(\sigma)$:
by decomposing $Z^{6V}_{n}[\theta]$ according to the position $\Ell$ where the uppermost path hits the right boundary,
we have
\begin{equation*}
\begin{split}
Z^{6V}_n[\theta]&=\sum_{\Ell=1}^n Z^{6V}_{n;\Ell}\left(\frac{b_1[\theta]}{b_1[1]}\right)^{\Ell-1}\  \frac{c_1[\theta]}{c_1[1]}\left(\frac{a_1[\theta)]}{a_1[1]}\right)^{n-\Ell}\\
&=\left(\frac{a_1[\theta]}{a_1[1]}\right)^n \frac{c_1[\theta]a_1[1]}{c_1[1]a_1[\theta]}\ \underbrace{\sum_{\Ell=1}^n Z^{6V}_{n;\Ell} (\sigma[\theta])^{\Ell-1}}_{=Z^{6V}_n(\sigma[\theta])}\\
\end{split}
\end{equation*}
with 
\begin{equation*}
\sigma[\theta]=\frac{b_1[\theta]a_1[1]}{b_1[1]a_1[\theta]}\ .
\end{equation*}
Similarly, decomposing $Z^{20V_{BC2}}_{n}[\theta]$ according to the position $\Ell$ where the uppermost path hits the right boundary and to whether
the step before the hitting point was horizontal or diagonal,
we have
\begin{equation*}
\begin{split}
Z^{20V_{BC2}}_n[\theta]&=\sum_{\Ell=1}^n Z^{20V_{BC2}\, \mbox{--}}_{n;\Ell} \left(\frac{\omega_1[\theta]}{\omega_1[1]}\right)^{\Ell-1}\  \frac{\omega_4[\theta]}{\omega_4[1]}
\left(\frac{\omega_0[\theta]}{\omega_0[1]}\right)^{n-\Ell}
+ \sum_{\Ell=1}^n Z^{20V_{BC2}\,  {\scriptscriptstyle{\diagdown}}}_{n;\Ell} \left(\frac{\omega_1[\theta]}{\omega_1[1]}\right)^{\Ell-1}\  \frac{\omega_2[\theta]}{\omega_2[1]}
\left(\frac{\omega_0[\theta]}{\omega_0[1]}\right)^{n-\Ell}\\
&=\left(\frac{\omega_0[\theta]}{\omega_0[1]}\right)^n \Bigg( \frac{\omega_4[\theta]\omega_0[1]}{\omega_4[1]\omega_0[\theta]}\ \underbrace{\sum_{\Ell=1}^n Z^{20V_{BC2}
\, \mbox{--}}_{n;\Ell} (\tau[\theta])^{\Ell-1}}_{=Z^{20V_{BC2}\,\mbox{--}}_n(\tau[\theta])}+ \frac{\omega_2[\theta]\omega_0[1]}{\omega_2[1]\omega_0[\theta]}\ \underbrace{\sum_{\Ell=1}^n Z^{20V_{BC2}\,  {\scriptscriptstyle{\diagdown}}}_{n;\Ell} (\tau[\theta])^{\Ell-1}}_{=
Z^{20V_{BC2}{\scriptscriptstyle{\diagdown}}}_n(\tau[\theta])}\Bigg)\\
\end{split}
\end{equation*}
with
\begin{equation*}
\tau[\theta]=\frac{\omega_1[\theta]\omega_0[1]}{\omega_1[1]\omega_0[\theta]}=\frac{(b_1[\theta]a_2[\theta]b_3[\theta])(a_1[1]a_2[1]a_3[1])}{(b_1[1]a_2[1]b_3[1])(a_1[\theta]a_2[\theta]a_3[\theta])}
=\frac{b_1[\theta]b_3[\theta]a_1[1]a_3[1]}{b_1[1]b_3[1]a_1[\theta]a_3[\theta]}\ .
\end{equation*}
Using
\begin{equation*}
\frac{\omega_0[\theta]}{\omega_0[1]}=\frac{a_1[\theta]a_2[\theta]a_3[\theta]}{a_1[1]a_2[1]a_3[1]}=\frac{a_1[\theta]a_3[\theta]}{a_1[1]a_3[1]}
\end{equation*}
since $a_2$ is not changed by the replacement $w\to w\theta$, we obtain from \eqref{eq:relwiththeta}
\begin{equation}
\left(\frac{a_2a_3}{t^{1/3}}\right)^{n^2}\ Z^{6V}_n(\sigma[\theta])=Z^{20V_{BC2}\,\mbox{--}}_n(\tau[\theta]) +g(\sigma[\theta])Z^{20V_{BC2}{\scriptscriptstyle{\diagdown}}}_n(\tau[\theta])
\label{eq:Therel}
\end{equation}
with 
\begin{equation*}
\begin{split}
g(\sigma[\theta])&=\frac{c_1[1]a_1[\theta]}{c_1[\theta]a_1[1]} \frac{\omega_2[\theta]\omega_0[1]}{\omega_2[1]\omega_0[\theta]}=
\frac{c_1[1]a_1[\theta]}{c_1[\theta]a_1[1]} \frac{(b_1[\theta]a_2[\theta]c_3[\theta])(a_1[1]a_2[1]a_3[1])}{(b_1[1]a_2[1]c_3[1])(a_1[\theta]a_2[\theta]a_3[\theta])}\\
& =
\frac{c_1[1]b_1[\theta]c_3[\theta]a_3[1]}{c_1[\theta]b_1[1]c_3[1]a_3[\theta]}\ .\\
\end{split}
\end{equation*}
Note the absence of prefactor in front of $Z^{20V_{BC2}\,\mbox{--}}_n(\tau[\theta])$ in \eqref{eq:Therel}, due to the identity
\begin{equation*}
\frac{c_1[1]a_1[\theta]}{c_1[\theta]a_1[1]} \frac{\omega_4[\theta]\omega_0[1]}{\omega_4[1]\omega_0[\theta]}=
\frac{c_1[1]a_1[\theta]}{c_1[\theta]a_1[1]} \frac{(c_1[\theta]a_2[\theta]a_3[\theta])(a_1[1]a_2[1]a_3[1])}{(c_1[1]a_2[1]a_3[1])(a_1[\theta]a_2[\theta]a_3[\theta])}=1\ .
\end{equation*}
In particular, if we wish to impose a strict proportionality relation between $Z^{20V_{BC2}}_n(\tau[\theta])$ and $Z^{6V}_n(\sigma[\theta])$, i.e. impose $g(\sigma[\theta])=1$,
we must impose $\frac{\omega_2[\theta]}{\omega_2[1]}=\frac{\omega_4[\theta]}{\omega_4[1]}$ for all $\theta$, i.e. $\mu=\lambda-5\eta$.
\medskip

The relation between $\tau$ and $\sigma$ and that between $g$ and $\sigma$ are obtained by eliminating $\theta$. We obtain
\begin{equation}
\begin{split} 
\tau&=\sigma\ 
\ \frac{\sigma\, \sin(\lambda-\eta)\sin\left(\frac{\lambda+3\eta-\mu}{2}\right)-\sin(\lambda+\eta)\sin\left(\frac{\lambda-\eta-\mu}{2}\right)
}{\sigma\,  \sin(\lambda-\eta) \sin\left(\frac{\lambda-\eta-\mu}{2}\right)-\sin(\lambda+\eta)\sin\left(\frac{\lambda-5\eta-\mu}{2}\right)} \times  \frac{\sin\left(\frac{\lambda+3\eta+\mu}{2}\right)}{\sin\left(\frac{\lambda-\eta+\mu}{2}\right)}\\
g(\sigma)&=\frac{ \sigma \sin(2\eta)\sin\left(\frac{\lambda+3\eta+\mu}{2}\right)}{\sigma\,  \sin(\lambda-\eta) \sin\left(\frac{\lambda-\eta-\mu}{2}\right)-\sin(\lambda+\eta)\sin\left(\frac{\lambda-5\eta-\mu}{2}\right)}\\
\end{split}
\label{eq:taug}
\end{equation}
which, with \eqref{eq:Therel}, is nothing but \eqref{eq:genrel}-\eqref{eq:taugofsigma}.

Using the original parametrization
\begin{equation*}
\sigma(\xi)=\frac{\sin(\lambda+\eta)\sin(\lambda-\eta+\xi)}{\sin(\lambda-\eta)\sin(\lambda+\eta+\xi)}\ ,
\end{equation*}
we get
\begin{equation}
\begin{split}
\tau(\xi)&= \frac{\sin(\lambda+\eta)\sin\left(\frac{\lambda+3\eta+\mu}{2}\right)\sin(\xi+\lambda-\eta)\sin\left(\xi+\frac{\lambda-\eta+\mu}{2}\right)}{\sin(\lambda-\eta)\sin\left(\frac{\lambda-\eta+\mu}{2}\right)\sin(\xi+\lambda+\eta)\sin\left(\xi+\frac{\lambda+3\eta+\mu}{2}\right)} \\
g(\sigma(\xi))&=\frac{\sin(\xi+\lambda-\eta)\sin\left(\frac{\lambda+3\eta+\mu}{2}\right)}{\sin(\lambda-\eta)\sin\left(\xi+\frac{\lambda+3\eta+\mu}{2}\right)}\ .\\
\end{split}
\label{eq:taugxi}
\end{equation}
\medskip
By attaching a different spectral parameter $z\tilde{\theta}$ (instead of $z$) to the top line, we get by a similar argument the
relation 
\begin{equation*}
\left(\frac{a_2a_3}{t^{1/3}}\right)^{n^2}\ Z^{6V}_n(\tilde{\sigma}[\tilde{\theta}])=\tilde{Z}^{20V_{BC2}\,\vert}_n(\tilde{\tau}[\tilde{\theta}]) +\tilde{g}(\tilde{\sigma}[\tilde{\theta}])\tilde{Z}^{20V_{BC2}{\scriptscriptstyle{\diagdown}}}_n(\tilde{\tau}[\tilde{\theta}])
\end{equation*}
where 
\begin{equation*}
\tilde{\sigma}[\tilde{\theta}]=\frac{b_1[\tilde{\theta}]a_1[1]}{b_1[1]a_1[\tilde{\theta}]}\ , \qquad
\tilde{\tau}[\tilde{\theta}]=\frac{b_1[\tilde{\theta}]b_2[\tilde{\theta}]a_1[1]a_2[1]}{b_1[1]b_2[1]a_1[\tilde{\theta}]a_2[\tilde{\theta}]}
\ , \qquad
\tilde{g}(\tilde{\sigma}[\tilde{\theta}])=\frac{c_1[1]b_1[\tilde{\theta}]c_2[\tilde{\theta}]a_2[1]}{c_1[\tilde{\theta}]b_1[1]c_2[1]a_2[\tilde{\theta}]}\ ,
\end{equation*}
i.e.\ expressions where the sub-lattices $2$ and $3$ have been exchanged.
By eliminating $\tilde{\theta}$, we obtain for $\tilde{\tau}$ and  $\tilde{g}$ the same relation \eqref{eq:taug} as above 
\emph{up to a simple substitution} $\mu\to -\mu$. This is nothing but the desired relations \eqref{eq:genrelbis}-\eqref{eq:taugofsigmabis}.
As a consequence of the above symmetry, 
taking for $\tilde{\sigma}$ the same parametrization as that for $\sigma$ above, this leads for $\tilde{\tau}$ and  $\tilde{g}$
to expressions identical to \eqref{eq:taugxi} with $\mu\to -\mu$. Note that the overall prefactor
\begin{equation*}
\left(\frac{a_2a_3}{t^{1/3}}\right)^{n^2}= \left(\sin\left(\frac{\lambda+3\eta-\mu}{2}\right)\sin\left(\frac{\lambda+3\eta+\mu}{2}\right)\right)^{n^2} 
\end{equation*}
is itself invariant under $\mu\to -\mu$.

\section{Enumeration of weighted Schr\"oder paths by transfer matrix}
\label{sec:trmat}
We wish to compute the partition function $Y^{20V}_{(n,L)\to (n+M,0)}=Y^{20V}_{(n,L)\to (n+M,0)}(\beta_1,\beta_2,\beta_3)$ for the escaping path, namely 
a weighted Schr\"oder path from $(n,L)$ to $(n+M,0)$ with (horizontal, diagonal and vertical) steps $(1,0)$, $(1,-1)$ and $(0,1)$ and weights corresponding to the 20V model at each vertex visited by the path. 
The parameters $\beta_1$, $\beta_2$ and $\beta_3$ denote suitably chosen weights for the first node of the path (the escape point), to be determined according to direction of the last step
before the escape point.  Note that the empty space around the path also receives a weight $\omega_0$ per empty vertex. Factoring those weights, the remaining weight is $\omega_i/\omega_0$ per vertex visited by the path, for which the local configuration corresponds to a vertex with weight $\omega_i$ in Figure \ref{fig:generalweights}. To compute $Y^{20V}_{(n,L)\to (n+M,0)}$,
we use a transfer matrix technique. We introduce the $3\times 3$ matrix $T$ with the following entries:
\begin{eqnarray*}
T&=& \quad
\left(
\begin{array}{ccc}
\begin{tikzpicture}
\clip(-0.25,-0.25)rectangle(0.25,0.25);
\draw (0,0)--(0,0.25);
\draw[red] (0,-0.25) -- (0,0) node[blue]{$\bullet$};
\end{tikzpicture} &

\begin{tikzpicture}
\clip(-0.25,-0.25)rectangle(0.25,0.25);
\draw (0,0)--(-0.25,0.25);
\draw[red] (0,-0.25) -- (0,0) node[blue]{$\bullet$};
\end{tikzpicture} & 

\begin{tikzpicture}
\clip(-0.25,-0.25)rectangle(0.25,0.25);
\draw (0,0)--(-0.25,0);
\draw[red] (0,-0.25) -- (0,0) node[blue]{$\bullet$};
\end{tikzpicture} \\

\begin{tikzpicture}
\clip(-0.25,-0.25)rectangle(0.25,0.25);
\draw (0,0)--(0,0.25);
\draw[red] (0.25,-0.25) -- (0,0) node[blue]{$\bullet$};
\end{tikzpicture} &

\begin{tikzpicture}
\clip(-0.25,-0.25)rectangle(0.25,0.25);
\draw (0,0)--(-0.25,0.25);
\draw[red] (0.25,-0.25) -- (0,0) node[blue]{$\bullet$};
\end{tikzpicture} & 

\begin{tikzpicture}
\clip(-0.25,-0.25)rectangle(0.25,0.25);
\draw (0,0)--(-0.25,0);
\draw[red] (0.25,-0.25) -- (0,0) node[blue]{$\bullet$};
\end{tikzpicture}\\

\begin{tikzpicture}
\clip(-0.25,-0.25)rectangle(0.25,0.25);
\draw (0,0)--(0,0.25);
\draw[red] (0.25,0) -- (0,0) node[blue]{$\bullet$};
\end{tikzpicture} &

\begin{tikzpicture}
\clip(-0.25,-0.25)rectangle(0.25,0.25);
\draw (0,0)--(-0.25,0.25);
\draw[red] (0.25,0) -- (0,0) node[blue]{$\bullet$};
\end{tikzpicture} & 

\begin{tikzpicture}
\clip(-0.25,-0.25)rectangle(0.25,0.25);
\draw (0,0)--(-0.25,0);
\draw[red] (0.25,0) -- (0,0) node[blue]{$\bullet$};
\end{tikzpicture}
\end{array}
\right)
\hspace{-110pt}
\begin{array}{c}
	\begin{tikzpicture}
	\clip(-0.25,-0.25)rectangle(0.25,0.25);
	\draw[red] (0,-0.25) -- (0,0) node[blue]{$\bullet$};
	\end{tikzpicture}\\ 
	
   \begin{tikzpicture}
   \clip(-0.25,-0.25)rectangle(0.25,0.25);
	\draw[red] (0.25,-0.25) -- (0,0) node[blue]{$\bullet$};
	\end{tikzpicture}\\
	
   \begin{tikzpicture}
	\clip(-0.25,-0.25)rectangle(0.25,0.25);
	\draw[red] (0.25,0) -- (0,0) node[blue]{$\bullet$};
	\end{tikzpicture}\\
\end{array}
\hspace{5.2pt}
\begin{array}{ccc}
	\begin{tikzpicture}
	\clip(-0.25,-0.25)rectangle(0.25,0.25);
	\draw (0,0.25) -- (0,0) node[blue]{$\bullet$};
	\end{tikzpicture} &
	
   \begin{tikzpicture}
	\clip(-0.25,-0.25)rectangle(0.25,0.25);
	\draw (-0.25,0.25) -- (0,0) node[blue]{$\bullet$};
	\end{tikzpicture} &
	
   \begin{tikzpicture}
	\clip(-0.25,-0.25)rectangle(0.25,0.25);
	\draw (-0.25,0) -- (0,0) node[blue]{$\bullet$};
	\end{tikzpicture}\\
	&&\\
	&&\\
	&&\\
	&&\\
	&&\\
\end{array}
\quad
=\frac{1}{\omega_0}\begin{pmatrix}
\omega_1 u & \omega_2 u & \omega_4 u\\
\omega_2 u v&\omega_3 u v & \omega_5 u v\\
\omega_4 v & \omega_5 v & \omega_6 v\end{pmatrix}
\end{eqnarray*}
Multiplication by $T$ on the left amounts to adding one extra step to paths, with the suitable 20V model normalized weights $\omega_i/\omega_0$ and some extra weights $u,u v,v$ per horizontal, diagonal
and vertical step respectively so as to keep track of the global vertical and horizontal shifts.
Then the generating function for the partition functions $Y^{20V}_{(n,L)\to (n+M,0)}$ reads: 
$${\mathcal Y}(u,v)=\sum_{L,M\geq 0} u^L v^M Y^{20V}_{(n,L)\to (n+M,0)}=1+(1 ,0 ,0)\  ({\mathbb I}-T)^{-1}\begin{pmatrix} \beta_1 u\\ \beta_2 uv\\ \beta_3 v\end{pmatrix} $$
The final state $(1,0,0)$ corresponds to a vertical final step, as in the geometry of Figure~\ref{fig:tgmethod1}. As for the initial state $(\beta_1 u ,\beta_2 uv ,\beta_3 v)^t$, 
it includes the special $\beta$ weights for the escape point depending on its local environment in the geometry at hand (in practice, the value of these weights is irrelevant as
they do not affect the large $n$ asymptotics, as shown below). The quantity ${\mathcal Y}(u,v)$ is a rational fraction ${\mathcal Y}(u,v)=g(u,v)/\Delta(u,v)$ where $g(u,v)$ is a polynomial 
which implicitly depends on the $\beta$ weights and
$$ \Delta(u,v)=\det({\mathbb I}-T)=1-\al_1 u-\al_2 v-\al_3 u v-\al_4 u^2 v-\al_5 u v^2 -\al_6 u^2v^2
$$
with
\begin{equation}
\begin{split}
\al_1&= \frac{\omega_1}{\omega_0} ,\quad \al_2=\frac{\omega_6}{\omega_0},\quad \al_3=\frac{\omega_0\omega_3+\omega_4^2-\omega_1\omega_6}{\omega_0^2}\\
\al_4&=\frac{\omega_2^2-\omega_1\omega_3}{\omega_0^2},\quad 
\al_5= \frac{\omega_5^2-\omega_6\omega_3}{\omega_0^2},\quad 
\al_6= \frac{2\omega_2\omega_4\omega_5+\omega_1\omega_6\omega_3-\omega_3\omega_4^2-\omega_1\omega_5^2-\omega_6\omega_2^2}{\omega_0^3}\ .\\
\end{split}
\label{eq:alphas}
\end{equation}
The large $n,L=\ell \,n, M=m\, n$ asymptotics of $Y^{20V}_{(n,L)\to (n+M,0)}$ are governed by $ \Delta(u,v)$ only. Indeed,
for large $L,M\propto n$:
\begin{eqnarray*}
&&Y^{20V}_{(n,L)\to (n+M,0)}=\oint \frac{du}{2{\rm i}\pi u^{L+1}} \frac{dv}{2{\rm i}\pi v^{M+1}} {\mathcal Y}(u,v)\\
&\propto& 
\oint \frac{du}{2{\rm i}\pi u^{L+1}} \frac{dv}{2{\rm i}\pi v^{M+1}} g(u,v) \sum_{p\geq 0} (\al_1 u+\al_2 v+\al_3 u v+\al_4 u^2 v+\al_5 u v^2+\al_6 u^2v^2)^p\\
&=& \oint \frac{du}{2{\rm i}\pi u^{L+1}} \frac{dv}{2{\rm i}\pi v^{M+1}} g(u,v) \, \sum_{L',M'}  u^{L'} v^{M'}\!\!\!\!\!\!\!\!\!\!\!\!\!\!\!\!\!\! \sum_{P_1,P_2,\dots,P_6\geq 0 \atop {P_1+P_3+2P_4+P_5+2P_6=L' \atop 
P_2+P_3+P_4+2P_5+2P_6=M'} } {P_1+P_2+P_3+P_4+P_5+P_6\choose P_1,P_2,P_3,P_4,P_5,P_6} 
\prod_{i=1}^6 \al_i^{P_i} \ .
\end{eqnarray*}
For large $n$, the integral selects values of $L'$ and $M'$ that differ from $L$ and $M$ by finite amounts bounded by the degree of 
the polynomial $g(u,v)$. Taking $L'=\ell\, n +O(1)$ and $M'=m\, n+O(1)$,  we obtain the leading behavior for large $P_i= n p_i$:
$$Y^{20V}_{(n,n \ell)\to (n(1+m),0)}\propto \int_0^1 dp_3dp_4dp_5dp_6  {\rm e}^{n\, S(\ell,m,p_3,p_4,p_5,p_6)}$$
where
\begin{equation}
\begin{split}
S(\ell,m,p_3,p_4,p_5,p_6)&=(\ell+m-p_3-2p_4-2p_5-3p_6){\rm Log}(\ell+m-p_3-2p_4-2p_5-3p_6)\\
&-(\ell-p_3-2p_4-p_5-2p_6){\rm Log}\left(\frac{\ell-p_3-2p_4-p_5-2p_6}{\al_1}\right)\\
&-(m-p_3-p_4-2p_5-2p_6){\rm Log}\left(\frac{m-p_3-p_4-2p_5-2p_6}{\al_2}\right)-\sum_{i=3}^6 p_i{\rm Log}\left(\frac{p_i}{\al_i}\right)\ . 
\label{eq:explS}
\end{split}
\end{equation}
A saddle point estimate then allows to write
$$Y^{20V}_{(n,n \ell)\to (n(1+m),0)}\propto   {\rm e}^{n\, S(\ell,m)}$$
where $S(\ell,m)$ is equal to $S(\ell,m,p_3,p_4,p_5,p_6)$ taken at the value of $p_3$, $p_4$, $p_5$ and $p_6$ which maximizes this latter quantity.

The above expression can be adapted to a situation where the $\alpha_i$ for $i\in I\subset \{3,\cdots,6\}$ vanish. In this case, the recipe consists
in simply dropping all the terms with indices $i\in I$.

Finally, the expression \eqref{eq:explS} can also be used to compute the asymptotics of the escape path partition function in the ``shear'' geometry,
i.e. the function $\bar{S}(\ell,m)$ defined via
$$\bar{Y}^{20V}_{(n,n+L-1)\to (n+M,2n)}\propto {\rm e}^{n\, \bar{S}(\ell,m)}\ .$$
Indeed, the change of geometry, which corresponds to an up-down symmetry together with a change of weights from the original values
$\omega_i$ to the inverted 20V model values $\bar{\omega}_i$ of Figure~\ref{fig:shearweightsgen} is entirely accounted for by changing in
the above expression for $S(\ell,m,p_3,p_4,p_5,p_6)$ both $\ell$ into $1-\ell$ 
and $\alpha_i$ into $\bar{\alpha}_i$, where the $\alpha_i$'s are obtained via \eqref{eq:alphas} with $\omega_i$ changed into  $\bar{\omega}_i$. 
From the equivalence Figure~\ref{fig:shearweightsgen} , the $\bar{\alpha}_i$'s are alternatively obtained from the $\alpha_i$'s by exchanging the
role of $\omega_0$ and $\omega_1$, and simultaneously that of $\omega_2$ and $\omega_4$, keeping $\omega_3$, $\omega_5$ and $\omega_6$ unchanged.
In terms of angular parameters, this change corresponds to the involution 
\begin{equation*}
(\eta,\lambda,\mu) \leftrightarrow \left(\eta, \pi-\frac{\lambda+\eta+\mu}{2},\pi-\frac{3\lambda+\eta-\mu}{2}\right)\ .
\end{equation*}
\bigskip

\bibliographystyle{amsalpha} 

\noindent{\bf Acknowledgments.}

\noindent  
BD acknowledges the financial support
of the Fonds Sp\'eciaux de Recherche (FSR) of the Universit\'e catholique de Louvain, 
the Fonds de la Recherche Scientifique (FNRS) and the Fonds Wetenschappelijk Onderzoek - Vlaanderen
(FWO) under EOS project no 30889451.
PDF is partially supported by the Morris and Gertrude Fine endowment and the NSF grant DMS18-02044. EG acknowledges the support of the grant ANR-14-CE25-0014 (ANR GRAAL).

\bibliography{20Varctic}

\providecommand{\bysame}{\leavevmode\hbox to3em{\hrulefill}\thinspace}
\providecommand{\MR}{\relax\ifhmode\unskip\space\fi MR }
\providecommand{\MRhref}[2]{%
  \href{http://www.ams.org/mathscinet-getitem?mr=#1}{#2}
}
\providecommand{\href}[2]{#2}
\begin{thebibliography}{BDFZJ13}

\bibitem[Agg18]{Agg}
Amol Aggarwal, \emph{Arctic boundaries of the ice model on three-bundle
  domains}, Preprint (2018), arXiv:1812.03847 [math.PR].

\bibitem[AR05]{Allison2005}
David Allison and Nicolai Reshetikhin, \emph{Numerical study of the 6-vertex
  model with domain wall boundary conditions}, Annales de l'Institut Fourier
  \textbf{55} (2005), 1847--1869, arXiv:cond-mat/0502314 [cond-mat.stat-mech].

\bibitem[Bax89]{Baxter}
Rodney~J. Baxter, \emph{Exactly solved models in statistical mechanics},
  Academic Press, Inc. [Harcourt Brace Jovanovich, Publishers], London, 1989,
  Reprint of the 1982 original. \MR{998375}

\bibitem[BDFZJ12]{BDFPZ1}
Roger~E. Behrend, Philippe Di~Francesco, and Paul Zinn-Justin, \emph{On the
  weighted enumeration of alternating sign matrices and descending plane
  partitions}, J. Combin. Theory Ser. A \textbf{119} (2012), no.~2, 331--363,
  arXiv:1103.1176 [math.CO]. \MR{2860598}

\bibitem[BDFZJ13]{BDFPZ2}
\bysame, \emph{A doubly-refined enumeration of alternating sign matrices and
  descending plane partitions}, J. Combin. Theory Ser. A \textbf{120} (2013),
  no.~2, 409--432, arXiv:1202.1520 [math.CO]. \MR{2995049}

\bibitem[BL15]{BL14}
Pavel Bleher and Karl Liechty, \emph{Six-vertex model with partial domain wall
  boundary conditions: Ferroelectric phase}, Journal of Mathematical Physics
  \textbf{56} (2015), no.~2, 023302.

\bibitem[CEP96]{CEP}
Henry Cohn, Noam Elkies, and James Propp, \emph{Local statistics for random
  domino tilings of the aztec diamond}, Duke Math. J. \textbf{85} (1996),
  no.~1, 117--166, arXiv:math/0008243 [math.CO].

\bibitem[CGP14]{Cugliandolo2015}
Leticia Cugliandolo, Giuseppe Gonnella, and Alessandro Pelizzola, \emph{Six
  vertex model with domain-wall boundary conditions in the bethe-peierls
  approximation}, Journal of Statistical Mechanics: Theory and Experiment
  \textbf{2015} (2014), 06008, arXiv:1501.00883 [cond-mat.stat-mech].

\bibitem[CP10a]{CP2009}
Filippo Colomo and Andrei~G. Pronko, \emph{The arctic curve of the domain-wall
  six-vertex model}, Journal of Statistical Physics \textbf{138} (2010), no.~4,
  662--700, arXiv:0907.1264 [math-ph].

\bibitem[CP10b]{CP2010}
\bysame, \emph{The limit shape of large alternating sign matrices}, SIAM J.
  Discrete Math. \textbf{24} (2010), no.~4, 1558--1571, arXiv:0803.2697
  [math-ph].

\bibitem[CPZJ10]{CPZ}
Filippo Colomo, Andrei~G. Pronko, and Paul Zinn-Justin, \emph{The arctic curve
  of the domain wall six-vertex model in its antiferroelectric regime}, J.
  Stat. Mech. Theory Exp. (2010), no.~3, L03002, 11, arXiv:1001.2189 [math-ph].
  \MR{2629602}

\bibitem[CS16]{COSPO}
Filippo Colomo and Andrea Sportiello, \emph{Arctic curves of the six-vertex
  model on generic domains: the tangent method}, J. Stat. Phys. \textbf{164}
  (2016), no.~6, 1488--1523, arXiv:1605.01388 [math-ph].

\bibitem[DFG18]{DFGUI}
Philippe Di~Francesco and Emmanuel Guitter, \emph{Arctic curves for paths with
  arbitrary starting points: a tangent method approach}, J. Phys. A: Math.
  Theor. \textbf{51} (2018), no.~35, 355201, arXiv:1803.11463 [math-ph].

\bibitem[DFG19a]{DFG3}
\bysame, \emph{The arctic curve for aztec rectangles with defects via the
  tangent method}, Journal of Statistical Physics \textbf{176} (2019), no.~3,
  639--678, arXiv:1902.06478 [math-ph].

\bibitem[DFG19b]{DFG2}
\bysame, \emph{A tangent method derivation of the arctic curve for q-weighted
  paths with arbitrary starting points}, Journal of Physics A: Mathematical and
  Theoretical \textbf{52} (2019), no.~11, 115205, arXiv:1810.07936 [math-ph].

\bibitem[DFG19c]{DFG20V}
\bysame, \emph{Twenty-vertex model with domain wall boundaries and domino
  tilings}, Preprint (2019), arXiv:1905.12387 [math.CO].

\bibitem[DFL18]{DFLAP}
Philippe Di~Francesco and Matthew~F. Lapa, \emph{Arctic curves in path models
  from the tangent method}, J. Phys. A: Math. Theor. \textbf{51} (2018),
  155202, arXiv:1711.03182 [math-ph].

\bibitem[DFSG14]{DFSG}
Philippe Di~Francesco and Rodrigo Soto-Garrido, \emph{Arctic curves of the
  octahedron equation}, J. Phys. A \textbf{47} (2014), no.~28, 285204, 34,
  arXiv:1402.4493 [math-ph]. \MR{3228361}

\bibitem[DGR19]{DGR}
Bryan Debin, Etienne Granet, and Philippe Ruelle, \emph{Concavity analysis of
  the tangent method}, Preprint (2019), arXiv:1905.11277 [math-ph].

\bibitem[DR18]{DR}
Bryan Debin and Philippe Ruelle, \emph{Tangent method for the arctic curve
  arising from freezing boundaries}, 2018, arXiv:1810.04909 [math-ph].

\bibitem[Joh03]{johansson2005}
Kurt Johansson, \emph{The arctic circle boundary and the airy process}, Annals
  of Probability \textbf{33} (2003), 1--30, arXiv:math/0306216 [math.PR].

\bibitem[JPS98]{JPS}
William Jockusch, James Propp, and Peter Shor, \emph{Random domino tilings and
  the arctic circle theorem}, arXiv:math/9801068 [math.CO] (1998).

\bibitem[{Kel}74]{Kel}
Stewart~B. {Kelland}, \emph{{Twenty-vertex model on a triangular lattice}},
  Australian Journal of Physics \textbf{27} (1974), 813--829.

\bibitem[KL17]{Keesman2017}
Rick Keesman and Jules Lamers, \emph{Numerical study of the $f$ model with
  domain-wall boundaries}, Phys. Rev. E \textbf{95} (2017), 052117,
  arXiv:1702.05474 [cond-mat.stat-mech].

\bibitem[KO06]{KO1}
Richard Kenyon and Andrei Okounkov, \emph{Planar dimers and harnack curves},
  Duke Math. J \textbf{131} (2006), no.~3, 499--524, arXiv:math/0311062
  [math.AG].

\bibitem[KO07]{KO2}
\bysame, \emph{Limit shapes and the complex burgers equation}, Acta Math.
  \textbf{199} (2007), no.~2, 263--302, arXiv:math-ph/0507007.

\bibitem[KOS06]{KOS}
Richard Kenyon, Andrei Okounkov, and Scott Sheffield, \emph{Dimers and
  amoebae}, Ann. Math. \textbf{163} (2006), 1019--1056, arXiv:math/0311062
  [math.AG].

\bibitem[KP13]{KP}
Richard Kenyon and Robin Pemantle, \emph{Double-dimers, the {I}sing model and
  the hexahedron recurrence}, 25th {I}nternational {C}onference on {F}ormal
  {P}ower {S}eries and {A}lgebraic {C}ombinatorics ({FPSAC} 2013), Discrete
  Math. Theor. Comput. Sci. Proc., AS, Assoc. Discrete Math. Theor. Comput.
  Sci., Nancy, 2013, arXiv:1308.2998 [math-ph], pp.~109--120. \MR{3090984}

\bibitem[Kra06]{KrattDPP}
Christian Krattenthaler, \emph{Descending plane partitions and rhombus tilings
  of a hexagon with a triangular hole}, European J. Combin. \textbf{27} (2006),
  no.~7, 1138--1146, arXiv:math/0310188 [math.CO]. \MR{2259946}

\bibitem[LKRV18]{Lyberg2017b}
Ivar Lyberg, Vladimir Korepin, G.~A.~P. Ribeiro, and Jacopo Viti, \emph{Phase
  separation in the six-vertex model with a variety of boundary conditions},
  Journal of Mathematical Physics \textbf{59} (2018), no.~5, 053301,
  arXiv:1711.07905 [cond-mat.stat-mech].

\bibitem[LKV17]{Lyberg2017a}
Ivar Lyberg, Vladimir Korepin, and Jacopo Viti, \emph{The density profile of
  the six vertex model with domain wall boundary conditions}, Journal of
  Statistical Mechanics: Theory and Experiment \textbf{2017} (2017), no.~5,
  053103, arXiv:1612.06758 [cond-mat.stat-mech].

\bibitem[SZ04]{Syljuasen2004}
Olav~F. Sylju\aa{}sen and M.~B. Zvonarev, \emph{Directed-loop monte carlo
  simulations of vertex models}, Phys. Rev. E \textbf{70} (2004), 016118,
  arXiv:cond-mat/0401491 [cond-mat.stat-mech].

\end{thebibliography}

\end{document}